\documentclass[11 pt]{article}
\usepackage[mathscr]{eucal}
\usepackage{amssymb,latexsym}
\usepackage{verbatim}

\usepackage{amsmath}
\usepackage{amsthm}
\usepackage{enumerate}
\usepackage{authblk}
\usepackage{color}
\usepackage{multicol}
\usepackage{tikz}
\usepackage{xypic}
\usepackage{caption}
\usepackage{cite}
\usepackage[makeroom]{cancel}

\newtheorem{thm}{Theorem}[section]
\newtheorem{lem}[thm]{Lemma}
\newtheorem{Def}[thm]{Definition}
\newtheorem{prop}[thm]{Proposition}

\renewcommand\l{\lambda}

\newcommand\Om{\Omega}

\newcommand\bbN{{\mathbb N}}
\newcommand\bbR{{\mathbb R}}
\newcommand\bbZ{{\mathbb{Z}}}

\newcommand\wt{\widetilde}

\renewcommand\Om{\Omega}
\newcommand\s{\sigma}

\newcommand\e{\varepsilon}
\renewcommand\b{\beta}

\renewcommand\l{\lambda}
\newcommand\g{\gamma}

\renewcommand\a{\alpha}

\renewcommand\th{\theta}

\newcommand\beq{\begin{equation}}
\newcommand\eeq{\end{equation}}
\newcommand\ben{\begin{enumerate}}
\newcommand\een{\end{enumerate}}
\newcommand\bit{\begin{itemize}}
\newcommand\eit{\end{itemize}}

\newcommand{\al} {\alpha}

\newcommand{\ga} {\gamma}

\newcommand{\ka} {\kappa}
\newcommand{\la} {\lambda}

\newcommand{\si} {\sigma}

\newcommand{\alSo} {\alpha_{\mbox{\textrm{\tiny{S}}}}}

\newcommand{\R}{\mathbb R}

\newcommand{\pd}{\partial}

\newcommand{\mc}{\mathcal}

\newcommand{\Z}{\mathbb{Z}}
\newcommand{\C}{\mathbb{C}}

\newcommand{\fa}{\mathfrak{a}}
\newcommand{\bq}{\mathbf{q}}
\newcommand{\bA}{\mathbf{A}}
\newcommand{\bF}{\mathbf{F}}

\newcommand{\bS}{\mathbf{S}}
\newcommand{\bL}{\mathbf{L}}
\newcommand{\bg}{\mathbf{g}}
\newcommand{\fh}{\mathfrak{h}}
\newcommand{\bAKN}{\mathbf{A}_{\textsc{KN}}}
\newcommand{\bAKNanom}{\mathbf{A}_{\textsc{KN}}^{\mbox{\tiny\textrm{gen}}}}

\newcommand{\cC}{\mathcal{C}}

\newcommand{\fD}{\mathfrak{D}}

\newcommand{\Sset}{\mathbb{S}}

\newcommand{\om}{\omega}
\newcommand{\De}{\varpi}
\newcommand{\half}{\frac{1}{2}}
\newcommand{\sgn}{\mbox{sgn}}

\newcommand{\bna}{\begin{eqnarray}}
\newcommand{\ena}{\end{eqnarray}}

\def\undertilde#1{\mathord{\vtop{\ialign{##\crcr
   $\hfil\displaystyle{#1}\hfil$\crcr\noalign{\kern1.5pt\nointerlineskip}
   $\hfil\tilde{}\hfil$\crcr\noalign{\kern1.5pt}}}}}

\newcounter{mnotecount}

\title{On the discrete Dirac spectrum of a point electron \\ in the zero-gravity Kerr-Newman spacetime}

\author{\normalsize\sc{Michael K.-H. Kiessling, Eric Ling, and A. Shadi Tahvildar-Zadeh}
\\ 
       \normalsize Department of Mathematics\\
	Rutgers, The State University of New Jersey\\
	110 Frelinghuysen Rd., Piscataway, NJ 08854
}
\vspace{-0.3cm}
\date{ \today}

\begin{document}
\date{}
\maketitle

\begin{abstract}
 The discrete spectrum of the Dirac operator for a point electron in the maximal analytically extended Kerr--Newman
spacetime is determined in the zero-$G$ limit (z$G$KN), under some restrictions on the electrical coupling constant and on the
radius of the ring-singularity of the z$G$KN spacetime.
 The spectrum is characterized by a triplet of integers, associated with winding numbers of orbits of dynamical systems on cylinders.
 A dictionary is established that relates the spectrum with the known hydrogenic Dirac spectrum. 
 Numerical illustrations are presented.
 Open problems are listed.
\end{abstract}

\newpage

\tableofcontents

\newpage

\section{Introduction}
   The discovery, by Chandrasekhar \cite{Chandra76a,Chandra76b}, Page \cite{Page76}, and Toop \cite{Toop76}, of the separability of
the Dirac equation for a charged spin-$\frac12$ point particle in the Kerr--Newman spacetimes has paved the ground for in depth studies 
of its solution properties; see 
\cite{BSa},
\cite{BSb},
\cite{BSW},
\cite{BelMar99},
\cite{ChandraBHbook}, 
\cite{FKSYinCPAM2000a},
\cite{FKSYinCPAM2000b},
\cite{FKSYinCMP2002},
\cite{FKSYinATMP2003},
\cite{GairPRIZE},
\cite{KalMil91},
\cite{KTZzGKNDa},
\cite{KTZzGKNDb},
\cite{KTZzGKNDc},
\cite{Schmid},
\cite{SufFacCos83},
\cite{WinYamA},
\cite{WinYamB}.
   Yet, despite such an impressively long list of studies, most of them rigorous, we are far away from what could be called a complete understanding.
   Namely, unlike the separable hydrogenic Dirac problem in Minkowski spacetime, for which closed form expressions of its general solution
and its discrete Sommerfeld fine-structure spectrum are available in terms of well-known functions, the system of ordinary differential equations 
(ODEs) into which the Dirac problem in Kerr--Newman spacetime separates does not seem to be completely solvable in closed form with the 
help of known functions.
   Only the dependence on the variables associated with the Killing vector fields $\partial_t$ 
for stationarity (outside the ergosphere) and $\partial_\varphi$ for axisymmetry, respectively, 
has the simple form $\exp(-i[\frac1\hbar Et - (k+\frac12)\varphi])$ in the separation product Ansatz, with $k\in\Z$ and $E\in\bbR$,
while the dependence on the remaining two spacelike variables requires careful discussions of the two pertinent ODEs. 

   Of these two, the sometimes so-called Chandrasekhar--Page angular equation depends parametrically on $E$, $k$, on an angular
momentum eigenvalue $\lambda$, and on the radius $a>0$ of the ring singularity of the Kerr--Newman spacetime, but not on
 its mass $M$ and charge $Q$.
   Thus its solution properties do not depend on whether the Dirac problem is considered on 
a Kerr--Newman black hole spacetime, or on the Kerr--Newman spacetime with a naked singularity, or on the analogous sectors of the Kerr spacetime 
family ($Q=0$).
 It has been very thoroughly and rigorously discussed in \cite{BSW}.
 There it was shown that for each $k\in\Z$ and $E\in\bbR$ there are countably many simple eigenvalues $\lambda$, and they depend analytically on $E$.
 And while the Chandrasekhar--Page angular equation is generally not solvable in terms of elementary or known special functions, it is worth 
mentioning that in the limit $a\to 0$ its square-integrable solutions are given by elementary functions of the angle variable $\theta$; 
see Appendix A of \cite{BSW}.

   The remaining so-called radial equation depends parametrically on $E$, $k$, and $\lambda$, so in concert with the $E$-dependence of $\lambda$ obtained 
in \cite{BSW} from the Chandrasekhar--Page angular equation, the radial equation determines which $E$ values are part of the spectrum of any pertinent 
self-adjoint extension of the Dirac hamiltonian.
   In contrast to the angular equation, the radial equation does depend parametrically not only on the radius 
$a$ of the ring singularity of the Kerr--Newman spacetime but also on its mass $M$ and charge $Q$, 
and so needs to be discussed separately for the black hole and for the naked singularity sectors. 
   The self-adjointness and associated spectral properties of the Dirac hamiltonian generally depend on which sector one considers.

   Rigorous and very thorough discussions of this radial equation obtained from the Dirac equation for bi-spinors supported on the positive mass region 
exterior to the event horizon of any Kerr--Newman black hole spacetime can be found in \cite{FKSYinATMP2003} and \cite{BSa}.
  There it was shown that Chandrasekhar's separation Ansatz yields a complete set of (generalized) eigenfunctions for the Hilbert space of bi-spinor wave functions
on that region of spacetime. 
   In particular, it was shown that this Dirac hamiltonian is essentially self-adjoint and that its essential $E$-spectrum consists of the whole real line, so its 
discrete $E$-spectrum is empty. 
   Interestingly, while in \cite{BSa} it was shown that in the subextremal black hole sector there are no eigenvalues whatsoever, in \cite{Schmid} it was 
found that on the borderline in parameter space between the black hole and naked singularity sectors, which features the extremal black hole
spacetimes, there are exceptional combinations of the mass and the angular momentum of the Kerr--Newman spacetime, given the 
mass and charge of the Dirac particle, for which there is a unique eigenvalue of the Dirac hamiltonian on the exterior spacetime of the extremal black hole, 
which is embedded in the continuum of the essential spectrum. 
It is part of the pure point spectrum, but obviously not part of the discrete spectrum.

   The Dirac equation for any spin-$\frac12$ particle inside the event horizon of the Kerr--Newman black hole spacetime is a very different problem, due to the 
 existence of closed timelike loops, a Cauchy horizon, and a timelike cylindrical singularity \cite{Car68}.
   In the spacetime of the naked singularity sector, 
 the event and Cauchy horizons are absent, but the closed timelike loops around the singularity persist \cite{Car68}.
   We are not aware of any definitive discussion of the Dirac equation in such problematic spacetimes. 
 
   However, any horizon and the closed timelike loops vanish in the zero-gravity (zero-$G$) limit of the maximal analytically extended Kerr--Newman spacetime.
   Its zero-$G$ limit yields a flat, static, double-sheeted spacetime with Zipoy topology \cite{Zipoy} supporting static electromagnetic Appell--Sommerfeld fields 
 \cite{Appell}, \cite{Som97} sourced by a sesqui-polar naked ring singularity; see \cite{TZzGKN} for a detailed discussion.
   We will refer to this double-sheeted electromagnetic spacetime as the z$G$KN spacetime.
   The Dirac equation for a charged spin-$\frac12$ particle in this z$G$KN spacetime was studied in \cite{KTZzGKNDa}, \cite{KTZzGKNDb}, and 
\cite{KTZzGKNDc}.
   In particular, in \cite{KTZzGKNDa} we showed that the Dirac hamiltonian associated with the Dirac equation for a charged spin-$\frac12$ particle of 
mass $m>0$ and charge $-e<0$ in the z$G$KN spacetime with charge parameter $Q\in \bbR \backslash\{0\}$ and dimensionless ring radius 
$a mc/\hbar =:\varsigma>0$ is essentially self-adjoint for all values of the dimensionless coupling constant $|Q|e/\hbar c \equiv Z\alSo =: \gamma$, 
where $\alSo$ is Sommerfeld's fine structure constant. 
   Furthermore, we showed that the spectrum is symmetric about zero.
   Moreover, we proved that its essential spectrum is $(-\infty,-mc^2]\cup[mc^2,\infty)$, with absolutely continuous interior $(-\infty,-mc^2)\cup(mc^2,\infty)$,
and that under the smallness conditions $\varsigma <\frac12$ and $\gamma < \sqrt{2\varsigma(1-2\varsigma)}$ 
a non-empty point spectrum exists in the gap $(-mc^2,mc^2)$.
 
  In this paper we study the discrete spectrum of the Dirac hamiltonian for a point charge on z$G$KN in more detail. 
   Compared to the conditions used in \cite{KTZzGKNDa}, we here use
the significantly weaker smallness condition $\gamma <\frac12$ on the electric coupling constant, paired with the slightly stronger 
smallness condition $\varsigma < 1-1/\sqrt{2}$ on the radius of the ring singularity, and
prove constructively that there are infinitely many energy eigenvalues $E$  in the gap $(-mc^2,mc^2)$ of the essential spectrum, 
each one of finite multiplicity.
  Our proof consists of constructing the pairs $(\lambda,E)$ jointly
 with the help of the coupled system of two dynamical evolution equations on finite cylinders that we introduced in \cite{KTZzGKNDa}.
  We show that these pairs $(\lambda,E)$ can be labeled with the topological winding numbers of particular pairs of solutions that ``wind around''
these two cylinders, as explained in detail in the next section.
  There we formulate the problem precisely and state our main results, which will then be proved in the ensuing sections. 
  Moreover, our constructive proof yields an algorithm that we have used to compute several eigenvalues for 
various combinations of the winding numbers of the solution pairs, as functions of the coupling constant and of the ring radius $a$.
  These numerical illustrations, which include some intriguing oscillatory structures not previously seen in Dirac spectra, 
are presented in section \ref{num}.

   Before coming to the detailed presentation of our mathematical results and their numerical illustration, we take the opportunity to 
offer a few thoughts concerning possible physical interpretations. 
   
    One way of looking at the Dirac equation for a point electron on z$G$KN is as a proxy model for hydrogen (or hydrogenic ions) that in 
addition to the charge also takes the magnetic moment of the proton (nucleus), and its finite size, into account in some effective manner. 
    The fact that these structures are a consequence of general relativity theory makes such an investigation less 
add-hoc than postulating some arbitrary structure for a finite size proton (nucleus) with electric current and charge densities.
  In this interpretation, different nuclei would not only be distinguished by their charge number $Z\in\bbN$, but also would each require
  a ring radius $a(Z)$ to match their magnetic moments.

  Incidentally, studies of Dirac's equation on the naked singularity sector of the Kerr--Newman spacetime in the zero-gravity limit as just such a proxy model
to hydrogen that takes the magnetic moment of the proton into account were previously attempted in \cite{Pek87} and \cite{GairPRIZE}. 
 However, different from our study these authors studied Dirac's equation on a \emph{one-sheeted truncation} of the z$G$KN spacetime; cf.
\cite{Israel}, \cite{LB}.
  This ad-hoc truncation creates an artificial ``disc singularity'' with non-integrable charge and current ``densities'' \cite{Kaiser}.\footnote{In the astrophysical 
context, integrable Kerr--Newman disc sources have been constructed in \cite{LedZofBic98}; however, these discs are not localized but infinitely extended.}
 Not only are these disc sources highly singular (cf. \cite{Kaiser}), the Dirac operator on such a singular spacetime is not essentially self-adjoint
but requires boundary conditions for the Dirac bi-spinors at the disc.
   By contrast, since the Dirac operator on the full z$G$KN spacetime is essentially self-adjoint and has a comparatively simple ring singularity with
distributional electric and magnetic sesqui-poles, no issues with ambiguous boundary conditions arise. 

   Another way of looking at the Dirac equation on z$G$KN is obtained by changing the perspective and to consider the point charge as the proton
(or nucleus) without magnetic moment, while thinking of the ring singularity as representing \emph{both} the electron (``seen'' in one sheet of the z$G$KN 
spacetime) \emph{and} the positron (``seen'' in the other), equipped with their anomalous magnetic moment, respectively,
which in magnitude is to very good precision given by $1/4\pi$ times the product of the 
elementary charge $e$ with the so-called ``classical radius of the electron,'' $e^2/mc^2$; see \cite{KTZzGKNDc}. 
 More precisely, in \cite{KTZzGKNDc} we noted  that the Bohr magneton, which is $1/4\pi$ times the product of the 
elementary charge $e$ with the Compton wave length of the electron, $h/mc$, features implicitly in Dirac's operator 
already for a point electron in external electromagnetic fields,  hence also in the Dirac operator for the z$G$KN  ``ring electron-positron" model,
so that the magnetic moment that is associated with the ring singularity has to be seen as an additional contribution, i.e.  the 
\emph{anomalous magnetic moment of the electron (positron)}.
 
 Since numerically the empirical magnetic moment of the proton and the anomalous magnetic moment of the electron are quite close in magnitude,
 the Dirac spectra for a z$G$KN ring singularity interacting with a point charge should only barely be distinguishable in the two interpretations of the
 ring singularity for such hydrogen parameters. 
 Moreover, for hydrogen parameters we expect (see below for why) that the discrete spectrum will be a small perturbation of the union of a positive and a negative 
Sommerfeld hydrogen energy spectrum, though with some of the high degeneracies of the Sommerfeld spectrum now resolved. 
  This resolution should result in hyperfine-\emph{like} and Lamb shift-\emph{like} effects, and it would be interesting to see whether for some choice 
of $a$ (in particular the one corresponding to the anomalous magnetic moment of the electron) these resemble the known effects in the empirical 
hydrogen spectrum at least qualitatively and to some extent also quantitatively, or not at all. 
  This will require extensive and highly accurate numerical evaluations that we plan to carry out in the future.
 
 The reason for why we expect a small perturbation of the union of a positive and a negative Sommerfeld hydrogen energy spectrum is that with 
$a$ matched to either the anomalous magnetic moment of the electron or the magnetic moment of the proton, the ring radius is about five orders
of magnitude smaller than the expected distance of electron and proton in the hydrogen ground state.
 Therefore the spectrum should not deviate much from its $a\to 0$ limit (assuming the limit exists).
 It was pointed out already in \cite{KTZzGKNDa} and \cite{KTZzGKNDb} that formally in the limit 
$a\to 0$ (which also switches off the z$G$KN magnetic moment) the problem decomposes into two independent copies of the familiar textbook 
Dirac hydrogen problem in Minkowski spacetime of a point proton without magnetic moment, except that in one of the two copies the proton
carries a positive charge and in the other a negative one (representing an anti-proton).
 Therefore,  in line with our result \cite{KTZzGKNDa} that the whole spectrum of the Dirac hamiltonian on z$G$KN is symmetric about $E=0$, 
the formal $a\to 0$ limit of the discrete spectrum would be a union of the well-known Sommerfeld fine-structure spectrum with a \emph{negative} copy thereof.
  We expect that this formal limit $a\to 0$ can be made entirely rigorous (and hope to present such a rigorous proof in some future publication). 
 
  As to the hydrogenic problem, the interpretation of the ring singularity as an electron-positron bi-particle has the simplifying advantage that
one can keep the ring radius $a$ and its charge fixed while changing only the charge of the point particle with which the ring singularity interacts to $Ze$. 
 With increasing $Z$ the expected distance between electron and nucleus in a hydrogenic ion shrinks $\propto 1/Z$, so that one should be inclined to
 expect that for all stable nuclei (i.e. $Z\leq 82$, except $Z=43$ and $Z=61$) the positive discrete Dirac spectrum is always only a small perturbation 
 of the hydrogenic Sommerfeld fine structure spectrum, with some degeneracies removed. 
 
  However, it is also of interest to let $Z$ grow beyond the range of stable nuclei (i.e. $Z\leq 82$); in fact, current efforts to create artificial nuclei in
  heavy ion collisions have reached $Z= 118$ and are hoped to go even beyond $137$.
   Of course, for large $Z$ also the size of the nucleus will no longer have only a small perturbative effect, so our study 
 of the hydrogenic z$G$KN spectra with large $Z$ point nuclei should not be seen as an attempt to accurately compute actual physical spectra, but 
 to compare the spectra with their Minkowski space counterparts.
   
   In \cite{KTZzGKNDa} we showed that this z$G$KN Dirac hamiltonian is essentially self-adjoint for all $Z>0$.
   By comparison, the Dirac hamiltonian of a point electron that interacts with a point nucleus in Minkowski spacetime is not essentially 
self-adjoint for $Z\alSo> \frac12\sqrt{3}$ and ceases to have a self-adjoint analytical extension for $Z\alSo>1$.
 Thus, when approaching 
$Z\alSo =1$ from the left,  the pertinent positive part of the discrete Dirac spectrum of a z$G$KN singularity in interaction with a point charge should deviate
from the Sommerfeld spectrum in a more and more pronounced way, and indeed this is what we found numerically.

 Also of interest in this hydrogenic setting is the comparison of the Dirac spectrum of a z$G$KN singularity in interaction with a point charge and
 the hydrogenic Dirac spectrum computed with two interacting point charges in Minkowski spacetime, one of which is given an anomalous 
 magnetic moment, cf. \cite{ThallerBOOK}.
  This Dirac operator on Minkowski spacetime is essentially self-adjoint for all $Z$, and thus allows an unambiguous comparison of its spectrum with 
  the spectrum of the Dirac operator on z$G$KN also for $Z\alSo >1$.
  What we found numerically was completely unexpected. 
   The z$G$KN Dirac spectrum shows oscillations when $Z$ is increased beyond $137$, not seen in the spectrum of the other Dirac operator, or
   any other Dirac operator so far --- to the best of our knowledge.
  
  It is very tempting to speculate whether these oscillations could indicate some real physical effects.
   The best we could come up with is by resorting to Dirac's interpretation of the negative energies. 
    Since the spectrum of the Dirac operator of a point particle on z$G$KN is symmetric about $E=0$, its oscillatory character means that the gap between
    positive and negative spectrum is oscillating with $Z$. 
     This could mean that in heavy ion collisions that temporarily create super-heavy nuclei, the amount of electron-positron pairs produced in such a 
     collision could perhaps vary in an oscillatory way with $Z$. 
  This should be noticeable in coincidence detection of $\gamma$ photon pairs with energies far below the electron rest energy (per photon), yet with an
  energy per photon that oscillates with $Z$. 
  
   We are now ready to report our mathematical and numerical results. 
   
   \section{Statement of problem and main result}\label{allTHEREis}
 \subsection{The z$G$KN spacetime and the Dirac equation}\label{zGKND}
 The electromagnetic zero-$G$ Kerr--Newman spacetime is static, double-sheeted with Zipoy topology \cite{Zipoy}, having
metric and four-potential (written in oblate spheroidal coordinates)
\beq
ds^2_{\bg} = c^2dt^2 - (r^2 + a^2)\sin^2\theta d\varphi^2 - \frac{r^2 + a^2\cos^2\theta}{r^2 + a^2}\big(dr^2 + (r^2 + a^2)d\theta^2\big) ,
\eeq
\beq\label{def:AKN}
\bAKN = - \frac{r}{r^2+a^2\cos^2\theta} \left(\textsc{q}dt - \frac{\textsc{q}a}{c} \sin^2\theta\, d\varphi\right).
\eeq
 Here, $t\in \bbR$, $r\in\bbR$, $\varphi\in[0,2\pi)$, and $\theta\in [0,\pi]$.
 The spacetime thus comes equipped with the electromagnetic field $\bF_{\textsc{KN}} =d\bAKN$ of the Kerr--Newman spacetime family, independently 
 of $G$ and therefore surviving intact in the zero-$G$ limit.
  
 Each spacelike slice of the z$G$KN spacetime is a double-sheeted Sommerfeld  space,
obtained by gluing two copies of (oriented) $\R^3$ along two congruous  circular disks, in such a way that the top of one disk is identified 
with the bottom of the other disk. 
 The resulting space will have a continuum of conical singularities along the common boundary of the two disks, which is referred to as the 
 {\em ring singularity} of the space.
 
 The field $\bF$ is singular on the same ring $\{r=0, \theta=\pi/2,\varphi\in[0,2\pi)\}$ as the metric; for $r$ very large and positive it exhibits an 
electric monopole of strength $\textsc{q}$ and a magnetic dipole moment of strength $\textsc{q}a$, while
for $r$ very large in absolute value and negative it exhibits an electric monopole of strength $-\textsc{q}$ and a magnetic dipole moment of strength 
$-\textsc{q}a$.
 The field $\bF$ satisfies the Maxwell--Lorentz field equations in the sense of distributions, with sesqui-poles supported by the ring singularity \cite{TZzGKN}.

The Dirac equation for a spin-$\frac12$ test particle of charge $\textsc{q}'$ and mass $m$ in the static z$G$KN background spacetime of charge $\textsc{q}$ (as seen from infinity in the $r>0$ sheet) and \lq\lq{}angular-momentum-per-unit-mass\rq\rq{} parameter\footnote{Recall that in the zero-gravity limit, when Newton\rq{}s constant of universal gravitation $G \to 0$, both the ADM mass and ADM angular momentum of Kerr-Newman spacetimes tend to zero, while their ratio remains constant.  Note that this spacetime is static, which is why \lq\lq{}angular momentum\dots\rq\rq{} is in quotes.} $a$  reads
\beq\label{eq:DirEqA}
{\tilde\ga}^\mu 
\left(-ic\hbar \nabla_\mu  - \textsc{q}' A_\mu\right)\Psi + mc^2 \Psi = 0.
\eeq
Here, the Dirac matrices $({\tilde\ga^\mu})_{\mu=0}^3$ satisfy $\tilde{\ga}^\mu \tilde{\ga}^\nu + \tilde{\ga}^\nu\tilde{\ga}^\mu = 2 g^{\mu\nu}$, with $g^{\mu\nu}$ the (inverse) metric coefficients of the z$G$KN metric,
and where the wave function $\Psi(t,\,.\,)$ and electromagnetic four-potential $\bA = A_\mu dx^\mu$ are evaluated at the location  $\bq_{\mathrm{pt}}$
of the point charge in the Sommerfeld space given by a spacelike slice of the z$G$KN spacetime.
 The same equation is obtained when the ring singularity is considered as test particle in the field of a fixed point charge, as long as only 
strictly stationary situations are considered, including the question of the discrete spectrum.

\subsection{Separation of variables}\label{sepVAR}

 Using Cartan's frame method (see \cite{BrillCohen66} and refs. therein) with a frame  well-adapted to oblate spheroidal coordinates,
 Chandrasekhar \cite{Chandra76a,Chandra76b}, Page \cite{Page76}, and Toop \cite{Toop76}
 transformed  the Dirac equation (\ref{eq:DirEqA}) into an equation for a bispinor $\hat{\Psi}= \fD^{-1} \hat{\Psi}$, 
 with $\fD = \fD(r,\theta,\varphi)$ an explicit diagonal matrix, that allows a clear separation of the $t$, $r$, $\theta$, and $\varphi$ derivatives, viz.
\beq\label{eq:DirSep}
(\hat{R} +\hat{A}) \hat{\Psi} = 0,
\eeq
with ($\hbar = c = 1$ units from now on)
\bna
\hat{R}& := & \left(\begin{array}{cccc} imr & 0 &D_- + i \textsc{q}'\textsc{q}\frac{r}{\varpi} & 0 \\
0 & imr & 0 & D_+ + i\textsc{q}'\textsc{q}\frac{r}{\varpi}\\
D_+ + i \textsc{q}'\textsc{q}\frac{r}{\varpi} & 0 & imr & 0 \\
0 & D_- +i\textsc{q}'\textsc{q}\frac{r}{\varpi} & 0 & imr \end{array} \right),\\
\hat{A} &:= & \left(\begin{array}{cccc} -m a \cos\theta & 0 & 0 & -L_- \\
0 & -ma\cos\theta & -L_+ &0 \\
0 & L_- & ma\cos\theta & 0 \\
L_+ & 0 & 0 & ma\cos\theta
\end{array}\right),
\ena
where
\beq\label{eq:DpmLpm}
D_\pm := \pm \De \pd_r + \left( \De \pd_t + \frac{a}{\De} \pd_\varphi\right),
\qquad 
L_\pm :=\pd_\theta  \pm i \left( a\sin\theta \pd_t+\csc\theta \pd_\varphi\right),
\eeq
and
\[
\varpi := \sqrt{r^2 + a^2}.
\]
 Once a solution $\hat{\Psi}$ to (\ref{eq:DirSep}) is found, the bispinor $\Psi := \fD\hat{\Psi}$
solves the original Dirac equation (\ref{eq:DirEqA}).
 The details of the transformation are also given in \cite{KTZzGKNDa}; $\fD$ are not needed in this paper.

  Separation of variables is now achieved with the Ansatz that a solution $\hat{\Psi}$ of (\ref{eq:DirSep}) is of the form
\beq\label{chandra-ansatz} 
\hat{\Psi} = e^{-i(Et-\kappa \varphi)} \left( \begin{array}{c}R_1S_1\\ R_2 S_2\\ R_2 S_1\\ R_1 S_2 \end{array}\right),
\eeq
with $E$ a yet to be found energy eigenvalue of the Dirac hamiltonian, $\kappa \in \bbZ + \frac12$, and with
 $R_k$ being complex-valued functions of $r$ alone, and $S_k$ real-valued functions of $\theta$ alone.  
Let
\beq
\vec{R} := \left(\begin{array}{c} R_1\\ R_2\end{array}\right),\qquad \vec{S} := \left(\begin{array}{c} S_1\\ S_2\end{array}\right).
\eeq
Plugging the Chandrasekhar Ansatz \eqref{chandra-ansatz} into \eqref{eq:DirSep} one easily finds that there must be 
$\la\in\C$ such that 
\beq\label{eq:rad} 
T_{rad}\vec{R} =  E\vec{R},
\eeq
\beq\label{eq:ang}
T_{ang}\vec{S} = \la \vec{S},
\eeq
where
\bna
T_{rad} & :=  \label{eq:Trad} 
& \left(\begin{array}{cc} d_- 
&m\frac{r}{\De} - i\frac{\la}{\De} \\ m\frac{r}{\De}+i\frac{\la}{\De} 
& -d_+ \end{array}\right)
\\
T_{ang}& := \label{eq:Tang} 
& \left(\begin{array}{cc}  ma\cos\theta & l_- \\
 -l_+ &-ma\cos\theta  \end{array}\right)
\ena
The operators $d_\pm$ and $l_\pm$ are ordinary differential operators in $r$ and $\theta$ respectively, 
with coefficients that depend on the unknown $E$, and parameters $a$, $\kappa$, and $\textsc{q}'\textsc{q}$:
\bna\label{opdefsA}
d_\pm & := & -i \frac{d}{dr} \pm \frac{-a\kappa - \textsc{q}'\textsc{q} r}{\De^2}\\
l_\pm & := & \frac{d}{d\theta} \pm \left( aE\sin\theta - \kappa \csc\theta\right)\label{opdefsB}
\ena
 
 \subsection{The coupled spectral problems for $T_{rad}$ and $T_{ang}$}\label{Eval}
 
 The angular operator $T_{ang}$ in (\ref{eq:ang}) is easily seen to be essentially self-adjoint on
 $(C^\infty_c((0,\pi),\sin\theta d\theta))^2 \subset (L^2((0,\pi),\sin\theta d\theta))^2$, and its self-adjoint extension (also denoted $T_{ang}$) has a
purely discrete spectrum $\la=\la_n(am,aE,\kappa)\in \R$, $n\in \Z\setminus 0$ (e.g. \cite{SufFacCos83,BSW}).

With $\la\in\R$ it then follows that the radial operator $T_{rad}$ is essentially self-adjoint on $(C^\infty_c(\R, dr))^2\subset (L^2(\R,dr))^2$;
its self-adjoint extension will also be denoted $T_{rad}$.
It is moreover easy to see that without loss of generality, one can take $R_1 = R_2^*$.  Thus, we can set
\beq
R_1 =\frac{1}{\sqrt{2}}( u-iv),\qquad R_2  =\frac{1}{\sqrt{2}}( u + iv)
\eeq
 for real funcions $u$ and $v$.
 This brings the radial system (\ref{eq:rad}) into the following standard (hamiltonian) form
\beq\label{eq:hamil}
(H_{rad} -E)\left(\begin{array}{c} u \\ v \end{array}\right) = \left(\begin{array}{c}0 \\ 0 \end{array}\right),
\eeq
where
\beq\label{eq:Hrad}
H_{rad} := \left(\begin{array}{cc} m \frac{r}{\De} + \frac{\ga r+a\kappa}{\De^2} & -\pd_r + \frac{\la}{\De} \\[20pt]
 \pd_r +\frac{\la}{\De} & -m\frac{r}{\De} + \frac{\ga r+a\kappa}{\De^2}  \end{array}\right),
\eeq
(cf. \cite{ThallerBOOK}, eq (7.105)) with
\beq
\ga := \textsc{q}'\textsc{q} <0.
\eeq  

 Using techniques of Weidmann it is straightforward to show that the essential spectrum of $H_{rad}$ consists of values $E\in (-\infty,-m]\cup[m,\infty)$, 
 and its interior is purely absolutely continuous; see  \cite{KTZzGKNDa}.
  The remaining task is to characterize the discrete spectrum in the gap, i.e. the eigenvalues $E\in (-m,m)$. 
  In \cite{KTZzGKNDa} it was shown that the spectrum is symmetric about $0$, hence it suffices to consider $E>0$.
  
   One possible strategy for obtaining the discrete energy spectrum is to take advantage of 
the complete characterization of the spectrum of the angular operator $T_{ang}$ in terms of the analytic expansion of $\la$ in powers of $E$,
   given $a,m,\kappa$, that was obtained in \cite{BSW}.
  When inserted for $\la$ into (\ref{eq:Hrad}), equation (\ref{eq:hamil}) turns into a closed equation for $E$ in which $E$ features in a 
  nonlinear manner. 
   Having to deal only with the $\la(E)$-infused (\ref{eq:hamil})  may offer numerical advantages for the computation of the energy eigenvalues. 
    In particular, in the small $a$ regime a linear approximation of $\lambda$ as function of $E$ should suffice for all practical purposes, 
   yielding a linear approximate problem for $E$ that closely, if not perfectly, resembles the familiar structure of a radial Dirac equation known 
   from problems with spherical symmetry. 
   
   We here follow a different strategy, designed in \cite{KTZzGKNDa}, which is to consider the pair (\ref{eq:rad}), (\ref{eq:ang}) as a coupled system 
  jointly for the eigenvalue pairs $(\la,E)$ that feature in a linear fashion as displayed in  (\ref{eq:rad}), (\ref{eq:ang}), with (\ref{opdefsB}).
   More precisely, we consider the pair  (\ref{eq:ang}), (\ref{eq:Hrad}), which we can partially decouple with the help of the Pr\"ufer transform \cite{Pru26}.

\subsection{The Pr\"ufer transformed system}
  Thus, following \cite{KTZzGKNDa} we transform the equations (\ref{eq:hamil}) and (\ref{eq:ang}) for the four unknowns $(u,v)$ and $(S_1,S_2)$
by defining four new unknowns $(R,\Om)$ and $(S,\Theta)$ via the Pr\"ufer transform 
\beq\label{eq:prufer}
u =\sqrt{2} R \cos\frac{\Om}{2},\quad v = \sqrt{2} R \sin\frac{\Om}{2},\quad S_1 = S \cos\frac{\Theta}{2},\quad S_2 = S \sin\frac{\Theta}{2}.
\eeq
Thus
\beq
R =\frac{1}{\sqrt{2}}\sqrt{u^2+v^2},\quad\Om = 2\tan^{-1}\frac{v}{u},\quad S = \sqrt{S_1^2+S_2^2},\quad \Theta = 2\tan^{-1}\frac{S_2}{S_1}.
\eeq
As a result, $R_1 =  Re^{-i\Om/2}$ and $R_2 = Re^{i\Om/2}$.  Hence $\hat{\Psi}$ can be re-expressed in terms of the Pr\"ufer variables as
\beq\label{ontology}
\hat{\Psi}(t,r,\theta,\varphi) = R(r)S(\theta)e^{-i(Et-\ka \varphi)} \left(\begin{array}{l}
 \cos(\Theta(\theta)/2)e^{-i\Om(r)/2}\\
\sin(\Theta(\theta)/2) e^{i\Om(r)/2}\\
\cos(\Theta(\theta)/2)e^{i\Om(r)/2}\\
\sin(\Theta(\theta)/2)e^{-i\Om(r)/2}\end{array}\right),
\eeq
and we obtain the following equations for the new unknowns, first
\bna
\frac{d}{dr}\Om    &=& 2 \frac{mr}{\De} \cos\Om + 2\frac{\la}{\De} \sin\Om +2\frac{a\kappa + \gamma r}{\De^2} - 2E ,\label{eq:Om}\\
\frac{d}{dr} \ln R &=& \frac{mr}{\De}\sin\Om - \frac{\la}{\De} \cos\Om ,\label{eq:R}
\ena
and second,
\bna
\frac{d}{d\theta}\Theta &=& -2ma\cos\theta\cos\Theta + 2\left(aE \sin\theta - \frac{\kappa}{\sin\theta}\right)\sin\Theta + 2\la,\label{eq:Theta}\\
\frac{d}{d\theta} \ln S &=& -ma \cos\theta\sin\Theta - \left(aE\sin\theta - \frac{\kappa}{\sin\theta}\right)\cos\Theta. \label{eq:S}
\ena
 Note that the $\Om$-equation (\ref{eq:Om}) is decoupled from $R$, and the $\Theta$-equation (\ref{eq:Theta}) is decoupled from $S$.
  Thus the pair (\ref{eq:Om}), (\ref{eq:Theta}) can be solved together independently of equations (\ref{eq:R}), (\ref{eq:S}), which in turn
  can be integrated subsequently by direct quadrature. 

 We can further simplify the analysis of these systems and reduce the number of parameters involved by noting that 
by defining the constants $a'=ma$, $E'=E/m$, and changing to the variable $r'=mr$, we eliminate $m$ from the system.
 Henceforth we therefore set $m=1$. 


\subsection{Transformation onto a coupled dynamical system on cylinders}
 
  Equations (\ref{eq:Theta}), (\ref{eq:Om}) exhibit the independent and the dependent variables explicitly. 
  It is more convenient to transform them to a parametrically coupled pair of autonomous two-dimensional dynamical systems, by introducing a new independent 
variable $\tau$, as follows.
  
 Equation (\ref{eq:Theta}) can be written as a smooth dynamical system in the $(\theta,\Theta)$ plane by introducing 
 $\tau$ such that $\frac{d\theta}{d\tau} = \sin\theta$.
Then, with dot representing differentiation in $\tau$,  we have, 
\beq\label{dynsysTh}
\left\{\begin{array}{rcl}
\dot{\theta} & = & \sin\theta\\
       \dot{\Theta} & = & -2a\sin\theta\cos\theta\cos\Theta+2aE\sin^2\theta\sin\Theta - 2\ka\sin\Theta + 2\la\sin\theta
       \end{array}\right.
\eeq
Identifying the line $\Theta=\pi$ with $\Theta=-\pi$, this becomes a dynamical system on a closed finite cylinder 
$\mc{C}_1=[0,\pi]\times\Sset^1$. 
 The only equilibrium points of the flow are on the two circular boundaries: 
Two on the left boundary: $S^- = (0,0)$, $N^- = (0,\pi)$; two on the right: $S^+ = (\pi,-\pi)$ and $N^+ = (\pi,0)$.  

For $\ka>0$, the linearization of the flow at the equilibrium points reveals that $S^-$ and $S^+$ are hyperbolic saddle points (with 
eigenvalues $\{1,-2\ka\}$ and $\{-1,2\ka\}$ respectively), while $N^-$ is a  source node (with  eigenvalues 1 and $2\ka$) 
and $N^+$ is a sink node (with eigenvalues $-1$ and $-2\ka$).  The situation with $\ka<0$ is entirely analogous, with the two
critical points on each boundary switching their roles.

Similarly, the $\Om$ equation (\ref{eq:Om}) can  be rewritten as a smooth dynamical system on a cylinder, in this case by setting 
$\tau := \frac{r}{a}$ as new independent variable, as well as introducing a new dependent variable
\beq
\xi := \tan^{-1}\frac{r}{a} = \tan^{-1}\tau.
\eeq
Then, with dot again representing differentiation in $\tau$, (\ref{eq:Om}) is equivalent to
\beq\label{dynsysOm}
\left\{\begin{array}{rcl}
\dot{\xi} & = & \cos^2\xi \\
\dot{\Om} & = & 2a\sin\xi\cos\Om+2\la\cos\xi\sin\Om + 2\ga\sin\xi\cos\xi+2\ka\cos^2\xi - 2aE
\end{array}\right.
\eeq
Once again, identifying $\Om=-\pi$ with $\Om  = \pi$ turns this into a smooth flow on the closed finite cylinder 
$\cC_2 := [-\frac{\pi}{2},\frac{\pi}{2}]\times \Sset^1$.  The only equilibrium points of the flow are on the two circular boundaries. 
For $E\in(0,1)$ there are two equilibria on each: $S^-_E = (-\frac{\pi}{2},-\pi+\cos^{-1}E)$ and $N^-_E=(-\frac{\pi}{2},\pi - \cos^{-1}E)$ 
on the left boundary, and $S^+_E = (\frac{\pi}{2}, -\cos^{-1}E)$ and $N^+_E = (\frac{\pi}{2},\cos^{-1}E)$ on the right boundary.
 $S^\pm_E$ are non-hyperbolic (degenerate) saddle-nodes, with eigenvalues $0$ and $\pm 2a\sqrt{1-E^2}$, while $N^-_E$ is a degenerate 
source-node and $N^+_E$ a degenerate sink-node \cite{QTPDS}.

In \cite{KTZzGKNDa} we showed that $E$ is an energy eigenvalue of the Dirac hamiltonian  and the corresponding $\Psi$ is a bound state, if and only if there exists a $\la\in \R$ such that each of the two dynamical systems above possesses a {\em saddles connector,} i.e. an orbit connecting the two saddle-nodes $S^-$ and $S^+$ in the $\Theta$-system (\ref{dynsysTh}) 
and an orbit connecting the two saddle-nodes $S^-_E$ and $S^+_E$ in the $\Omega$-system (\ref{dynsysOm}). 

Given a dynamical system on a cylinder, there corresponds an integer known as the \emph{winding number} which describes how many times an orbit in the dynamical system winds around the cylinder before terminating at an equilibrium point. 
For the $\Omega$ system, saddles connectors with different winding numbers correspond to different energy values (with energy increasing as the winding number increases). In \cite{KTZzGKNDa} we showed that a bound state $\Psi$ exists corresponding to winding number $N_\Theta = 0$ for the $\Theta$ system and winding number $N_\Omega = 0$ for the $\Omega$ system (and $\kappa = \half$).

\subsection{Statement of main result}

 In this paper, we prove the following characterization of the discrete energy spectrum.

\medskip
\medskip

\begin{thm}\label{main thm}\:
Set $a_{\rm{max}} = 1 - \frac{1}{\sqrt{2}}$ and $\gamma_{\rm{min}} = -\half$. Fix $a \in (0, a_{\rm{max}})$, $\gamma \in (\gamma_{\rm{min}}, 0)$, and $\kappa \in \Z +\half$. Assume $\Psi$ is of the form \emph{(\ref{ontology})} constructed from solutions of \emph{(\ref{eq:Om})} - \emph{(\ref{eq:S})}.

\begin{itemize}
\item[$\bullet$] Suppose $N_\Theta \geq 0$ is an integer. For all integers $N_\Omega \geq 0$, there is a bound state $\Psi$ such that the $\Theta$ system and $\Omega$ system have winding numbers $N_\Theta$ and $N_\Omega$, respectively. There are no bound states with $N_\Omega \leq - 1$. 

\item[$\bullet$] Suppose $N_\Theta \leq -1$ is an integer. For all integers $N_\Omega \geq 1$, there is a bound state $\Psi$ such that the $\Theta$ system and $\Omega$ system have winding numbers $N_\Theta$ and $N_\Omega$, respectively. There are no bound states with $N_\Omega \leq 0$.
\end{itemize}
\end{thm}

\medskip
\medskip

As to our conditions, 
when restoring units and the mass $m$, we have $a_{\rm{max}} = (1 - \frac{1}{\sqrt{2}})\frac{\hbar}{mc}$ and $\gamma_{\text{min}} = -\half \hbar c.$ 
 If, with the hydrogenic problem in mind, we set  
 $\gamma = -Ze^2$, then $\gamma \in (\gamma_{\rm min}, 0)$ implies $\frac{Ze^2}{\hbar c} < \frac{1}{2}$, that is, $Z < \frac{137.036}{2}$. This condition on $Z$ is used in our proof to ensure that there are no bound states with $N_\Omega \leq -1$ for $N_\Theta \geq 0$ and no bound states with $N_\Omega \leq 0$ 
 for $N_\Theta \leq -1$. 
 
Now compare this with the conditions on $Z$ for the familiar hydrogenic Dirac operator on a Minkowski background with a Coulomb potential, for which
essential self-adjointness breaks down for $Z > 118$. For the Dirac operator on z$G$KN, there is no condition for essential self-adjointness \cite{KTZzGKNDa}.
So our  conditions are probably not optimal. 

We now come to our proof.
\medskip
\medskip

\section{Proof of Theorem 2.1}\label{THEproof}

\subsection{Flow on a finite cylinder}\label{assumptions on flow section}

\medskip

Let $\mc{C} = [x_-, x_+] \times \Sset^1$ be a finite cylinder (i.e. $-\infty < x_- < x_+ < \infty$). We denote its universal cover by $\wt{\mc{C}} = [x_-, x_+] \times \R$. Let $f\colon [x_-, x_+] \to \R$ be smooth. Let $g \colon \mc{C} \times I \to \R$ be smooth where $I \subset \R$ is an interval (i.e. a nonempty connected subset of $\R$ with nonempty interior). For each $\mu \in I$, define $g_\mu \colon \mc{C} \to \R$ by $g_\mu(x,y) := g(x,y,\mu)$. For each $\mu \in I$, consider the dynamical system on $\mc{C}$ given by 
\[  \left\{
 \begin{array}{ll}
      \dot{x} \,=\,  f(x)   \\
      \dot{y} \,=\,  g_\mu(x,y).      
\end{array} 
\right. \]
We will use $\tau$ to denote independent variable (e.g. $\dot{x} = \frac{dx}{d \tau}$). 
Note that the dynamical system on $\mc{C}$ induces a dynamical system on the universal cover $\wt{\mc{C}}$. We make five assumptions for the dynamical system on $\mc{C}$ labeled (a) - (e). 

\medskip
\medskip

\noindent{\bf Assumptions:}

\medskip

\begin{enumerate}

\item[\bf{(a)}] $f(x_-) = f(x_+) = 0$ and $f(x) > 0$ for all $x \in (x_-, x_+)$. Consequently, equilibrium points can only occur at the boundary of the cylinder (i.e. at $x = x_\pm$), and the boundaries $x = x_-$ and $x = x_+$ are images of orbits of the dynamical system. An orbit whose image is not completely contained in one of the boundaries $x = x_-$ or $x = x_+$ will be referred to as a \emph{non-boundary orbit}. 

Since $f(x) > 0$ for $x \in (x_-, x_+)$, the flow of the dynamical system ``points to the right" (i.e. for any non-boundary orbit $\big(x(\tau), y(\tau)\big)$, we have $x(\tau)$ is strictly increasing in $\tau$).

We will assume that the one-sided derivatives of $f$ at $x_-$ and $x_+$ satisfy 
\[
f'(x_-) \,\geq\, 0 \:\:\:\: \text{ and } \:\:\:\: f'(x_+) \,\leq\,0.
\]

\item[\bf{(b)}] For all $\mu \in I$, there are exactly two equilibrium points on each of the boundaries. That is there are exactly two distinct points $n^-_\mu, s^-_\mu\in\Sset^1$ and exactly two distinct points $n^+_\mu,s^+_\mu \in \Sset^1$ such that 
\[
0 \,=\, g_\mu(x_-, n^-) \,=\, g_\mu(x_-, s^-)\,=\, g_\mu(x_+, n^+) \,=\, g_\mu(x_+, s^+). 
\]
Let $N^{\pm}_\mu = (x_{\pm}, n^{\pm}_\mu)$ and $S^{\pm}_\mu = (x_{\pm}, s^{\pm}_\mu)$ denote the four equilibrium points on $\mc{C}$. We assume

\[
\frac{\pd g_\mu}{\pd y}(N^-_\mu) \,>\, 0, \:\:\:\: \frac{\pd g_\mu}{\pd y}(S^-_\mu) \,<\, 0, \:\:\:\: \frac{\pd g_\mu}{\pd y}(N^+_\mu) \,<\, 0, \:\:\:\: \frac{\pd g_\mu}{\pd y}(S^+_\mu) \,>\, 0.
\]

The jacobian matrix of the system is

\[
\begin{pmatrix}
f' & 0 \\
\frac{\pd g_\mu}{\pd x} & \frac{\pd g_\mu}{\pd y} \\
\end{pmatrix}.
\]
When $f'(x_{\pm}) \neq 0$, the equilibrium points are hyperbolic and so the Hartman-Grobman theorem can be used to characterize their local behavior. We have 

\begin{itemize}
\item[-] $N^-_\mu$ is a source
\item[-] $N^+_\mu$ is a sink
\item[-] $S^{\pm}_\mu$ are saddle points.
\end{itemize}
By the stable manifold theorem, there is a unique one-dimensional manifold emanating from $S^{\pm}_\mu$ called the \emph{unstable manifold} which is the image of an orbit of the dynamical system.  

When $f'(x_{\pm}) = 0$, the equilibrium points are nonhyperbolic and so one cannot apply the stable manifold theorem. In this case the orbits emanatings from $S^\pm_\mu$ are center manifolds and so they may not be unique. However, we will assume conditions on $f$ and $g_\mu$ so that these center manifolds are unique, and we will still refer to them as unstable manifolds. Therefore it always makes sense to speak of \emph{the} unstable manifold emanating from $S^{\pm}_\mu$.

\item[\bf{(c)}] There is a fundamental domain $\mc{C}_* = [x_-, x_+] \times [y_0,\,y_0 + 2\pi)$ of the universal cover $\wt{\mc{C}} = [x_-, x_+] \times \R$ such that for all $\mu \in I$ the following hold
\[
 y_0 \leq n^-_\mu < s^-_\mu < y_0 + 2\pi \quad \text{ and } \quad y_0 \leq s^+_\mu < n^+_\mu < y_0 + 2\pi.
\]
Here we have identified the points $n^\pm_\mu, s^\pm_\mu \in \Sset^1$ with points in $[y_0,\, y_0 + 2\pi)$. We will also make this identification with the equilibrium points so that we can think of $N^\pm_\mu, S^\pm_\mu$ as being points in either $\mc{C}$ or $\mc{C}_*$. In general we can make this identification for all points $(x,y) \in \mc{C}$.

\item[\bf{(d)}] For all $(x,y) \in \mc{C}$, we have $\frac{\pd}{\pd \mu}g_\mu(x,y) \leq 0$. 

\item[\bf{(e)}] Every non-boundary orbit is a heteroclinic orbit. That is, the $\a$ and $\omega$-limit sets of any non-boundary orbit are precisely one-point sets containing equilibrium points on the boundary $x = x_-$ and $x= x_+$, respesctively.  Therefore we will refer to the $\alpha$ and $\omega$-limit sets as $\a$ and $\omega$-limit \emph{points}.

\end{enumerate}

\noindent\emph{Remark.} The ordering of the equilibrium points in assumption (c) is only a technical assumption. We conjecture that the theorems in the next section hold even without the specific orderings imposed in assumption (c). Nevertheless, the systems of interest (e.g. the $\Theta$ system (\ref{dynsysTh}) and the $\Omega$ system (\ref{dynsysOm})) satisfy assumption (c).

\medskip
\medskip

\begin{figure}[h]
\[
\begin{tikzpicture}[scale = 0.55]

\draw[ultra thick] (-3,-6) -- (-3, 6);
\draw[ultra thick] (3,-6) -- (3,6);

\draw[ultra thick, dashed] (-3,6) -- (-3,6.9);
\draw[ultra thick, dashed] (3,6) -- (3,6.9);
\draw[ultra thick, dashed] (-3,-6) -- (-3,-6.9);
\draw[ultra thick, dashed] (3,-6) -- (3,-6.9);

\draw[ultra thick] (-3,-2) -- (3,-2);
\draw[ultra thick] (-3,2) -- (3,2);
\draw[ultra thick] (-3,-6) -- (3,-6);
\draw[ultra thick] (-3,6) -- (3,6);

\node [scale = .50] [circle, draw, fill = black] at (-3,-1.25)  {};
\node [scale = .50] [circle, draw, fill = black] at (-3,1.25)  {};
\node [scale = .50] [circle, draw, fill = black] at (3,-.75)  {};
\node [scale = .50] [circle, draw, fill = black] at (3,.75)  {};

\node [scale = .50] [circle, draw, fill = black] at (-3,-5.25)  {};
\node [scale = .50] [circle, draw, fill = black] at (-3,-2.75)  {};
\node [scale = .50] [circle, draw, fill = black] at (3,-4.75)  {};
\node [scale = .50] [circle, draw, fill = black] at (3,-3.25)  {};

\node [scale = .50] [circle, draw, fill = black] at (-3,2.75)  {};
\node [scale = .50] [circle, draw, fill = black] at (-3,5.25)  {};
\node [scale = .50] [circle, draw, fill = black] at (3,3.25)  {};
\node [scale = .50] [circle, draw, fill = black] at (3,4.75)  {};

\draw (-3, -7.8) node [scale = .85] {$x_-$};
\draw (3, -7.8) node [scale = .85] {$x_+$};

\draw (-3.8,1.2) node [scale =.85] {$S^-_\mu$};
\draw (-3.8,-1.3) node [scale =.85] {$N^-_\mu$};
\draw (3.9,.70) node [scale =.85] {$N^+_\mu$};
\draw (3.9,-.80) node [scale =.85] {$S^+_\mu$};

\draw (-5.75,-2) node [scale =.85] {$y_0$};
\draw (-5.75,2) node [scale =.85] {$y_0 + 2\pi$};
\draw (-5.75,-6) node [scale =.85] {$y_0 - 2\pi$};
\draw (-5.75,6) node [scale =.85] {$y_0 + 4\pi$};

\draw [->] [thick] (4,4.4) arc [start angle=100, end angle=180, radius=130pt];
\draw (4.725,4.65) node [scale = 1]{\small{$\mc{C}_*$}};

\end{tikzpicture}
\]
\captionsetup{format=hang}
\caption{\small{The universal cover $\wt{\mc{C}}$ of the finite cylinder $\mc{C}$. The fundamental domain $\mc{C}_*$ and the equilibrium points are $N^{\pm}_\mu$ and $S^{\pm}_\mu$ are shown. By assumption (b), $N^-_\mu$ is a source, $N^+_\mu$ is a sink, and $S^{\pm}_\mu$ are saddle points. }}
\end{figure}
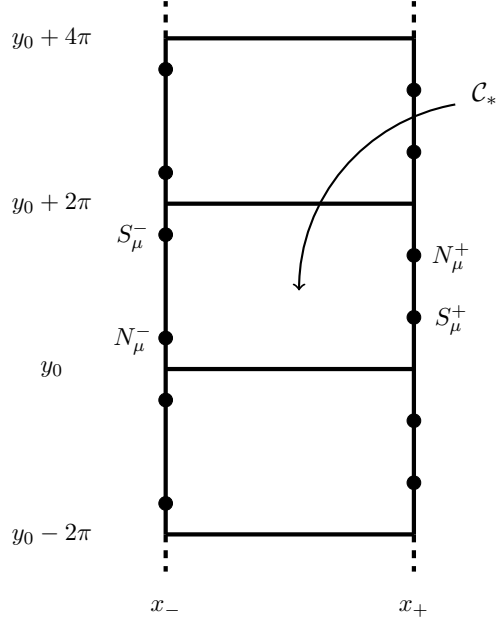

\subsection{Consequences of assumptions (a) - (e)} 

\medskip
\medskip

In this section we list some consequences of assumptions (a) - (e). The goal of this section is to prove Proposition \ref{saddles conn prop} which shows how to guarantee the existence of saddles connectors. By assumption (e), we know that all orbits have $\a$ and $\omega$-limit points on the boundaries.

\medskip
\medskip

\begin{itemize}

\item [$\bullet$] Let $\mc{W}^-_\mu$ denote the unique (up to a constant) orbit of the flow on $\mc{C}$ whose $\alpha$-limit point is the saddle $S^-_\mu$.  
Let $\wt{\mc{W}}^-_\mu$ denote the lift of $\mc{W}^-_\mu$ into the universal cover $\wt{\mc{C}}$ whose $\a$-limit point is $S^-_\mu\in \mc{C}_*$ (we are identifying the point $S^-_\mu \in \mc{C}$ with the corresponding point in $\mc{C}_*$; see assumption (c)). The $\omega$-limit point of $\mc{W}^-_\mu$ is either $N^+_\mu$ or $S^+_\mu$. 

\smallskip
\begin{itemize}

\item[-] Assume the $\omega$-limit point of $\mc{W}^-_\mu$ is $N^+_\mu$. Then the $\omega$-limit point of $\wt{\mc{W}}^-_\mu$ is some copy of $N^+_\mu$ in the universal cover, say $\wt{N}^+_\mu \in \wt{\mc{C}}$. Then $\wt{N}^+_\mu = (x_+,\, n^+_\mu - 2\pi k^+_\mu)$ for some $k^+_\mu \in \Z$.  

\smallskip

\item[-] Assume the $\omega$-limit point of $\mc{W}^-_\mu$ is $S^+_\mu$. The $\omega$-limit point of $\wt{\mc{W}}^-_\mu$ is some copy of $S^+_\mu$ in $\wt{\mc{C}}$. Hence it's $(x_+, \, s^+_\mu - 2\pi k^+_\mu)$ for some $k^+_\mu \in \Z$.

\end{itemize}

\medskip

\item[$\bullet$] Let $\mc{W}^+_\mu$ denote the unique (up to a constant) orbit of the flow on $\mc{C}$ whose $\omega$-limit point is the saddle $S^+_\mu$. Let $\wt{\mc{W}}^+_\mu$ denote the lift of $\mc{W}_\mu^+$ in the universal cover $\wt{\mc{C}}$ whose $\omega$-limit point is $S^+_\mu \in \mc{C}_*$. The $\a$-limit point of $\mc{W}^+_\mu$ is either $N^-_\mu$ or $S^-_\mu$. 

\smallskip

\begin{itemize}

\item[-] Assume the $\a$-limit point of $\mc{W}^+_\mu$ is $N^-_\mu$. Then the $\a$-limit point of $\wt{\mc{W}}^-_\mu$ is some copy of $N^-_\mu$ in the universal cover, say $\wt{N}^+_\mu \in \wt{\mc{C}}$. Then $\wt{N}^-_\mu = (x_-,\, n^-_\mu + 2\pi k^-_\mu)$ for some $k^-_\mu \in \Z$. 

\smallskip
\item[-] Assume the $\a$-limit point of $\mc{W}^+_\mu$ is $S^-_\mu$. The $\a$-limit point of $\wt{\mc{W}}^+_\mu$ is some copy of $S^-\mu$ in $\wt{\mc{C}}$. Hence it's $(x_-,\, s^-_\mu  + 2\pi k^-_\mu)$ for some $k^-_\mu \in \Z$. 

\end{itemize}

\medskip

\item[$\bullet$] If the $\omega$-limit point of $\mc{W}^-_\mu$ is $S^+_\mu$, then $\mc{W}^-_\mu = \mc{W}^+_\mu$. In this case, we call $\mc{W}^-_\mu$ a \emph{saddles connector}.
\end{itemize}
\medskip
\medskip

\begin{figure}[h]
\[
\begin{tikzpicture}[scale = 0.50]

\draw[ultra thick] (-13,-6) -- (-13, 6);
\draw[ultra thick] (-7,-6) -- (-7,6);

\draw[ultra thick, dashed] (-13,6) -- (-13,6.9);
\draw[ultra thick, dashed] (-7,6) -- (-7,6.9);
\draw[ultra thick, dashed] (-13,-6) -- (-13,-6.9);
\draw[ultra thick, dashed] (-7,-6) -- (-7,-6.9);

\draw[ultra thick] (-13,-2) -- (-7,-2);
\draw[ultra thick] (-13,2) -- (-7,2);
\draw[ultra thick] (-13,-6) -- (-7,-6);
\draw[ultra thick] (-13,6) -- (-7,6);

\draw [ ultra thick, blue] (-13,5.25) .. controls (-10,3.5) .. (-7,4.75);

\draw (-10,3.1) node [scale = .75] {$\wt{\mc{W}}^-_{\mu_0}$};

\node [scale = .50] [circle, draw, fill = black] at (-13,-1.25)  {};
\node [scale = .50] [circle, draw, fill = black] at (-13,1.25)  {};
\node [scale = .50] [circle, draw, fill = black] at (-7,-.75)  {};
\node [scale = .50] [circle, draw, fill = black] at (-7,.75)  {};

\node [scale = .50] [circle, draw, fill = black] at (-13,-5.25)  {};
\node [scale = .50] [circle, draw, fill = black] at (-13,-2.75)  {};
\node [scale = .50] [circle, draw, fill = black] at (-7,-4.75)  {};
\node [scale = .50] [circle, draw, fill = black] at (-7,-3.25)  {};

\node [scale = .50] [circle, draw, fill = black] at (-13,2.75)  {};
\node [scale = .50] [circle, draw, fill = black] at (-13,5.25)  {};
\node [scale = .50] [circle, draw, fill = black] at (-7,3.25)  {};
\node [scale = .50] [circle, draw, fill = black] at (-7,4.75)  {};

\draw (-13.8,5.2) node [scale = .8] {$S^-_{\mu_0}$};

\draw (-15.5,2) node [scale = .8] {$y_0$};
\draw (-15.5,6) node [scale = .8] {$y_0 + 2\pi$};
\draw (-15.5,-2) node [scale = .8] {$y_0 - 2\pi$};
\draw (-15.5,-6) node [scale = .8] {$y_0 - 4\pi$};


\draw[ultra thick] (13,-6) -- (13, 6);
\draw[ultra thick] (7,-6) -- (7,6);

\draw[ultra thick, dashed] (13,6) -- (13,6.9);
\draw[ultra thick, dashed] (7,6) -- (7,6.9);
\draw[ultra thick, dashed] (13,-6) -- (13,-6.9);
\draw[ultra thick, dashed] (7,-6) -- (7,-6.9);

\draw[ultra thick] (13,-2) -- (7,-2);
\draw[ultra thick] (13,2) -- (7,2);
\draw[ultra thick] (13,-6) -- (7,-6);
\draw[ultra thick] (13,6) -- (7,6);

\draw [ ultra thick, blue] (7,4.75) .. controls (10.75,-2.5) .. (13,-5.25);

\draw (11,-4) node [scale = .75] {$\wt{\mc{W}}^-_{\mu_2}$};

\node [scale = .50] [circle, draw, fill = black] at (7,-.75)  {};
\node [scale = .50] [circle, draw, fill = black] at (7,.75)  {};
\node [scale = .50] [circle, draw, fill = black] at (13,-1.25)  {};
\node [scale = .50] [circle, draw, fill = black] at (13,1.25)  {};

\node [scale = .50] [circle, draw, fill = black] at (7,-4.75)  {};
\node [scale = .50] [circle, draw, fill = black] at (7,-3.25)  {};
\node [scale = .50] [circle, draw, fill = black] at (13,-5.25)  {};
\node [scale = .50] [circle, draw, fill = black] at (13,-2.75)  {};

\node [scale = .50] [circle, draw, fill = black] at (7,3.25)  {};
\node [scale = .50] [circle, draw, fill = black] at (7,4.75)  {};
\node [scale = .50] [circle, draw, fill = black] at (13,2.75)  {};
\node [scale = .50] [circle, draw, fill = black] at (13,5.25)  {};

\draw (6.2,4.8) node [scale = .8] {$S^-_{\mu_2}$};


\draw[ultra thick] (-3,-6) -- (-3, 6);
\draw[ultra thick] (3,-6) -- (3,6);

\draw[ultra thick, dashed] (-3,6) -- (-3,6.9);
\draw[ultra thick, dashed] (3,6) -- (3,6.9);
\draw[ultra thick, dashed] (-3,-6) -- (-3,-6.9);
\draw[ultra thick, dashed] (3,-6) -- (3,-6.9);

\draw[ultra thick] (-3,-2) -- (3,-2);
\draw[ultra thick] (-3,2) -- (3,2);
\draw[ultra thick] (-3,-6) -- (3,-6);
\draw[ultra thick] (-3,6) -- (3,6);

\draw [ ultra thick, blue] (-3,5) .. controls (0.75,0.5) .. (3,1);

\draw (.5,.25) node [scale = .75] {$\wt{\mc{W}}^-_{\mu_1}$};

\node [scale = .50] [circle, draw, fill = black] at (-3,-1)  {};
\node [scale = .50] [circle, draw, fill = black] at (-3,1)  {};
\node [scale = .50] [circle, draw, fill = black] at (3,-1)  {};
\node [scale = .50] [circle, draw, fill = black] at (3,1)  {};

\node [scale = .50] [circle, draw, fill = black] at (-3,-5)  {};
\node [scale = .50] [circle, draw, fill = black] at (-3,-3)  {};
\node [scale = .50] [circle, draw, fill = black] at (3,-5)  {};
\node [scale = .50] [circle, draw, fill = black] at (3,-3)  {};

\node [scale = .50] [circle, draw, fill = black] at (-3,3)  {};
\node [scale = .50] [circle, draw, fill = black] at (-3,5)  {};
\node [scale = .50] [circle, draw, fill = black] at (3,3)  {};
\node [scale = .50] [circle, draw, fill = black] at (3,5)  {};

\draw (-3.8,5) node [scale = .8] {$S^-_{\mu_1}$};


\end{tikzpicture}
\]
\captionsetup{format=hang}
\caption{\small{The dynamical system for three different parameter values $\mu \in \{\mu_0, \mu_1, \mu_2\}$ are shown. Note that the equilibrium points $S^\pm_\mu$ and $N^\pm_\mu$ vary as $\mu$ varies, but they always remain in the fixed fundamental domain. The winding numbers for $\mu_0$, $\mu_1$, $\mu_2$ are $0$, $1$, $2$, respectively. Note that $\mc{W}^-_{\mu_2}$ is a saddles connector while $\mc{W}^-_{\mu_0}$ and $\mc{W}^-_{\mu_1}$ are not. }}
\end{figure}
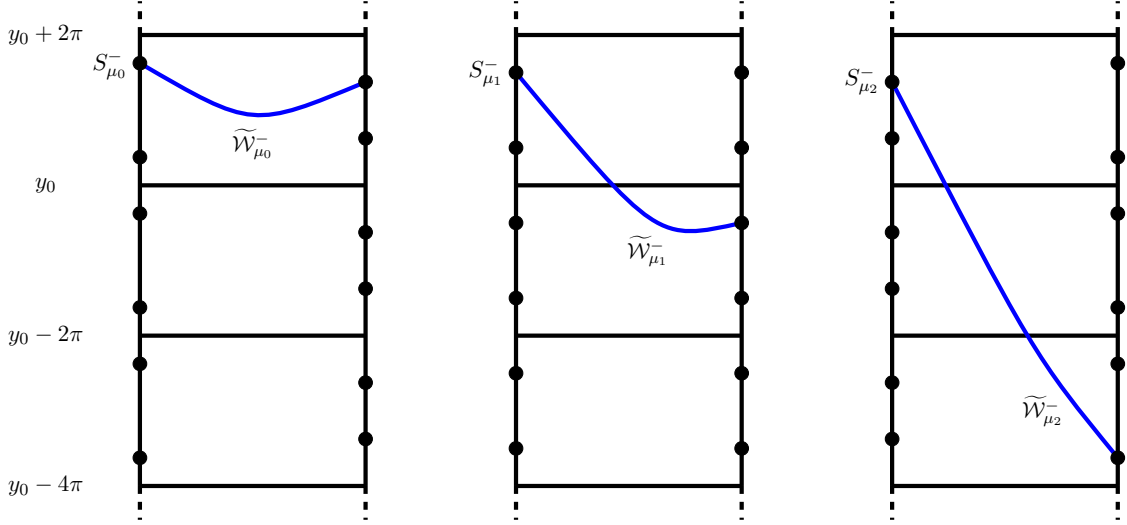

\begin{prop}\label{eq k prop}
$k^+_\mu = k^-_\mu.$
\end{prop}

\proof Assume first that the $\omega$-limit point of $\mc{W}^-_\mu$ is $N^+_\mu$. In the universal cover $\wt{\mc{C}}$, consider two copies of $\mc{W}^-_\mu$: one which begins at $(x_-,\, s^-_\mu + 2\pi k^+_\mu)$ and ends at $N^+_\mu = (x_+, n^+_\mu)$ and another copy which begins at $\big(x_-, \, s^-_\mu + 2\pi (k^+_\mu - 1)\big)$ and ends at $(x_+, n^+_\mu - 2\pi)$. Since $n^+_\mu - 2\pi < s^+_\mu < n^+_\mu$, these two copies of $\mc{W}^-_\mu$ are barriers for $\wt{\mc{W}}^+_\mu$. Therefore the $\a$-limit point of $\wt{\mc{W}}^+_\mu$ must be $(x_-, n^-_\mu + 2\pi k^+_\mu)$. Hence $k^+_\mu = k^-_\mu$. A similar proof is used when $\mc{W}^-_\mu$ is a saddles connector. 
\qed

\medskip
\medskip

\begin{Def}\label{winding number def}
\emph{
The integer $w_\mu := k^+_\mu = k^-_\mu$ is called the \emph{winding number} of $\mc{W}^-_\mu$ and $\mc{W}^+_\mu$. 
}
\end{Def}

\medskip
\medskip

\noindent\emph{Remark.}
We can give an invariant definition of winding number without relying on the fundamental domain $\mc{C}_*$. Note that $\mc{W}^-_\mu$ has winding number $N$ if and only if $\mc{W}^-_\mu$ is fixed-point homotopic to the curve formed by concatenating the curves $c_1$ and $c_2$ where $c_1$ is the minimizing geodesic from $S^-_\mu$ to the $\omega$-limit point of $\mc{W}^-_\mu$ and $c_2$ is the curve which winds around the right boundary $N$ times. If $N > 0$ then $c_2$ winds around clockwise (when facing the boundary $x_+$ away from the cylinder), while if $N < 0$ then $c_2$ winds around counterclockwise. If $N = 0$, then $c_2$ is just the constant curve.

\medskip
\medskip

The existence of saddles connectors is established in Proposition \ref{saddles conn prop}, but first we need the following lemma.

\medskip
\medskip

\begin{lem}\label{lem 7.3 in JMP paper}
Let $\s^\pm_\mu$ given by  $\big\{ \s^\pm_\mu(\tau) = \big(x(\tau), y_\mu^\pm(\tau)\big) \mid \tau \in \R\big\}$ be any lift of the orbits $\mc{W}^\pm_\mu$ into the universal cover $\wt{\mc{C}}$. Under assumption \emph{(d)}, we have $y_\mu^\pm$ are monotone in $\mu$. Specifically,
\[
\mu_1 \,<\, \mu_2 \:\:\:\: \Longrightarrow \:\:\:\: y^-_{\mu_1}(\tau) \,\geq\, y^-_{\mu_2}(\tau) \:\:\: \text{ and } \:\:\: y^+_{\mu_1}(\tau) \,\leq\, y^+_{\mu_2}(\tau) \:\: \text{ for all } \:\: \tau \in \R. 
\]
\end{lem}

\proof
We prove it for $\s^-_\mu$; the proof for $\s^+_\mu$ is analogous.  Define
\[
z(\tau) \,=\, \frac{\pd}{\pd \mu}y^-_\mu(\tau).
\]
It suffices to show $z(\tau) \leq 0$ for all $\tau$. Since $y^-_\mu$ solves $\dot{y} = g_\mu(x,y)$, we have 
\[
\dot{z} \,=\, \frac{\pd}{\pd \mu}\big(g_\mu(x, y^-_\mu)\big) \,=\, z\frac{\pd g_\mu}{\pd y}(x, y^-_\mu) + \frac{\pd g_\mu}{\pd \mu}(x, y^-_\mu). 
\]
Set $P(\tau) = \frac{\pd g_\mu}{\pd y}\big(x(\tau), y^-_\mu(\tau)\big)$. By assumption (d), we have $\dot{z}(\tau) \leq z(\tau)P(\tau)$. Therefore the differential form of Gr{\"o}nwall's inequality gives for any $\tau_0 \in \R$
\[
z(\tau) \,\leq\, z(\tau_0)e^{\int_{\tau_0}^\tau P(s)ds}.
\]
Since $\pd g_{\mu}/\pd y |_{S^-_\mu} < 0$ by assumption (b), we have $P(-\infty) < 0$. Therefore $\int_{-\infty}^\tau P(s) ds = -\infty$. So by taking $\tau_0 \to -\infty$ in the above inequality, we have $z(\tau) \leq 0$. Note we used $|z(-\infty)| = |ds_\mu^-/d\mu| < \infty$ which follows by $C^1$ dependence of solutions on parameters; see Theorem 2 in Section 2.3 of \cite{Perko}.
\qed

\medskip
\medskip

\begin{prop}\label{saddles conn prop}
Let $N$ be an integer. Suppose there exists two parameter values $\mu' < \mu''$ such that 
\[
w_{\mu'} \,\leq\, N \:\:\:\: \text{ and } \:\:\:\: w_{\mu''} \,\geq\, N + 1.
\]
 Then there exists a  $\mu \in [\mu', \mu'')$ such that $\mc{W}^-_\mu$ is a saddles connector with winding number $w_\mu = N$. 

\end{prop}

\proof
For each $\mu$, let $\sigma_\mu$ denote the unique lift of $\mc{W}^+_\mu$ into the universal cover $\wt{\mc{C}}$ such that the $\omega$-limit point of $\sigma_\mu$ is $(x_+,\, s^+_\mu-2\pi N)$. Define $\mc{K}_\mu$ as the open domain in the universal cover $\wt{\mc{C}}$ such that both $\wt{\mc{W}}^-_\mu$ and $\sigma_\mu$ lie on the boundary $\pd\mc{K}_\mu$. Orient $\pd\mc{K}_\mu$ so that the orientation induced on $\wt{\mc{W}}^-_\mu$ coincides with the direction of the flow. The \emph{signed area} of $\mc{K}_\mu$ is defined via Green's theorem.
\[
a(\mu) \,=\, \oint_{\pd\mc{K}_\mu}(-y)dx \,=\, \int_{x_-}^{x^+}\big(y^+_\mu - y_\mu^-\big)dx.
\]
In the second expression, $y_\mu^-$ denotes the $y$-component of $\wt{\mc{W}}^-_\mu$ while $y_\mu^+$ denotes the $y$-component of $\s_\mu$. Since orbits can't intersect without coinciding, we have either $y^+_\mu - y^-_\mu \geq 0$ or $y^+_\mu - y^-_\mu \leq 0$. Therefore, if there exists a $\mu$ such that $a(\mu) = 0$, then $y^+_\mu = y^-_\mu$ and so the orbits coincide yielding a saddles connector with winding number $N$. Thus we want to show there exists a $\mu \in [\mu', \mu'')$ such that $a(\mu) = 0$.

If $\mc{W}_{\mu'}^-$ is a saddles connector and $w_{\mu'} = N$, then we just take $\mu = \mu'$. Otherwise the $\omega$-limit point of $\mc{W}_{\mu'}^-$ is $N^+_\mu$ or $w_{\mu'} < N$. In either case, we have $\wt{\mc{W}}^-_{\mu'}$ lies above $\s_{\mu'}$ and so $a(\mu') < 0$. If $w_{\mu''}\geq N +1$, then $\wt{\mc{W}}^-_{\mu''}$ lies below $\s_{\mu''}$ and so $a(\mu'') > 0$. Therefore, it suffices to show that $a(\mu)$ is a continuous function of $\mu$ within the interval $[\mu', \mu'')$, since then the intermediate value theorem implies the existence of a $\mu \in [\mu', \mu'')$ such that $a(\mu) = 0$.  

To show continuity, let $\mu_n \in [\mu', \mu'')$ be any sequence such that $\mu_n \to \mu$. Since the orbits of the flow depend continuously on the parameter $\mu$, we have $y^\pm_{\mu_n} \to y^\pm_{\mu}$ pointwise. Break up $y^-_{\mu_n}$ into its positive and negative parts. By Lemma \ref{lem 7.3 in JMP paper}, we can bound the positive part by the positive part of $y^-_{\mu'}$, and we can bound the negative part by the negative part of $y^-_{\mu''}$. Therefore $|y^-_{\mu_n}|$ is bounded uniformly. Likewise $|y^+_{\mu_n}|$ is bounded uniformly. Thus $a(\mu_n) \to a(\mu)$ by Lebesgue's dominated convergence theorem. 
\qed

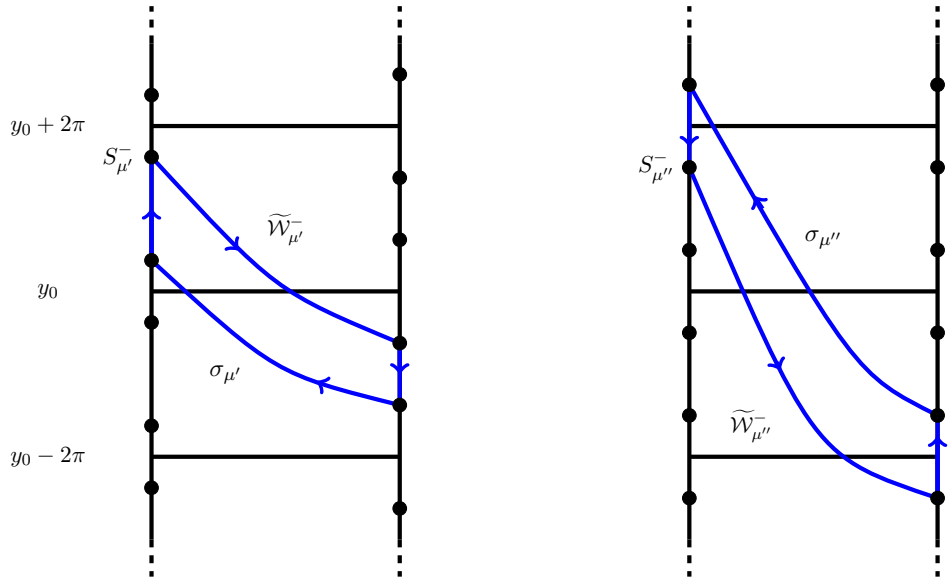
\begin{figure}[h]
\[
\begin{tikzpicture}[scale = 0.55]

\draw[ultra thick] (-13,-4) -- (-13, 8);
\draw[ultra thick] (-7,-4) -- (-7,8);

\draw[ultra thick, dashed] (-13,8) -- (-13,8.9);
\draw[ultra thick, dashed] (-7,8) -- (-7,8.9);
\draw[ultra thick, dashed] (-13,-4) -- (-13,-4.9);
\draw[ultra thick, dashed] (-7,-4) -- (-7,-4.9);

\draw[ultra thick] (-13,-2) -- (-7,-2);
\draw[ultra thick] (-13,2) -- (-7,2);
\draw[ultra thick] (-13,6) -- (-7,6);


\draw [ ultra thick, blue] (-13,5.25) .. controls (-10,2) .. (-7,.75);
\draw[ ->, ultra thick, blue] (-11,3.1) -- (-10.9, 3);

\draw [ultra thick, blue] (-13,2.75) .. controls (-10,0) .. (-7,-.75);
\draw[ <-, ultra thick, blue] (-9,-0.175) -- (-8.9, -.225);

\draw [ultra thick, blue] (-7, -.75) -- (-7, .75);
\draw [->,ultra thick, blue ](-7, .75) -- (-7, 0);

\draw [ultra thick, blue] (-13, 5.25) -- (-13, 2.75); 
\draw [->,ultra thick, blue ](-13, 2.75) -- (-13, 4);

\draw (-9.75,3.5) node [scale = .75] {$\wt{\mc{W}}^-_{\mu'}$};
\draw (-11.2,0) node [scale = .85] {$\s_{\mu'}$};

\node [scale = .50] [circle, draw, fill = black] at (-13,-1.25)  {};
\node [scale = .50] [circle, draw, fill = black] at (-13,1.25)  {};
\node [scale = .50] [circle, draw, fill = black] at (-7,-.75)  {};
\node [scale = .50] [circle, draw, fill = black] at (-7,.75)  {};

\node [scale = .50] [circle, draw, fill = black] at (-13,6.75)  {};
\node [scale = .50] [circle, draw, fill = black] at (-7,7.25)  {};

\node [scale = .50] [circle, draw, fill = black] at (-13,-2.75)  {};
\node [scale = .50] [circle, draw, fill = black] at (-7,-3.25)  {};

\node [scale = .50] [circle, draw, fill = black] at (-13,2.75)  {};
\node [scale = .50] [circle, draw, fill = black] at (-13,5.25)  {};
\node [scale = .50] [circle, draw, fill = black] at (-7,3.25)  {};
\node [scale = .50] [circle, draw, fill = black] at (-7,4.75)  {};

\draw (-13.8,5.2) node [scale = .8] {$S^-_{\mu'}$};

\draw (-15.5,2) node [scale = .8] {$y_0$};
\draw (-15.5,6) node [scale = .8] {$y_0 + 2\pi$};
\draw (-15.5,-2) node [scale = .8] {$y_0 - 2\pi$};


\draw[ultra thick] (-0,-4) -- (-0, 8);
\draw[ultra thick] (6,-4) -- (6,8);

\draw[ultra thick, dashed] (-0,8) -- (-0,8.9);
\draw[ultra thick, dashed] (6,8) -- (6,8.9);
\draw[ultra thick, dashed] (-0,-4) -- (-0,-4.9);
\draw[ultra thick, dashed] (6,-4) -- (6,-4.9);

\draw[ultra thick] (-0,-2) -- (6,-2);
\draw[ultra thick] (-0,2) -- (6,2);
\draw[ultra thick] (-0,6) -- (6,6);


\draw [ultra thick, blue] (0,5) .. controls (3,-2) .. (6,-3);
\draw[->, ultra thick, blue] (2.1,.3) -- (2.2,.1);

\draw[ultra thick, blue] (0,7) .. controls (4,0) .. (6,-1);
\draw[->, ultra thick, blue] (1.77,4.0) -- (1.57,4.3);

\draw[ultra thick, blue] (6,-1) -- (6,-3);
\draw[->, ultra thick, blue] (6,-3) -- (6,-1.5);

\draw[ultra thick, blue] (0,5) -- (0,7); 
\draw[->, ultra thick, blue] (0,7) -- (0,5.5);

\draw (1.5,-1.25) node [scale = .75] {$\wt{\mc{W}}^-_{\mu''}$};
\draw (3.25,3.25) node [scale = .85] {$\s_{\mu''}$};

\node [scale = .50] [circle, draw, fill = black] at (-0,-1)  {};
\node [scale = .50] [circle, draw, fill = black] at (-0,1)  {};
\node [scale = .50] [circle, draw, fill = black] at (6,-1)  {};
\node [scale = .50] [circle, draw, fill = black] at (6,1)  {};

\node [scale = .50] [circle, draw, fill = black] at (-0,7)  {};
\node [scale = .50] [circle, draw, fill = black] at (-0,-3)  {};
\node [scale = .50] [circle, draw, fill = black] at (6,7)  {};
\node [scale = .50] [circle, draw, fill = black] at (6,-3)  {};

\node [scale = .50] [circle, draw, fill = black] at (-0,3)  {};
\node [scale = .50] [circle, draw, fill = black] at (-0,5)  {};
\node [scale = .50] [circle, draw, fill = black] at (6,3)  {};
\node [scale = .50] [circle, draw, fill = black] at (6,5)  {};

\draw (-0.8,5) node [scale = .8] {$S^-_{\mu''}$};


\end{tikzpicture}
\]
\captionsetup{format=hang}
\caption{\small{Proof of Proposition \ref{saddles conn prop}. At $\mu'$ the winding number is $w_{\mu'} = N$ (the figure on the left shows the case $N = 1$). At $\mu''$ the winding number is $w_{\mu''} = N + 1$ (the figure on the right shows the case $N + 1 = 2$). The signed area of the corridor at $\mu'$ and $\mu''$ are $a(\mu') < 0$ and $a(\mu'') > 0$. The intermediate value theorem implies that there exists a $\mu \in [\mu', \mu'')$ such that $a(\mu) = 0$ at which we have a saddles connector with winding number $w_\mu = N$.  }}
\end{figure}

\medskip
\medskip

\subsection{Using nullclines to verify assumption (e)}

\medskip

In general in it is not too difficult to verify assumptions (a) - (d) in the previous section. Verifying assumption (e) requires some work. In this section we show how nullclines can be used to verify assumption (e). 

Assume we have a finite cylinder $\mc{C} = [x_-, x_+] \times \Sset^1$ satisfying assumptions (a) - (d). The nullclines are defined as the set of points $\Gamma = \big\{(x, y) \in \mc{C} \mid g_\mu(x, y) = 0\big\}$. The nullclines $\Gamma$ separate the cylinder via $\mc{C} = \Gamma \cup \mc{N} \cup \mc{P}$ where $\mc{N}$ and $\mc{P}$ correspond to the points in $\mc{C}$ where $\dot{y} < 0$ and $\dot{y} > 0$, respectively. Note that $N^\pm_\mu$ and $S^\pm_\mu$ are points in $\Gamma$.
 
 Define the sets $\mc{A} = \big\{(x,y) \in \Gamma \mid \frac{\pd g_\mu}{\pd x}(x, y) = 0 \big\}$ and $\mc{B} = \big\{(x, y) \in \Gamma \mid \frac{\pd g_\mu}{\pd y}(x, y) = 0\big\}$. The set $\Gamma \setminus (\mc{A} \cap \mc{B})$ forms a smooth one-dimensional manifold \cite[page 20]{LeeSmoothMan}.  For any point $(x,y) \in \Gamma \setminus \mc{B}$, the implicit function theorem implies that we can locally graph $\Gamma$ as a smooth function of $x$. If in addition we have $(x,y) \in \mc{A}$, then the slope of this graph vanishes at $(x,y)$.

 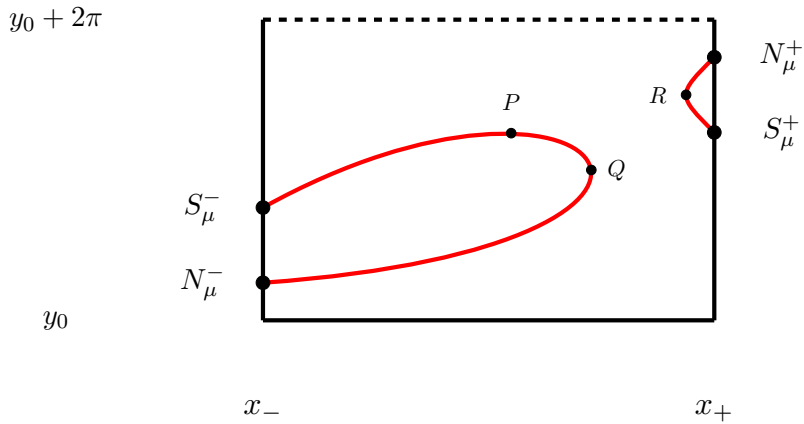
\begin{figure}[h]
\[
\begin{tikzpicture}[scale = 1.0]


\draw[ultra thick] (-3,-2) -- (-3, 2);
\draw[ultra thick] (3,-2) -- (3,2);
\draw[ultra thick] (-3,-2) -- (3,-2);
\draw[ultra thick, dashed] (-3,2) -- (3,2);

\draw (-3, -3.2) node [scale = 1.10] {$x_-$};
\draw (3, -3.2) node [scale = 1.10] {$x_+$};

\draw (-5.75,-2) node [scale =1.0] {$y_0$};
\draw (-5.75,2) node [scale =1.0] {$y_0 + 2\pi$};


\draw[ultra thick, red] (-3,-.5) .. controls (1.5,2) and (4,-1) .. (-3,-1.5);
\draw[ultra thick, red] (3,.5) .. controls (2.5, 1) .. (3, 1.5);


\node [scale = .35] [circle, draw, fill = black] at (2.625,1) {};
\node [scale = .35] [circle, draw, fill = black] at (1.365,0) {};
\node [scale = .35] [circle, draw, fill = black] at (.3,.49) {};

\draw (2.25,1) node [scale =.8] {$R$};
\draw (1.7,0) node [scale =.8] {$Q$};
\draw (.3,.9) node [scale =.8] {$P$};


\node [scale = .50] [circle, draw, fill = black] at (-3,-1.5)  {};
\node [scale = .50] [circle, draw, fill = black] at (-3,-.5)  {};
\node [scale = .50] [circle, draw, fill = black] at (3,1.5)  {};
\node [scale = .50] [circle, draw, fill = black] at (3,.5)  {};

\draw (-3.8,-.5) node [scale =1.0] {$S^-_\mu$};
\draw (-3.8,-1.5) node [scale =1.0] {$N^-_\mu$};
\draw (3.9,1.5) node [scale =1.0] {$N^+_\mu$};
\draw (3.9,.5) node [scale =1.0] {$S^+_\mu$};

\end{tikzpicture}
\]
\captionsetup{format=hang}
\caption{\small{The nullclines $\Gamma$ depicted in the fundamental domain $\mc{C}_* = [x_-, x_+] \times [y_0, y_0 + 2\pi)$. In this picture we have $P \in \mc{A}$ and $Q,R \in \mc{B}$.}}
\end{figure}
 
 \medskip
 \medskip
 
 \begin{thm}\label{verification of (e) thm}
 Let $\mc{C}$ be a finite cylinder satisfying assumptions \emph{(a) - (d)} in the previous section. Suppose that for each $\mu \in I$, we have the following two conditions
 \begin{itemize}
 \item[\emph{(1)}] $N^\pm_\mu, S^\pm_\mu \notin \mc{B}$ 
 
 \item[\emph{(2)}] the cardinality of $\mc{A} \cup \mc{B}$ is finite.
 \end{itemize}
 Then $\mc{C}$ satisfies assumption \emph{(e)}. 
 \end{thm}

\proof
We will prove that the $\omega$-limit set of any non-boundary orbit is precisely a one-point set containing an equilibrium point on $x = x_+$. The corresponding proof for the $\a$-limit set is analogous. We will work in the fundamental domain $\mc{C}_*$ as the proof is easier to see geometrically in $\mc{C}_*$.

By condition (1), we can locally graph $\Gamma$ around $N^+_\mu$ and $S^+_\mu$ as a smooth function of $x$. Call these graphs $f_N(x)$ and $f_S(x)$, respectively. Let $L$ denote the vertical line given by $x = x_+ - \e$ where $\e > 0$ is chosen small enough so that $L$ intersects $f_N$ and $f_S$ and such that $f_N(x) \neq f_S(x)$ for all $x \in [x_+ - \e, x_+]$. Let $R := \{(x,y) \in \mc{C} \mid x_+ - \e < x < x_+\}$ denote the region between $L$ and $x = x_+$. Choosing $\e$ even smaller, we can assume that the minimum of $f_N(x)$ is strictly greater than the maximum of $f_S(x)$. Then $f_N$ and $f_S$ divide the region $R$ into two open sets given by $R \cap \mc{N}$ and $R \cap \mc{P}$.

To visualize $R \cap \mc{N}$ and $R \cap \mc{P}$, consider the fundamental domain $\mc{C}_*$. The nullclines divide region $R$ within $\mc{C}_*$ into three open sets I, II, and III where region I corresponds to points above $f_N$, region II corresponds to points between $f_N$ and $f_S$, and region III corresponds to points below $f_S$. Consider the slope field along the boundary $x = x_+$. Above $N^+_\mu$ we have $\dot{y} < 0$, between $N^+_\mu$ and $S^+_\mu$ we have $\dot{y} > 0$, and below $S^+_\mu$ we have $\dot{y} < 0$. Consequently, we have $\dot{y} < 0$ in regions I and III and $\dot{y} > 0$ in region II. Therefore, within the fundamental domain $\mc{C}_*$, we have $\text{I} \cup \text{III} = R \cap \mc{N}$ and $\text{II} = R \cap \mc{P}$.

 By condition (2), we can choose $\e > 0$ even smaller so that $\mc{A} \cup \mc{B}$ does not intersect $R$. Therefore $f'_N$ is either positive or negative for all $x \in (x_+ -\e, x_+)$. Likewise with $f_S'$. Therefore there are four cases to consider depending on the signs of $f'_N$ and $f'_S$. We will verify assumption (e) in all four cases. Fix a non-boundary orbit $\sigma(\tau) = \big(x(\tau), y(\tau)\big)$ in $\mc{C}$. Within the fundamental domain $\mc{C}_*$, either $\sigma$ enters region I, or region II, or region III. 
 
Suppose first that $f'_N$ is positive and $f'_S$ is negative. If $\sigma$ enters region II, then it cannot cross $f_N$ nor $f_S$ since the sign of their slopes forbid it. Therefore, if it enters region II, then it must terminate at $N^+_\mu$. If it enters region I, then either it terminates at $N^+_\mu$ or it crosses $f_N$ and enters region II in which case it terminates at $N^+_\mu$. If it enters region III, then either it terminates at $S^+_\mu$ or it winds around the cylinder entering region I in which case it terminates at $N^+_\mu$. 
 
Now suppose that $f'_N$ is negative and $f'_S$ is positive. If $\sigma$ enters region I, then it must terminate at $N^+_\mu$ since it cannot cross $f_N$. Likewise, if it enters region III, then it winds around the cylinder and terminates at $N^+_\mu$. Suppose it enters region II. If it crosses $f_N$, then it enters region I and hence terminates at $N^+_\mu$. If it crosses $f_S$, then it enters region III and hence winds around the cylinder and terminates at $N^+_\mu$. If it crosses neither $f_N$ nor $f_S$, then it remains in region II and so either terminates at $N^+_\mu$ or $S^+_\mu$. 
 
Now suppose that $f'_N$ and $f'_S$ are both negative.  If $\sigma$ enters region I, then it must terminate at $N^+_\mu$. If it enters region III, then it either terminates at $S^+_\mu$ or winds around the cylinder and terminate at $N^+_\mu$. If it enters region II, then it can cross $f_N$ but not $f_S$. If it crosses $f_N$, then it enters region I and terminates at $N^+_\mu$. If it does not cross $f_N$, then it remains in region II and terminates at $N^+_\mu$. 

Lastly, suppose that $f'_N$ and $f'_S$ are both positive. If $\sigma$ enters region I, then it either terminates at $N^+_\mu$ or crosses $f_N$ in which case it enters region II. If it enters region II, then it either terminates at $N^+_\mu$ or $S^+_\mu$ or crosses $f_S$ and enters region III. If it enters region III, then it winds around the cylinder and enters region I. Hence it suffices to show that when it enters region II, it cannot cross $f_S$. Indeed recall that we chose $L$ so close to $x = x_+$ such that the minimum of $f_N$ is strictly greater than the maximum of $f_S$. This implies that once it enters region II, it cannot cross $f_S$. 
\qed

\medskip
\medskip

\subsection{The $\Theta$ system}\label{th sec}

In this section we establish existence of saddles connectors for the $\Theta$ system (\ref{dynsysTh}). 

Let $a > 0$, $E \in [0,1]$, and $\kappa \in \Z + \frac{1}{2}$ be constants. Let $\l \in \R$ be a parameter. The system we are interested in is 
\[  \left\{
 \begin{array}{ll}
      \dot{\theta} \,=\, f(\theta) \,=\, \sin \theta   \\
      \dot{\Theta} \,=\, g_\lambda(\theta,\Theta) \,=\, -2a\sin \theta \cos\theta \cos \Theta + 2aE \sin^2 \theta \sin \Theta - 2\kappa \sin \Theta + 2\lambda \sin \theta.      
\end{array} 
\right. \]

The finite cylinder for this system is $\mc{C} = [0, \pi] \times \Sset^1$. The universal cover is $\wt{\mc{C}} = [0, \pi] \times \R$. The Jacobian for the system evaluated at $\theta = 0$ or $\pi$ is given by
\[
\begin{pmatrix}
\pm 1 & 0
\\
-2a\cos \Theta \pm 2 \l & -2\kappa \cos \Theta
\end{pmatrix}.
\]
Since $\kappa \neq 0$, we can use the Hartman-Grobman Theorem to characterize the equilibrium points. 

For $\kappa > 0$, we take the fundamental domain to be $\mc{C}_* = [0,\pi] \times [-\pi, \pi)$. The four equilibrium points occur at 
\[
S^- \,=\, (0, 0), \:\:\:\: N^-  \,=\, (0,-\pi), \:\:\:\: S^+ \,=\, (\pi, -\pi), \:\:\:\: N^+ \,=\, (\pi, 0).
\]
For $\kappa < 0$, we take the fundamental domain to be $\mc{C}_* = [0,\pi] \times [0, 2\pi)$ and the four equilibrium points occur at
\[
S^- \,=\, (0, \pi), \:\:\:\: N^-  \,=\, (0,0), \:\:\:\: S^+ \,=\, (\pi, 0), \:\:\:\: N^+ \,=\, (\pi, \pi).
\]

\medskip
\noindent\emph{Remarks.} The equilibrium points don't depend on $\lambda$ which is why we write $S^-$ instead of $S^-_\lambda$, etc. Also, the roles of $S$ and $N$ switch depending on the sign of $\kappa$. The fundamental domains where chosen in each case so that the ordering of the equilibrium point agrees with assumption (c) from the previous section. 

\medskip
\medskip

The linearization at $S^-$ for $\kappa > 0$ and $\kappa < 0$ is, respectively,
\[
\begin{pmatrix}
1 & 0 \\
-2a + 2\l & -2\kappa \\
\end{pmatrix} \:\:\:\: \text{ and } \:\:\:\: \begin{pmatrix} 
1 & 0 \\
2a + 2\l & 2\kappa \\
\end{pmatrix}.
\]
Therefore the tangent of the unstable manifold $\wt{\mc{W}}^-_\lambda$ at $S^-$ lies in the span of the eigenvector $(1, b)^T$ with eigenvalue $1$ where $b = (\l - a)/(\frac{1}{2} + \kappa)$ for $\kappa > 0$ and $b = (\l + a)/(\frac{1}{2} - \kappa)$ for $\kappa < 0$. 

The linearization at $S^+$ for $\kappa > 0$ and $\kappa < 0$ is, respectively, 
\[
\begin{pmatrix}
-1 & 0 \\
2a - 2\l & 2\kappa \\
\end{pmatrix} \:\:\:\: \text{ and } \:\:\:\: \begin{pmatrix} 
-1 & 0 \\
-2a - 2\l & -2\kappa \\
\end{pmatrix}.
\]
Therefore the tangent of the stable manifold $\wt{\mc{W}}^+_\lambda$ at $S^+$ lies in the span of the eigenvector $(1,b)^T$ with eigenvalue $-1$ where $b = (\l - a)/(\frac{1}{2} + \kappa)$ for $\kappa > 0$ and $b = (\l + a)/(\frac{1}{2} - \kappa)$. 

\medskip
\medskip

\noindent\emph{Remark.} The slope of the unstable manifold at $S^-$ equals the slope of the stable manifold at $S^+$. 

\medskip
\medskip

It is easy to verify that this system satisfies assumptions (a), (b), (c), and (d) from Section \ref{assumptions on flow section}.  Note that this system is special since the equilibria are independent of the parameter $\l$.  Assumption (d) is satisfied with parameter $\mu = -\lambda$.   We will denote the winding number of the system by $w_\lambda$ instead of $w_{-\lambda}$. We verify assumption (e) in the next section; it follows from an analysis of the nullclines. 

\medskip
\medskip

\subsection{Analysis of nullclines for the $\Theta$ system}

\medskip

 Recall that the nullclines are defined as the set of points $\Gamma = \big\{(\theta, \Theta) \in \mc{C} \mid g_\lambda(\xi, \Omega) = 0\big\}$ and separate the fundamental domain via $\mc{C} = \Gamma \cup \mc{N} \cup \mc{P}$ where $\mc{N}$ and $\mc{P}$ correspond to the points in $\mc{C}$ where $\dot{\Theta} < 0$ and $\dot{\Theta} > 0$, respectively. Define the sets $\mc{A} = \big\{(\theta,\Theta) \in \Gamma \mid \frac{\pd g_\lambda}{\pd \theta}(\theta, \Theta) = 0 \big\}$ and $\mc{B} = \big\{(\theta, \Theta) \in \Gamma \mid \frac{\pd g_\lambda}{\pd \Theta}(\theta, \Theta) = 0\big\}$. Recall that the set $\Gamma \setminus (\mc{A} \cap \mc{B})$ forms a smooth one-dimensional manifold.  If we imagine the nullclines as a graph over $\theta$, then, by the implicit function theorem, $\mc{A} \setminus \mc{B}$ corresponds to points where the slope of this graph is zero, while $\mc{B} \setminus \mc{A}$ corresponds to points where the slope of this graph is infinite. A quick computation establishes the following proposition. 

\medskip
\medskip

\begin{prop}\label{equi points for Th prop}
$N^\pm, S^\pm \notin \mc{B}$. 
\end{prop}

\medskip
\medskip

Verification of assumption (e) for the $\Theta$ system follows from applying Proposition \ref{equi points for Th prop} and the following lemma to Theorem \ref{verification of (e) thm}. 

\medskip
\medskip

\begin{lem}\label{resultant lem for Th}
The cardinality of $\mc{A} \cup \mc{B}$ is $\leq 64$.
\end{lem}

\proof
We first look at the cardinality of $\mc{B}$. Note that a point $(\theta, \Theta)$ belongs to $\mc{B}$ so long as $g_\lambda = \frac{\pd g_\lambda}{\pd \theta} = 0$ at $(\theta, \Theta)$. We have
\begin{align*}
g_\lambda(\theta, \Theta) \,&=\, -2a\sin\theta\cos\theta\cos\Theta + 2aE \sin^2\theta \sin \Theta -2\kappa \sin \Theta + 2\lambda \sin \theta
\\
\frac{\pd g_\lambda}{\pd \Theta}(\theta, \Theta) \,&=\, 2a\sin\theta\cos\theta\sin\Theta + 2aE\sin^2\theta\cos\Theta + 2\kappa\cos\Theta
\end{align*}
Introduce the variable $T = \tan \frac{\Theta}{2}$ so that $\cos\Theta = \frac{1-T^2}{1+T^2}$ and $\sin \Theta = \frac{2T}{1+T^2}$.  
Then 
\[
g_\lambda \,=\, \frac{2p}{1 + T^2} \:\:\:\: \text{ and } \:\:\:\: \frac{\pd g_E}{\pd \xi} \,=\, \frac{2q}{1 + T^2}
\]
where
\begin{align*}
p(T, \theta) \,&=\, \a(\theta)T^2 + b(\theta)T + c(\theta)
\\
q(T,\theta) \,&=\, d(\theta)T^2 + e(\theta)T + f(\theta)
\end{align*}
where
\begin{align*}
\a(\theta) \,&=\, a\sin\theta\cos\theta + \lambda \sin \theta  &&d(\theta) \,=\, -aE\sin^2\theta -\kappa
\\
b(\theta) \,&=\, 2aE\sin^2\theta -2\kappa  &&e(\theta) \,=\, 2a\sin\theta\cos\theta
\\
c(\theta) \,&=\, -a\sin\theta\cos\theta + \lambda\sin\theta  &&f(\theta) \,=\, aE\sin^2\theta + \kappa
\end{align*}
(We use the Greek letter $\a$ instead of $a$ in our notation since since we are already using $a$ to denote the ring radius in z$G$KN). Note that a point $(\theta, \Theta)$ belongs to $\mc{B}$ so long as $p = q = 0$ at $(\theta, \Theta)$. Make the substitution $t = \tan\frac{\theta}{2}$ so that $\cos\th = \frac{1 - t^2}{1 + t^2}$ and $\sin \th = \frac{2t}{1 + t^2}$. Then
\begin{align*}
A \,&:=\, \a(1+t^2)^2 \,=\, 2at(1-t^2) + 2\lambda t(1+t^2)  && D \,:=\, d(1+t^2) \,=\, -4aEt^2 -\kappa (1+t^2)^2
\\
B \,&:=\, b(1+t^2)^2 \,=\, 8aEt^2 -2\kappa(1+t^2)^2
&& E \,:=\, e(1+t^2)^2 \,=\,4a t(1-t^2)
\\
C \,&:=\, c(1+t^2)^2 \,=\, -2at(1-t^2) + 2\lambda t(1+t^2)
&& F \,:=\, f(1+t^2)^2 \,=\, 4aEt^2 -\kappa (1+t^2)^2.
\end{align*}
Define 
\begin{align*}
P \,&:=\, p(1+t^2)^2 \,=\, AT^2 + BT + C
\\
Q\,&:=\, q(1+t^2)^2 \,=\, D T^2 + E T + F.
\end{align*} 
Note that $p = q = 0$ if and only if $P = Q =0$. This holds if and only if the resultant of $P$ and $Q$ equals zero where the resultant is given by
\[
\text{res}_T(P,Q) \,=\, \det \begin{pmatrix}
A & B & C & 0
\\
0 & A & B & C
\\
D & E & F & 0
\\
 0 & D & E & F
\end{pmatrix}.
\]
Note that $\text{res}_T(P,Q)$ is a polynomial in $t$ with degree at most 16. Since $P$ and $Q$ are polynomials in $T$ with degree $2$, it follows that the cardinality of $\mc{B}$ is $\leq 32$. 

An analogous argument for $\mc{A}$ shows that its cardinality is $\leq 32$ as well. Hence the cardinality of $\mc{A} \cup \mc{B}$ is $\leq 64$.
\qed

\subsection{Obtaining saddles connectors for the $\Theta$ system}

\medskip

In the next couple sections we prove existence of saddles connectors. Uniqueness (with respect to $\l$) follows from the following proposition.

\medskip
\medskip

\begin{prop}\label{unique theta prop}
Fix an integer $N \geq 0$. Suppose $\wt{\mc{W}}_\l^-$ and $\wt{\mc{W}}_{\l'}^-$ are saddles connectors with winding numbers $w_\l = w_{\l'} = N$. Then $\l = \l'$. 
\end{prop}

\proof
We prove it for $\kappa > 0$. The proof is analogous for $\kappa < 0$. Seeking a contradiction, suppose $\l < \l'$. Let $\Theta_{\l}$ and $\Theta_{\l'}$ denote the corresponding $\Theta$-components of the saddles connectors. Since $-\l' < -\l$, Lemma \ref{lem 7.3 in JMP paper} implies that $\Theta_{\l'}(\tau) \geq \Theta_{\l}(\tau)$ for all $\tau \in \R$.  Therefore $\wt{\mc{W}}^-_{\l'}$ lies above $\wt{\mc{W}}^-_{\l}$. However, from the linearization taken at the right endpoint $S^+_N := (\pi, -\pi -2\pi N)$, we find
\[
\frac{d\Theta_{\l'}}{d\theta}\bigg|_{S^+_N} \,=\, \frac{\l' -a}{\frac{1}{2} + \kappa} \,>\, \frac{\l-a}{\frac{1}{2} + \kappa} \,=\, \frac{d\Theta_{\l}}{d\theta}\bigg|_{S^+_N}.
\]
Therefore there exists a point sufficiently close to $S^+_N$ such that $\wt{\mc{W}}^-_{\l'}$ lies below $\wt{\mc{W}}^-_{\l}$ at this point which is a contradiction.
\qed

\medskip
\medskip

In the next two sections we establish existence of saddles connectors. The first section proves existence of saddles connectors with nonnegative winding number; the second section proves existence of saddles connectors with negative winding number. 

\medskip
\medskip

\subsubsection{Obtaining saddles connectors with nonnegative winding number for the $\Theta$ system}

\medskip

We will use barrier arguments along with Proposition \ref{saddles conn prop} to find saddles connectors with nonnegative winding number for the $\Theta$ system.

\medskip
\medskip

We note the following solutions when $E = 1$.

\medskip

\begin{prop}\label{sol for E = 1 and la = -1 + a prop}
Set $E = 1$. 
\begin{itemize}
\item[$\bullet$] Suppose $\kappa > 0$. If $\l = -\frac{1}{2} - \kappa + a$, then $\Theta(\tau) = - \theta(\tau)$ is an orbit.

\item[$\bullet$] Suppose $\kappa < 0$. If $\l = -\frac{1}{2} + \kappa - a$, then $\Theta(\tau) = - \theta(\tau) + \pi$ is an orbit.
\end{itemize}
\end{prop}

\proof
Consider $\kappa > 0$. We have $\dot{\Theta} = -\dot{\theta} = -\sin \theta$. Plugging $E = 1$, $\l = -\frac{1}{2} - \kappa + a$, and $\Theta = - \theta$ into $g_\l$, we find $\dot{\Theta} = -\sin \theta$. The proof for $\kappa < 0$ is analogous. 
\qed

\medskip
\medskip

\begin{prop}\label{pos wind num theta prop 1}\: $\phantom{nix}$
\begin{itemize}
\item[$\bullet$] Suppose $\kappa > 0$. If $\l > -\frac{1}{2} - \kappa + a$, then $w_\l \leq 0$. 
\item[$\bullet$] Suppose $\kappa < 0$. If $\l > -\frac{1}{2} + \kappa - a$, then $w_\l \leq 0$. 
\end{itemize} 
\end{prop}

\proof
Consider $\kappa > 0$. We will show that the line $L = \{(\theta, \Theta) \mid \Theta = -\theta,\: \theta \in [0,\pi]\}$ is a barrier for $\wt{\mc{W}}^-_\l$. Set $\l_0 = - \frac{1}{2} - \kappa + a$. First note that $\l > \l_0 > -1$ implies that the unstable manifold starts off above $L$. Let $g_{E,\l}$ denote $\dot{\Theta}$ for the Omega system with arbitrary $E \in [0,1]$ and $\l > \l_0$.  We have
\[
g_{E,\l}(\theta, \Theta) \,=\, g_{1, \l_0}(\theta, \Theta) + 2a(E - 1)\sin^2\theta \sin\Theta + 2\big(\l-\l_0\big)\sin\theta.
\]
Comparing $g_{E,\l}$ with $g_{1, \l_0}$ on $L$ we find $g_{E, \l}(\theta, -\theta) > g_{1,\l_0}(\theta, -\theta)$  for all $\theta \in (0,\pi)$. By Proposition \ref{sol for E = 1 and la = -1 + a prop}, $\Theta = - \theta$ is a solution to the system for $E = 1$ and $\l = \l_0$ and hence the vector field of the flow is tangent to $L$. Therefore, for $E \leq 1$ and $\l > \l_0$, the flow crosses $L$ from below. Hence $L$ is a barrier for $\wt{\mc{W}}^-_\l$. 

For $\kappa < 0$ take $L$ to be the line $\Theta = - \theta + \pi$ and mimic the above proof. 
\qed

\medskip
\medskip

\begin{prop}\label{pos wind num theta prop 2}
Let $N \geq 0$ be an integer.
\begin{itemize}
\item[$\bullet$] Suppose $\kappa > 0$. If $\lambda < -(2N +1)(\frac{1}{2} + \kappa) - 2a$, then $w_\lambda \geq N + 1$.

\item[$\bullet$] Suppose $\kappa < 0$. If $\lambda < -(2N +1)(\frac{1}{2} - \kappa) - 2a$, then $w_\lambda \geq N + 1$.
\end{itemize}
\end{prop}

\proof
Consider $\kappa > 0$. Set $m = 2N + 1$. Fix $\l < -m(\frac{1}{2} + \kappa) - 2a$.  Consider the line $L = \{ (\theta, \Theta) \mid \Theta = -m\theta\}$ which connects $S^-$ to the $n\text{th}$ copy of $S^+$ in the universal cover $\wt{\mc{C}}$. We want to show that $L$ is a barrier for $\wt{\mc{W}}^-_\lambda$. Since the tangent of the unstable manifold at $S^-$ is $\big(1, (\l -a)/(\frac{1}{2} + \kappa)\big)^T$, the assumption $\l < -m(\frac{1}{2} + \kappa) - 2a$ implies $(\l -a)/(\frac{1}{2} + \kappa) < -m$. Therefore the unstable manifold starts off below $L$. 

Calculating the slope field along $L$, we find
\begin{align*}
\frac{g_\l(\theta, -m\theta)}{\sin \theta} \,&=\, -2a \cos\theta \cos m\theta - 2aE \sin \theta \sin m\theta + 2\kappa\frac{\sin m\theta}{\sin \theta} + 2\lambda      
\\
&\leq\, 2a + 2a + 2\kappa\frac{\sin m\theta }{\sin \theta} + 2\l
\\
&\leq\, 4a + 2\kappa m + 2\l
\\
&<\, -m.
\end{align*}
Therefore $\wt{\mc{W}}^-_\l$ cannot cross $L$. Hence we have shown $w_\l \geq N$. It suffices to show $w_\l \neq N$. Seeking a contradiction, suppose $w_\l = N$. Then $\wt{\mc{W}}^-_\l$ is a saddles connector with winding number $N$. Fix $\l'$ such that $\l < \l' < -m(\frac{1}{2} + \kappa) -2a$. Then $\wt{\mc{W}}^-_{\l'}$ lies below $L$. Since $g_{\l'} = g_\l + 2(\l' - \l)\sin \theta > g_\l$, we have $\wt{\mc{W}}^-_{\l'}$ lies above $\wt{\mc{W}}^-_\l$. But since saddles connectors are unique (Proposition \ref{unique theta prop}), the $\omega$-limit point of $\wt{\mc{W}}^-_{\l'}$ is strictly greater than the $\omega$-limit point of $\wt{\mc{W}}^-_{\l}$. But this implies $\wt{\mc{W}}^-_{\l'}$ crosses $L$ which is a contradiction.

For $\kappa < 0$ take $L$ to be the line $\Theta = -m\theta + \pi$ and mimic the above proof. 
\qed

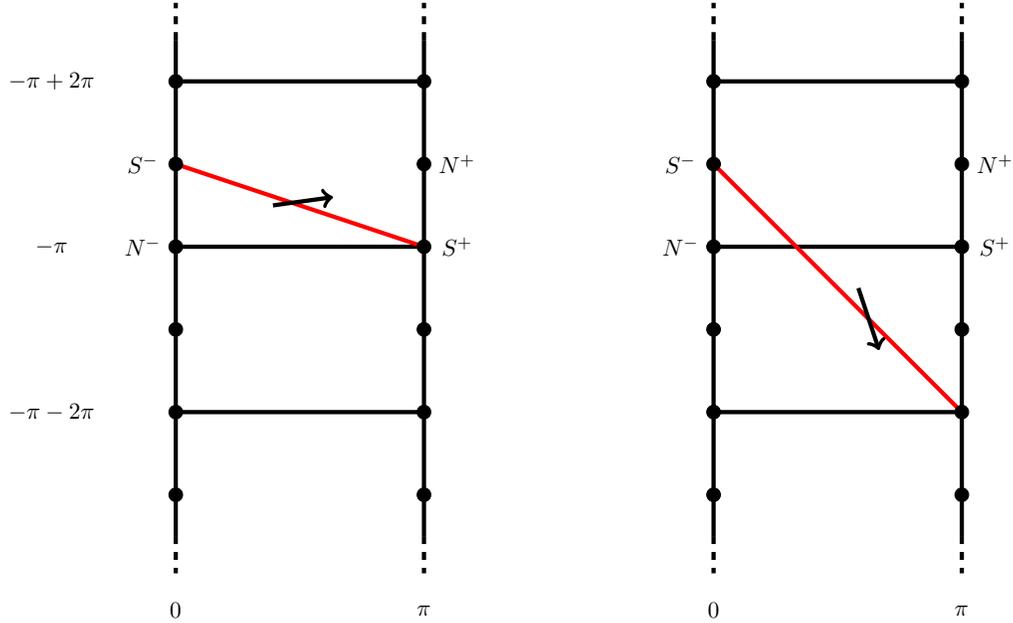
\begin{figure}[h]
\[
\begin{tikzpicture}[scale = 0.55]


\draw[ultra thick] (-3,-4) -- (-3, 8);
\draw[ultra thick] (3,-4) -- (3,8);

\draw[ultra thick, dashed] (-3,8) -- (-3,8.9);
\draw[ultra thick, dashed] (3,8) -- (3,8.9);
\draw[ultra thick, dashed] (-3,-4) -- (-3,-4.9);
\draw[ultra thick, dashed] (3,-4) -- (3,-4.9);

\draw[ultra thick] (-3,-1) -- (3,-1);
\draw[ultra thick] (-3,3) -- (3,3);
\draw[ultra thick] (-3,7) -- (3,7);

\draw[ultra thick, red] (-3,5) -- (3,3);

\draw[->, ultra thick] (-.65,4) -- (.8,4.2);

\node [scale = .50] [circle, draw, fill = black] at (-3,-1)  {};
\node [scale = .50] [circle, draw, fill = black] at (-3,1)  {};
\node [scale = .50] [circle, draw, fill = black] at (3,-1)  {};
\node [scale = .50] [circle, draw, fill = black] at (3,1)  {};

\node [scale = .50] [circle, draw, fill = black] at (-3,7)  {};
\node [scale = .50] [circle, draw, fill = black] at (-3,-3)  {};
\node [scale = .50] [circle, draw, fill = black] at (3,7)  {};
\node [scale = .50] [circle, draw, fill = black] at (3,-3)  {};

\node [scale = .50] [circle, draw, fill = black] at (-3,3)  {};
\node [scale = .50] [circle, draw, fill = black] at (-3,5)  {};
\node [scale = .50] [circle, draw, fill = black] at (3,3)  {};
\node [scale = .50] [circle, draw, fill = black] at (3,5)  {};

\draw (-3.8,5) node [scale = .8] {$S^-$};
\draw (-3.8,3) node [scale = .8] {$N^-$};
\draw (3.8,5) node [scale = .8] {$N^+$};
\draw (3.8,3) node [scale = .8] {$S^+$};

\draw (-3, -5.8) node [scale = .8] {$0$};
\draw (3, -5.8) node [scale = .8] {$\pi$};

\draw (-6,3) node [scale = .8] {$-\pi$};
\draw (-6,7) node [scale = .8] {$-\pi + 2\pi$};
\draw (-6,-1) node [scale = .8] {$-\pi -2\pi$};


\draw[ultra thick] (10,-4) -- (10, 8);
\draw[ultra thick] (16,-4) -- (16,8);

\draw[ultra thick, dashed] (10,8) -- (10,8.9);
\draw[ultra thick, dashed] (16,8) -- (16,8.9);
\draw[ultra thick, dashed] (10,-4) -- (10,-4.9);
\draw[ultra thick, dashed] (16,-4) -- (16,-4.9);

\draw[ultra thick] (10,-1) -- (16,-1);
\draw[ultra thick] (10,3) -- (16,3);
\draw[ultra thick] (10,7) -- (16,7);

\draw[ultra thick, red] (10,5) -- (16,-1);

\draw[->, ultra thick] (13.5,2) -- (14,0.5);

\node [scale = .50] [circle, draw, fill = black] at (10,-1)  {};
\node [scale = .50] [circle, draw, fill = black] at (10,1)  {};
\node [scale = .50] [circle, draw, fill = black] at (16,-1)  {};
\node [scale = .50] [circle, draw, fill = black] at (16,1)  {};

\node [scale = .50] [circle, draw, fill = black] at (10,7)  {};
\node [scale = .50] [circle, draw, fill = black] at (10,-3)  {};
\node [scale = .50] [circle, draw, fill = black] at (16,7)  {};
\node [scale = .50] [circle, draw, fill = black] at (16,-3)  {};

\node [scale = .50] [circle, draw, fill = black] at (10,3)  {};
\node [scale = .50] [circle, draw, fill = black] at (10,5)  {};
\node [scale = .50] [circle, draw, fill = black] at (16,3)  {};
\node [scale = .50] [circle, draw, fill = black] at (16,5)  {};

\draw (9.2,5) node [scale = .8] {$S^-$};
\draw (9.2,3) node [scale = .8] {$N^-$};
\draw (16.8,5) node [scale = .8] {$N^+$};
\draw (16.8,3) node [scale = .8] {$S^+$};

\draw (10, -5.8) node [scale = .8] {$0$};
\draw (16, -5.8) node [scale = .8] {$\pi$};

\end{tikzpicture}
\]
\captionsetup{format=hang}
\caption{\small{The figure on the left represents the barrier for Proposition \ref{pos wind num theta prop 1}, while the figure on the right represents the barrier for Proposition \ref{pos wind num theta prop 2}. Both figures are for $\kappa > 0$.  }} 
\end{figure}

\medskip
\medskip

\begin{thm}\label{exist theta system saddle connectors thm}
For any integer $N \geq 0$, there is a unique $\l_N$ such that $\wt{\mc{W}}^-_{\l_N}$ is a saddles connector with winding number $w_{\l_N} = N$ . Moreover $\l_N \geq \l_{N+1}$ and
\begin{itemize}
\item[$\bullet$] If $\kappa > 0$, then $\frac{1}{2} + \kappa - a \,\leq\, -\l_N \,\leq\, (2N + 1)\left(\frac{1}{2}+\kappa\right) + 2a$.
\item[$\bullet$] If $\kappa < 0$, then $\frac{1}{2} - \kappa + a \,\leq\, -\l_N \,\leq \,(2N + 1)\left(\frac{1}{2}-\kappa\right) + 2a$.
\end{itemize}

\end{thm}

\proof
Consider $\kappa > 0$. Set $-\l' = \frac{1}{2} +\kappa - a$ and $-\l'' = (2N + 1)(\frac{1}{2} +\kappa) + 2a$. Then Propositions \ref{pos wind num theta prop 1}, \ref{pos wind num theta prop 2}, and \ref{saddles conn prop} imply that for every $\e > 0$, there is a $-\l_N \in [-\l' - \e, -\l'' + \e)$ such that $\wt{\mc{W}}^-_{\l_N}$ is a saddles connector with winding number $N$. Taking $\e \to 0$, it follows that  $-\l_N \in [-\l', -\l'']$. Uniqueness follows from Proposition \ref{unique theta prop}. 

Now we prove the monotonicity condition.  Seeking a contradiction, suppose $\l_1 > \l_0$. Then $-\l_1 < -\l_0$. Therefore Lemma \ref{lem 7.3 in JMP paper} implies that $\Theta_{\l_1}(\tau) \geq \Theta_{\l_0}(\tau)$ for all $\tau \in \R$. But this contradicts the fact that $\Theta_{\l_1}(\tau) \to - 3\pi$ and $\Theta_{\l_0}(\tau) \to -\pi$ as $t \to \infty$. The same argument shows $\l_N \geq \l_{N+1}$ for each $N \geq 0$. 

The proof for $\kappa < 0$ is analogous.
\qed

\medskip
\medskip

\subsubsection{Obtaining saddles connectors with negative winding for the $\Theta$ system}

\medskip
\medskip

\begin{prop}\label{neg wind num theta prop 1}\:  $\phantom{nix}$
\begin{itemize}
\item[$\bullet$] Suppose $\kappa > 0$. If $\l < \frac{1}{2} + \kappa - a$, then $w_\l \geq 0$.

\item[$\bullet$] Suppose $\kappa < 0$. If $\l < \frac{1}{2} - \kappa-a$, then $w_\l \geq 0$. 
\end{itemize}
\end{prop}

\proof
Consider $\kappa > 0$. We will show that the line $L = \{(\theta, \Theta) \mid \Theta = \theta, \theta \in [0, \pi]\}$ is a barrier for $\wt{\mc{W}}_\l^-$. Calculating the slope field along $L$, we find 
\begin{align*}
\frac{g_\l(\theta, \theta)}{\sin \theta} \,&=\, -2a\cos^2\theta + 2aE \sin^2 \theta - 2\kappa + 2\l
\\
&\leq \, 0 + 2a - 2\kappa + 2\l
\\
&<\, 1.
\end{align*}
Since the slope of the unstable manifold $\mc{\wt{W}}_\l^-$
at $S^-$ is given by $(\l-a)/(\frac{1}{2} + \kappa) < 1$, we see that $L$ is a barrier for $\wt{\mc{W}}_\l^-$. It follows that $w_\l \geq -1$. It suffices to show that $w_\l \neq -1$. Seeking a contradiction, suppose $w_\l = -1$. Then $\wt{\mc{W}}^-_\l$ is a saddles connector with winding number $-1$. Fix $\l'$ such that $\l < \l' < -\frac{1}{2} + \kappa -a$. Then $\wt{\mc{W}}^-_{\l'}$ lies below $L$. Since $g_{\l'} = g_\l + 2(\l' - \l)\sin \theta > g_\l$, we have $\wt{\mc{W}}^-_{\l'}$ lies above $\wt{\mc{W}}^-_\l$. But since saddles connectors are unique (Proposition \ref{unique theta prop}), the $\omega$-limit point of $\wt{\mc{W}}^-_{\l'}$ is strictly greater than the $\omega$-limit point of $\wt{\mc{W}}^-_{\l}$. But this implies $\wt{\mc{W}}^-_{\l'}$ crosses $L$ which is a contradiction.

For the case $\kappa < 0$, one repeats the same argument above but takes $L$ to be given by the line $\Theta = \theta + \pi$. 
\qed

\medskip
\medskip

\begin{prop}\label{neg wind num theta prop 2}
Let $N \leq -1$ be an integer. 
\begin{itemize}
\item[$\bullet$] Suppose $\kappa > 0$. 
If $\l > -(2N+1)(\frac{1}{2} +\kappa) + 2a$, then $w_\l \leq N$. 

\item[$\bullet$] Suppose $\kappa < 0$. If $\l > -(2N+1)(\frac{1}{2} -\kappa) + 2a$, then $w_\l \leq N$. 
\end{itemize}
\end{prop}

\proof
Consider $\kappa > 0$. Set $m = -2N - 1$. Fix $\l > m(\frac{1}{2} +\kappa) + 2a$.  Consider the line $L = \{ (\theta, \Theta) \mid \Theta = m\theta\}$ which connects $S^-$ to the $N\text{th}$ copy of $S^+$ in the universal cover $\wt{\mc{C}}$. We want to show that $L$ is a barrier for $\wt{\mc{W}}^-_\lambda$. Since the tangent of the unstable manifold at $S^-$ is $\big(1, (\l -a)/(\frac{1}{2} + \kappa)\big)^T$, the assumption $\l > m(\frac{1}{2} + \kappa) + 2a$ implies $(\l -a)/(\frac{1}{2} + \kappa) > m$. Therefore the unstable manifold starts off above $L$. 

Calculating the slope field along $L$, we find
\begin{align*}
\frac{g_\l(\theta, m\theta)}{\sin \theta} \,&=\, -2a \cos\theta \cos m\theta - 2aE \sin \theta \sin m\theta + 2\kappa\frac{\sin m\theta}{\sin \theta} + 2\lambda      
\\
&\geq\, -2a - 2a + 2\kappa\frac{\sin m\theta }{\sin \theta} + 2\l
\\
&\geq\, -4a - 2\kappa m + 2\l
\\
&>\, m.
\end{align*}
Therefore $\wt{\mc{W}}^-_\l$ cannot cross $L$. Hence we have shown $w_\l \leq N$. 

For $\kappa < 0$ take $L$ to be the line $\Theta = m\theta + \pi$ and mimic the above proof. 
\qed

\medskip
\medskip

\begin{thm}\label{exist theta system saddle connectors thm for neg wind num}
For any integer $N \leq -1$, there is a unique $\l_{N}$ such that $\wt{\mc{W}}^-_{\l_{N}}$ is a saddles connector with winding number $N$. Moreover  $\l_{N} \geq \l_{N + 1}$ and
\begin{itemize}
\item[$\bullet$] If $\kappa > 0$, then $(2N+1)(\frac{1}{2}+\kappa) - 2a \,\leq \, -\l_{n} \,\leq\, -\frac{1}{2} - \kappa + a$.
\item[$\bullet$] If $\kappa < 0$, then  $(2N+1)(\frac{1}{2}-\kappa) - 2a \,\leq \, -\l_{n} \,\leq\, -\frac{1}{2} + \kappa + a$.
\end{itemize}
\end{thm}

\proof
The proof is similar to the proof of Theorem \ref{exist theta system saddle connectors thm} but uses Propositions \ref{neg wind num theta prop 1} and \ref{neg wind num theta prop 2} in place of Propositions \ref{pos wind num theta prop 1} and \ref{pos wind num theta prop 2}, respectively.
\qed

\medskip
\medskip

\subsection{The $\Omega$ system}\label{Om sec}

In this section we establish existence of saddles connectors for the $\Omega$ system (\ref{dynsysOm}). 

Set $a_{\text{max}} = 1 - \frac{1}{\sqrt{2}}$ and $\gamma_{\text{min}} =-\frac{1}{2}$.
Fix
\[
E \in (0,1), \:\:\:\: a \in (0,\, a_{\text{max}}),  \:\:\:\:  \g \in (\gamma_{\text{min}},\,0), \:\:\:\: \kappa \in \Z + \half,
\]
and 
\[ \l \,\leq\, \left\{
  \begin{array}{ll}
      -\frac{1}{2} - \kappa + a  & \text{ if } \kappa > 0   \\\\
      -\frac{1}{2} + \kappa -a & \text{ if } \kappa < 0    
\end{array} 
\right. \:\:\:\:\:\: \text{ or } \:\:\:\:\:\: \l \,\geq\, \left\{
  \begin{array}{ll}
      \frac{1}{2} + \kappa - a  & \text{ if } \kappa > 0   \\\\
      \frac{1}{2} - \kappa -a & \text{ if } \kappa < 0    
\end{array} 
\right. \]

\medskip
\medskip

\noindent\emph{Remark.} The negative values of $\l$ will correspond to winding numbers $N \geq 0$ for the $\Theta$-system, while the positive values of $\l$ will correspond to winding numbers $N \leq -1$ for the $\Theta$-system. These values for $\lambda$ are determined by Theorems \ref{exist theta system saddle connectors thm} and \ref{exist theta system saddle connectors thm for neg wind num}.

\medskip
\medskip

 \noindent The system we are interested in is
\[  \left\{
 \begin{array}{ll}
      \dot{\xi} \,=\, f(\xi) \,=\, \cos^2 \xi   \\
      \dot{\Omega} \,=\, g_E(\xi,\Omega) \,=\, 2a\sin \xi \cos \Omega + 2\lambda \cos \xi \sin \Omega + 2\gamma\sin\xi\cos\xi + 2\kappa\cos^2 \xi - 2aE    
\end{array} 
\right. \]
The finite cylinder for this system is $\mc{C} = [-\pi/2, \pi/2] \times \Sset^1$. The universal cover is $\wt{\mc{C}} = [-\pi/2, \pi/2] \times \R$. The Jacobian at $\xi = -\frac{\pi}{2}$ and $\xi = \frac{\pi}{2}$ is given by 
\[
\begin{pmatrix}
0 & 0
\\
\pm 2\lambda \sin \Omega - 2\gamma & \pm 2a \sin \Omega
\end{pmatrix}
\]
where `$+$' corresponds to $\xi = -\frac{\pi}{2}$ and `$-$' corresponds to $\xi = \frac{\pi}{2}$. Note that the Jacobian does not depend on $\kappa$ and so the quality of the equilibrium points does not change with $\kappa$; this is unlike what happens in the $\Theta$-system.

The four equilibrium points occur at $S^\pm_E = (\pm \frac{\pi}{2},\, s^\pm_E)$ and $N^\pm_E = (\pm \frac{\pi}{2}, \, s^\pm_E)$ 
\[
s_E^- \,=\, -\pi + \cos^{-1}E, \:\:\:\: n_E^- \,=\, -\pi - \cos^{-1}E, \:\:\:\: s_E^+ \,=\, -\cos^{-1}E,\:\:\:\: n_E^+ \,=\, \cos^{-1}E. 
\]

We choose the principal branch $0 \leq \cos^{-1}x \leq \pi$ for $\cos^{-1}$. Since $E \in (0,1)$, we have $0 < \cos^{-1} E < \pi/2$.  We choose our fundamental domain of the universal cover $\wt{\mc{C}}$ to be $\mc{C}_* = [-\pi/2, \pi/2] \times [-3\pi/2, \pi/2)$. Note that the equilibrium points depend on the parameter $E$; this is unlike what happens in the $\Theta$-system. 

Assumptions (a), (b), and (c) from section \ref{assumptions on flow section} are easy to verify. Note that assumption (d) is satisfied with parameter $\mu = E$. Now we verify assumption (d). We have $f'(-\frac{\pi}{2}) = f'(\frac{\pi}{2}) = 0$ and so the equilibrium points are non hyperbolic. For example the linearization at $S^-_E$ is 
\[
\begin{pmatrix}
0 & 0
\\
-2(\g + \l\sqrt{1 - E^2}) & -2a\sqrt{1 - E^2}
\end{pmatrix}.
\]
The eigenvalues are $0$ and $-2a\sqrt{1 - E^2}$.  The eigenvector for $0$ is 
\[
\begin{pmatrix}
1
\\
\frac{-\g - \l\sqrt{1 - E^2}}{a\sqrt{1 - E^2}}
\end{pmatrix}.
\]

The equilibrium points $S^\pm_E$ and $N^\pm_E$ correspond to saddle-nodes. Their local behavior is determined by Theorem 2.19(iii) in \cite{QTPDS}. Their phase portraits are depicted in Figure 2.13(c) in \cite{QTPDS}. The uniqueness of the unstable manifolds emanating from $S^\pm_E$ follows from this theorem. For example the unstable manifold emanating from $S^-_E$ will have the eigenvector corresponding to eigenvalue 0 lie in its tangent space at $S^-_E$. This completes the verification of assumption (d). 

\medskip
\medskip

\noindent\emph{Remark.} Note that Theorem 2.19(iii) in \cite{QTPDS} only applies when $E < 1$. When $E = 1$ the linearization at the equilibrium point changes. This will be important in the next section when we establish the existence of saddle-connectors. 

\medskip
\medskip

We verify assumption (e) in the next section; it follows from an analysis of the nullclines. 

\medskip
\medskip

\begin{subsection}{Analysis of nullclines for the $\Omega$ system}

\end{subsection}

  Recall that the nullclines are defined as the set of points $\Gamma = \big\{(\xi, \Omega) \in \mc{C} \mid g_E(\xi, \Omega) = 0\big\}$ and separate the fundamental domain via $\mc{C} = \Gamma \cup \mc{N} \cup \mc{P}$ where $\mc{N}$ and $\mc{P}$ correspond to the points in $\mc{C}$ where $\dot{\Omega} < 0$ and $\dot{\Omega} > 0$, respectively. Define the sets $\mc{A} = \big\{(\xi,\Omega) \in \Gamma \mid \frac{\pd g_E}{\pd \xi}(\xi, \Omega) = 0 \big\}$ and $\mc{B} = \big\{(\xi, \Omega) \in \Gamma \mid \frac{\pd g_E}{\pd \Omega}(\xi, \Omega) = 0\big\}$. Recall that the set $\Gamma \setminus (\mc{A} \cap \mc{B})$ forms a smooth one-dimensional manifold.  If we imagine the nullclines as a graph over $\xi$, then, by the implicit function theorem, $\mc{A} \setminus \mc{B}$ corresponds to points where the slope of this graph is zero, while $\mc{B} \setminus \mc{A}$ corresponds to points where the slope of this graph is infinite.
 
 Although $E$ takes values in $(0,1)$ for our cylinder $\mc{C}$, we will examine the nullclines for $E = 1$ as well since it will be used to help prove the existence of saddles connectors. A quick computation establishes the following proposition. 

\medskip
\medskip

\begin{prop}\label{equi points for Om prop}\: $\phantom{nix}$
\begin{itemize}
\item[$\bullet$] If $E \in (0,1)$, then $N^\pm_E, S^\pm_E \notin \mc{B}$.
\item[$\bullet$] If $E = 1$, then $N^\pm_E = S^\pm_E \in \mc{B} \setminus \mc{A}$. 
\end{itemize}
\end{prop}

\medskip
\medskip

Verification of assumption (e) for the $\Omega$ system follows from applying the first bullet point of Proposition \ref{equi points for Om prop} and the following lemma to Theorem \ref{verification of (e) thm}. 

\medskip
\medskip

\begin{lem}\label{resultant lem}
The cardinality of $\mc{A} \cup \mc{B}$ is $\leq 64$ for all $E \in (0,1]$.  
\end{lem}

\proof
We first look at the cardinality of $\mc{A}$. The proof for the set $\mc{B}$ is similar. Note that a point $(\xi, \Omega)$ belongs to $\mc{A}$ so long as $g_E = \frac{\pd g_E}{\pd \xi} = 0$ at $(\xi, \Omega)$. We have
\begin{align*}
g_E(\xi, \Omega) \,&=\, 2a\sin \xi \cos \Omega + 2\lambda \cos \xi \sin\Omega + 2\gamma \sin \xi \cos \xi + 2\kappa\cos^2\xi -2aE
\\
\frac{\pd g_E}{\pd \xi}(\xi, \Omega) \,&=\, 2a\cos\xi\cos \Omega -2\lambda\sin \xi \sin \Omega + 2\gamma (\cos^2\xi - \sin^2\xi)  - 4\kappa\sin \xi \cos \xi. 
\end{align*}
Introduce the variable $T = \tan \frac{\Omega}{2}$ so that $\cos\Omega = \frac{1-T^2}{1+T^2}$ and $\sin \Omega = \frac{2T}{1+T^2}$. Then 
\[
g_E \,=\, \frac{2p}{1 + T^2} \:\:\:\: \text{ and } \:\:\:\: \frac{\pd g_E}{\pd \xi} \,=\, \frac{2q}{1 + T^2}
\]
where
\begin{align*}
p(T, \xi) \,&=\, \a(\xi)T^2 + b(\xi)T + c(\xi)
\\
q(T,\xi) \,&=\, \a'(\xi)T^2 + b'(\xi)T + c'(\xi)
\end{align*}
where
\begin{align*}
\a(\xi) \,&=\, \gamma \sin \xi \cos \xi + \kappa\cos^2\xi - aE - a\sin \xi
\\
b(\xi) \,&=\, 2\lambda \cos \xi 
\\
c(\xi) \,&=\, \gamma \sin \xi \cos \xi + \kappa\cos^2\xi - aE + a\sin\xi
\end{align*}
and $\a'$, $b'$, and $c'$ are the derivatives with respect to $\xi$ (we use the Greek letter $\a$ instead of $a$ in our notation since since we are already using $a$ to denote the ring radius in z$G$KN). 

Note that a point $(\xi, \Omega)$ belongs to $\mc{A}$ so long as $p = q = 0$ at $(\xi, \Omega)$. Make the substitution $t = \tan\frac{\xi}{2}$ so that $\cos\xi = \frac{1 - t^2}{1 + t^2}$ and $\sin \xi = \frac{2t}{1 + t^2}$. Then 
\begin{align*}
A \,&:=\, \a(1+t^2)^2 \,=\, 2\g t(1-t^2) + \kappa (1-t)^2 - aE(1 + t)^2 -2at(1+t^2)
\\
B \,&:=\, b(1+t^2)^2 \,=\, 2\l(1 - t^4)
\\
C \,&:=\, c(1+t^2)^2 \,=\, 2\gamma t(1-t^2) + \kappa(1-t^2)^2 -aE (1+t^2)^2 + 2at(1+t^2)
\\
A' \,&:=\, \a'(1+t^2)^2 \,=\, \gamma(1 - 6t^2 + t^4) -4\kappa t(1-t^2) -a(1-t^4)
\\
B' \,&:=\, b'(1+t^2)^2 \,=\,-4\lambda t(1+t^2)
\\
C' \,&=\, c'(1+t^2)^2 \,=\, \gamma(1 -6t^2 + t^4) -4\kappa t(1-t^2) + a(1-t^4).
\end{align*}
Define 
\begin{align*}
P \,&:=\, p(1+t^2)^2 \,=\, AT^2 + BT + C
\\
Q\,&:=\, q(1+t^2)^2 \,=\, A' T^2 + B' T + C'.
\end{align*} 
Note that $p = q = 0$ if and only if $P = Q =0$. This holds if and only if the resultant of $P$ and $Q$ equals zero where the resultant is given by
\[
\text{res}_T(P,Q) \,=\, \det \begin{pmatrix}
A & B & C & 0
\\
0 & A & B & C
\\
A' & B' & C' & 0
\\
 0 & A' & B' & C'
\end{pmatrix}.
\]
Note that $\text{res}_T(P,Q)$ is a polynomial in $t$ with degree at most 16. Since $P$ and $Q$ are polynomials in $T$ with degree $2$, it follows that the cardinality of $\mc{A}$ is $\leq 32$. 

An analogous argument shows that the cardinality of $\mc{B}$ is $\leq 32$. Hence the  cardinality of $\mc{A} \cup \mc{B}$ is $\leq 64$. 
\qed

\medskip
\medskip

The following proposition will be used when we establish the existence of saddles connectors in the next section.

\medskip
\medskip

\begin{prop}\label{monotone prop}
Let $E = 1$. Then there is an $\e > 0$ such that the region $\frac{\pi}{2} - \e < \xi < \frac{\pi}{2}$ in the fundamental domain $\mc{C}_*$ is contained in $\mc{N}$.
\end{prop}

\proof

Let $R$ denote the region $\frac{\pi}{2} - \e < \xi < \frac{\pi}{2}$ in the fundamental domain $\mc{C}_*$. We first show that there is an $\e > 0$ such that the nullclines $\Gamma$ do not intersect $R$. 

By Lemma \ref{resultant lem}, we can choose $\e > 0$ small enough so that $\mc{A} \cup \mc{B}$ does not intersect the region $R$. Therefore if the nullclines $\Gamma$ did intersect $R$, then they can be locally parameterized as smooth graphs over $\xi$ with nonvanishing derivative. Let $P$ denote the point $P := N^+_1 = S^+_1$. Since $P$ is the only equilibrium point on $\xi = \frac{\pi}{2}$, it follows that any graph of $\Gamma$ must terminate at $P$. Since $\Omega = 0$ at $P$, it suffices to show that there is a $\delta > 0$ such that the nullclines do not intersect the region $R' := \{(\xi, \Omega) \in R \mid -\delta < \Omega < \delta\}$. 

By the second bullet point of Proposition \ref{equi points for Om prop}, we have $P \notin \mc{A}$. Therefore we can locally graph the nullcline as a smooth graph over $\Omega$. A calculation shows that the second derivative of this graph at $P$ is $-\frac{a}{\gamma} > 0$; hence $P$ is a local minimum of the graph. Therefore there is a $\delta > 0$ such that the graph does not intersect $R'$.

Thus we have shown that the nullclines do not intersect $R$. Since $\dot{\Omega} < 0$ for points on $\xi = \frac{\pi}{2}$ away from $P$ and the nullclines do intersect $R$, it follows that $\dot{\Omega} < 0$ in $R$, i.e. $R \subset \mc{N}$. 
\qed

\medskip
\medskip

 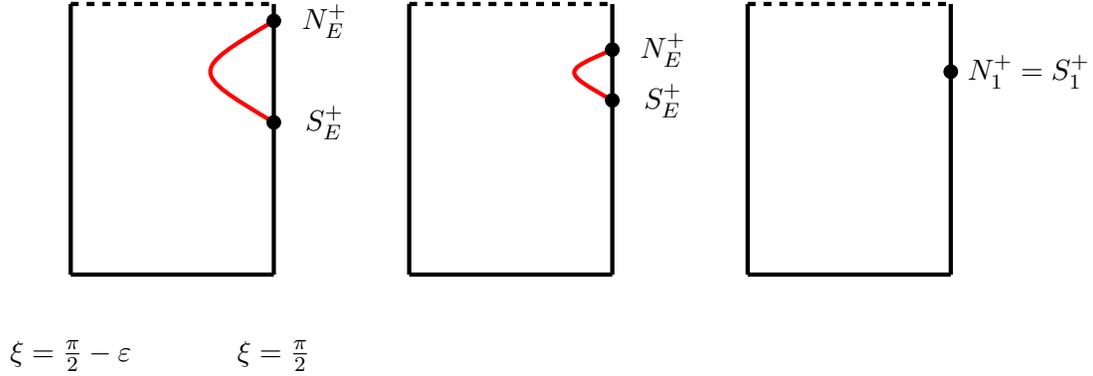
\begin{figure}[h]
\[
\begin{tikzpicture}[scale = 0.9]


\draw[ultra thick] (0,-2) -- (0, 2);
\draw[ultra thick] (3,-2) -- (3,2);
\draw[ultra thick] (0,-2) -- (3,-2);
\draw[ultra thick, dashed] (0,2) -- (3,2);

\draw (0, -3.2) node [scale = 1.0] {$\xi = \frac{\pi}{2} - \e$};
\draw (3, -3.2) node [scale = 1.0] {$\xi = \frac{\pi}{2}$};



\draw[ultra thick, red] (3,.25) .. controls (1.75, 1) .. (3, 1.75);


 {};
\node [scale = .50] [circle, draw, fill = black] at (3,1.75)  {};
\node [scale = .50] [circle, draw, fill = black] at (3,.25)  {};

\draw (3.75,1.75) node [scale =1.0] {$N^+_E$};
\draw (3.75,.25) node [scale =1.0] {$S^+_E$};



\draw[ultra thick] (5,-2) -- (5, 2);
\draw[ultra thick] (8,-2) -- (8,2);
\draw[ultra thick] (5,-2) -- (8,-2);
\draw[ultra thick, dashed] (5,2) -- (8,2);


\draw[ultra thick, red] (8,1.325)  .. controls (7.25, 1) .. (8,.575) ;


 {};
\node [scale = .50] [circle, draw, fill = black] at (8,1.325)  {};
\node [scale = .50] [circle, draw, fill = black] at (8,.575)  {};

\draw (8.75,1.325) node [scale =1.0] {$N^+_E$};
\draw (8.75,.575) node [scale =1.0] {$S^+_E$};



\draw[ultra thick] (10,-2) -- (10, 2);
\draw[ultra thick] (13,-2) -- (13,2);
\draw[ultra thick] (10,-2) -- (13,-2);
\draw[ultra thick, dashed] (10,2) -- (13,2);


 {};
\node [scale = .50] [circle, draw, fill = black] at (13,1)  {};

\draw (14.15,1) node [scale =1.0] {$N^+_1 = S^+_1$};
\

\end{tikzpicture}
\]
\captionsetup{format=hang}
\caption{\small{The region $R$ in Proposition \ref{monotone prop}. $E$ increases from left to right until $E = 1$. The nullclines are sketched in each case; they vanish when $E = 1$. Hence $R \subset \mc{N}$. }}
\end{figure}

\medskip
\medskip

\subsection{Obtaining saddles connectors for the $\Omega$ system}

We set out to establish existence and uniqueness (with respect to $E$) of saddles connectors for the $\Omega$ system. Uniqueness follows from the following proposition.

\medskip
\medskip

\begin{prop}\label{unique omega prop}
Fix an integer $N \geq 0$. Suppose $\wt{\mc{W}}^-_E$ and $\wt{\mc{W}}^-_{E'}$ are saddles connectors with winding numbers $w_E = w_{E'} = N$. Then $E = E'$. 
\end{prop}

\proof
Seeking a contradiction, suppose $\wt{\mc{W}}^-_E$ and $\wt{\mc{W}}^-_{E'}$ are saddles connectors with $E < E'$. Then $s^-_{E'} < s^-_E$. Since $g_{E'} = g_E - 2a(E' - E) < g_E$, it follows that $\wt{\mc{W}}^-_E$ is a barrier for $\wt{\mc{W}}^-_{E'}$. However, $s^+_{E'} > s^+_E$. Therefore $\wt{\mc{W}}^-_{E'}$ must have crossed $\wt{\mc{W}}^-_{E}$ which is a contradiction. 
\qed

\medskip
\medskip

For the remainder of this section we break up the propositions/theorems into two cases: $\l < 0$ and $\l > 0$. Note that the cases $\l < 0$ and $\l > 0$ imply, respectively, 
\[ \l \,\leq\, \left\{
  \begin{array}{ll}
      -\frac{1}{2} - \kappa + a  & \text{ if } \kappa > 0   \\\\
      -\frac{1}{2} + \kappa -a & \text{ if } \kappa < 0    
\end{array} 
\right. \:\:\:\:\:\: \text{ and } \:\:\:\:\:\: \l \,\geq\, \left\{
  \begin{array}{ll}
      \frac{1}{2} + \kappa - a  & \text{ if } \kappa > 0   \\\\
      \frac{1}{2} - \kappa -a & \text{ if } \kappa < 0    
\end{array} 
\right. \]

We will apply Proposition \ref{saddles conn prop} to prove the existence of a saddles connector. The following two propositions involving barriers will be used. The values of $a_{\rm{max}}$ and $\gamma_{\rm{min}}$ were chosen so that the next two propositions hold; they are probably not optimal. 

\medskip
\medskip

\begin{prop}\label{horiz barr prop}\: $\phantom{nix}$
\begin{itemize}
\item[$\bullet$] If $\l < 0$, then $\dot{\Omega}< 0$ for all $\xi$ on the horizontal line $\Omega = \frac{\pi}{2}$.
\item[$\bullet$] If $\l > 0$, then $\dot{\Omega}<0$ for all $\xi$ on the horizontal line $\Omega = -\frac{\pi}{2}$. 
\end{itemize}
\end{prop}

\proof
We prove it for $\l < 0$. The case for $\l > 0$ follows the exact same proof. Whether or not $\kappa$ is positive, we have $\l \leq -\frac{1}{2} -|\kappa| + a$. Evaluating $\dot{\Omega}$ at $\Omega = \pi/2$ gives
\begin{align*}
j(\xi) \,:=\, \dot{\Omega}(\xi, \pi/2) \,&=\, 2\l \cos \xi + 2 \gamma \sin \xi \cos\xi + 2\kappa\cos^2 \xi - 2aE
\\
&\leq\, 2\left(-\frac{1}{2} - |\kappa| + a\right) \cos\xi + 2\gamma \sin\xi \cos\xi + 2\kappa\cos^2 \xi - 2aE.
\\
&=:\, h(\xi)
\end{align*}
Thinking of $h$ as a function of $a, E, \gamma,$ and $\kappa$, we have $\frac{\pd h}{\pd E} = -2a < 0$. Therefore 
\begin{align*}
h(\xi) \,&\leq\, \cos\xi\big(-(1+2|\kappa|) + 2\gamma \sin\xi + 2\kappa\cos\xi + 2a\big)
\\
&=:\, \ell(\xi).
\end{align*}
Then $\frac{\pd \ell}{\pd a}= 2\cos \xi \geq 0$ on $[-\frac{\pi}{2}, \frac{\pi}{2}]$. Therefore
\begin{align*}
\ell(\xi) \,\leq\, \cos \xi\big( -(1+2|\kappa|) + 2\gamma \sin\xi + 2\kappa\cos\xi + 2a_{\text{max}}\big)
\end{align*}
Consider $\kappa > 0$. Then $\frac{\pd \ell}{\pd \kappa} = \cos\xi(-2 + 2\cos \xi) \leq 0$ on $[-\frac{\pi}{2}, \frac{\pi}{2}]$. Therefore, for $\kappa > 0$, we have
\begin{align*}
\ell(\xi) \,&\leq\, \cos\xi\big( -2 + 2\gamma \sin \xi +  \cos \xi + 2a_{\text{max}}\big)
\\
&=:\, m(\xi).
\end{align*}
We have $\frac{\pd m}{\pd \g} = 2\cos \xi \sin \xi$. First consider the interval $[-\frac{\pi}{2}, 0]$. Then $\frac{\pd m}{\pd \g} \leq 0$ on $[-\frac{\pi}{2}, 0]$. Therefore 
\begin{align*}
m(\xi) \,&\leq\, \cos\xi\big(-2 + \sin \xi + \cos \xi + 2a_{\text{max}}\big)
\\
&=\, \cos\xi\big(-2 + \sqrt{2}\sin(\pi/4 - x) + 2a_{\text{max}} \big).
\end{align*}
Since $a_{\text{max}} = 1 - \frac{1}{\sqrt{2}}$, we have $m(\xi) \leq 0$ on $[-\frac{\pi}{2}, 0]$. For the interval $[0, \frac{\pi}{2}]$, we have $\frac{\pd m}{\pd \gamma} \geq 0$ and so $m(\xi) \leq \cos\xi(-2 + \cos \xi + 2a_{\text{max}})$, and so the result follows on this interval as well. 

Now consider $\kappa < 0$. Then $\frac{\pd \ell}{\pd\kappa} = \cos\xi(2 + 2\cos \xi) \geq 0$ on $[-\frac{\pi}{2}, \frac{\pi}{2}]$. Therefore, for $\kappa < 0$, we have
\begin{align*}
\ell(\xi) \,\leq \, \cos\xi\big(-2 + 2\gamma \sin \xi - \cos \xi + 2a_{\text{max}}). 
\end{align*}
An analogous argument for the $\kappa > 0$ case shows that $\ell(\xi) \leq 0$. 
\qed

\medskip
\medskip

\noindent\emph{Remark.} Proposition \ref{horiz barr prop} shows that that the winding for the $\Omega$-system satisfies $w_E \geq 0$ for all $E \in (0,1)$ if $\l < 0$, while $w_E \geq 1$ for all $E \in (0,1)$ if $\l > 0$.

\medskip
\medskip

\begin{prop}\label{Omega winding leq 0 prop}\: $\phantom{nix}$
\begin{itemize}
\item[$\bullet$] Let $\l < 0$. There exists an $E' \in (0,1)$ such that for all $E \in (0, E')$, we have $w_E = 0$.
\item[$\bullet$] Let $\l > 0$. There exists an $E' \in (0,1)$ such that for all $E \in (0, E')$, we have $w_E = 1$.
\end{itemize}
\end{prop}

\proof
By Proposition \ref{horiz barr prop}, we have $w_E \geq 0$ if $\l < 0$ and $w_E \geq 1$ if $\l > 0$. Therefore it suffices to show that $w_E \leq 0$ if $\l < 0$ and $w_E \leq 1$ if $\l < 0$ for all sufficiently small $E$. 

We break up the proof into four cases:

\begin{itemize}
\item[(1)] $\lambda < 0$ and $\kappa > 0$
\item[(2)] $\lambda < 0$ and $\kappa < 0$
\item[(3)] $\lambda > 0$ and $\kappa > 0$
\item[(4)] $\lambda > 0$ and $\kappa < 0$. 
\end{itemize}

For cases (1) and (2), we want to construct a barrier for $\wt{\mc{W}}^-_E$ so that the $\omega$-limit point of $\wt{\mc{W}}^-_E$ lies above the point $S^+_E = (\frac{\pi}{2},\, -\cos^{-1}E)$. For cases (3) and (4), we want to construct a barrier for $\wt{\mc{W}}^-_E$ so that the $\omega$-limit point of $\wt{\mc{W}}^-_E$ lies above the point $(\frac{\pi}{2},\, - \cos^{-1}E-2\pi)$. For cases (1) and (2), we will use the barrier given by the line $\Omega = \frac{\xi}{2}-\frac{\pi}{2}$, while for cases (3) and (4), we will use the barrier given by the line $\Omega = -\frac{\xi}{2} -\frac{\pi}{2}$. We prove the cases in this order: (2), (4), (1), and lastly (3). 

Consider case (2): $\l < 0$ and $\kappa < 0$. We  use the barrier given by the line $\Omega = \frac{\xi}{2} - \frac{\pi}{2}$. We want to find an $E'$ such that for all $E \in (0,E')$, the slope field along this line is $> \frac{1}{2}$. Hence for all sufficiently small $E$, it suffices to show that $j_E(\xi) > 0$ for all $\xi$ where $j_E(\xi)$ is given by   
\begin{align*}
j_{E}(\xi) \,&:=\, g_E\left(\xi, \frac{\xi}{2} - \frac{\pi}{2}\right) - \frac{1}{2}\cos^2\xi
\\
&=\, 2a\sin\xi\sin\frac{\xi}{2} -2\l\cos\xi\cos\frac{\xi}{2} + 2\gamma \sin \xi \cos \xi + (2\kappa - \frac{1}{2})\cos^2\xi - 2aE.
\end{align*}
Thinking of $j_E$ as a function of $a, \l, \g, \kappa$, we have $\frac{\pd j_E}{\pd \l}(\xi) = -2\cos \xi \cos \frac{\xi}{2} \leq 0$ on $[-\frac{\pi}{2}, \frac{\pi}{2}]$. Therefore $j_E$ is greater than the maximum value $\lambda$ can take. Since $\kappa < 0$, we have $\lambda \leq -\frac{1}{2} + \kappa -a $. Therefore
\begin{align*}
j_E(\xi) \,&\geq\, 2a\sin\xi\sin\frac{\xi}{2} + (1 - 2\kappa + 2a)\cos\xi\cos\frac{\xi}{2} + 2\gamma \sin \xi \cos \xi + (2\kappa - \frac{1}{2})\cos^2\xi - 2aE
\\
&=:\, h_E(\xi).
\end{align*}
Since $\frac{\pd h_E}{\pd \kappa}(\xi) = -2\cos \xi \cos \frac{\xi}{2} + 2 \cos^2\xi \leq 0$ on $[-\frac{\pi}{2}, \frac{\pi}{2}]$, we have 
\begin{align*}
j_E(\xi) \,&\geq\, 2a\sin \xi \sin \frac{\xi}{2} + 2(1+a)\cos \xi \cos \frac{\xi}{2} + 2\gamma \sin \xi \cos \xi - \frac{3}{2}\cos^2\xi - 2aE
\\
&=:\,\ell_E(\xi).
\end{align*}
We have $\frac{\pd \ell_E}{\pd a} = 2(\sin\xi \sin\frac{\xi}{2} + \cos\xi\cos\frac{\xi}{2} - E)$. It's not hard to see that for all $E$ sufficiently small, we have $\frac{\pd \ell_E}{\pd a} \geq 0$. Therefore, for these $E$, we have 
\begin{align*}
j_E(\xi) \,&\geq\,  2\cos\xi \cos\frac{\xi}{2} + 2\gamma \sin \xi \cos \xi - \frac{3}{2}\cos^2\xi
\\
&=:\, m(\xi). 
\end{align*}

Consider first the interval $[-\frac{\pi}{2}, 0]$. We have $\frac{\pd m}{\pd \gamma}(\xi) = 2\sin \xi \cos \xi \leq 0$ on $[-\frac{\pi}{2}, 0]$. Therefore, on $[-\frac{\pi}{2}, 0]$, we have
\begin{align*}
j_E(\xi) \,&\geq\, 2\cos\xi \cos\frac{\xi}{2} - \frac{3}{2}\cos^2\xi
\\
&=: 
n(\xi).
\end{align*}
We have $n(\xi) > 0$ on $(-\frac{\pi}{2},0]$ and $j_E(-\frac{\pi}{2}) = 2a(\sin\frac{\pi}{4}-E) > 0$ for all $E$ sufficiently negative. This proves the proposition for case (2) for the interval $[-\frac{\pi}{2}, 0]$.

Now consider the interval $[0, \frac{\pi}{2}]$. We have $\frac{\pd m}{\pd \g}(\xi) \geq 0$ on $[0, \frac{\pi}{2}]$. Since $\g_{\text{min}} = -\frac{1}{2}$, we have
\begin{align*}
j_E(\xi) \,&\geq\,2\cos\xi \cos\frac{\xi}{2} - \sin \xi \cos \xi - \frac{3}{2}\cos^2\xi
\\
&=:\, o(\xi). 
\end{align*}
We have $o(\xi) > 0$ on $[0, \frac{\pi}{2})$ and $j_E(\frac{\pi}{2}) = 2a(\sin\frac{\pi}{4}-E) > 0$ for all $E$ sufficiently small. This proves case (2) for the interval $[0, \frac{\pi}{2}]$. Thus the proposition is proved for case (2).

We now consider case (4): $\l > 0$ and $\kappa < 0$. We  use the barrier given by the line $\Omega = -\frac{\xi}{2} - \frac{\pi}{2}$. We want to find an $E'$ such that for all $E \in [0,E')$, the slope field along this line is $>- \frac{1}{2}$. Hence for all sufficiently small $E$, it suffices to show that $j_E(\xi) > 0$ for all $\xi$ where $j_E(\xi)$ is given by    
\begin{align*}
j_E(\xi) \,&:=\, g_E\left(\xi, -\frac{\xi}{2} - \frac{3\pi}{2}\right) + \frac{1}{2}\cos^2\xi
\\
&=\, 2a\sin\xi \sin \frac{\xi}{2} + 2\l \cos \xi \cos \frac{\xi}{2} + 2\gamma\sin \xi \cos\xi + (2\kappa + \frac{1}{2})\cos^2\xi -2aE.
\end{align*}
Since $\kappa < 0$, we have $\l \geq \frac{1}{2} - \kappa - a$.  Therefore
\begin{align*}
j_E(\xi) \,&\geq\, 2a\sin \xi \sin \frac{\xi}{2} + (1 - 2\kappa - 2a) \cos \xi \cos \frac{\xi}{2} + 2\gamma \sin \xi \cos \xi + (2\kappa + \frac{1}{2})\cos^2 \xi - 2aE
\\
&=:\, h_E(\xi). 
\end{align*}
Since $\frac{\pd h_E}{\pd \kappa}(\xi) = -2\cos \xi \cos \frac{\xi}{2} + 2 \cos^2\xi \leq 0$ on $[-\frac{\pi}{2}, \frac{\pi}{2}]$, we have 
\begin{align*}
j_E(\xi) \,&\geq\, 2a\sin \xi \sin \frac{\xi}{2} + 2(1-a)\cos \xi \cos \frac{\xi}{2} + 2\gamma \sin \xi \cos \xi - \frac{1}{2}\cos^2\xi - 2aE
\\
&=:\,\ell_E(\xi).
\end{align*}
We have $\frac{\pd \ell_{E}}{\pd a}(\xi) = 2\sin\xi \sin \frac{\xi}{2} -2\cos\xi\cos\frac{\xi}{2} - 2E \leq 0$ on $[-\frac{\pi}{3}, \frac{\pi}{3}]$ for all $E$. Therefore, on $[-\frac{\pi}{3}, \frac{\pi}{3}]$, we have $j_E(\xi) \geq h_E(\xi)|_{a = a_\text{max}}$. Since $a_\text{max} < \frac{1}{2}$, we have $j_E(\xi) \geq \ell_E(\xi)|_{a = 1/2}$. Therefore
\begin{align*}
j_E(\xi) \,&\geq\, \sin \xi \sin \frac{\xi}{2} + \cos \xi \cos \frac{\xi}{2} + 2\gamma \sin \xi \cos \xi - \frac{1}{2}\cos^2\xi - E
\\
&=:\, m_E(\xi). 
\end{align*}
Similar to case (2), we need break up the interval $[-\frac{\pi}{2}, \frac{\pi}{3}]$ into two intervals $[-\frac{\pi}{3},0]$ and $[0, \frac{\pi}{3}]$. Evaluate $m_E$ on the former with $\g = 0$ and the latter with $\g = \g_{\text{min}} = -\frac{1}{2}$; in both cases we have $j_E(\xi) > 0$ for all sufficiently small $E$. This proves case (4) for the interval $[-\frac{\pi}{3}, \frac{\pi}{3}]$.

Now we work on the intervals $[-\frac{\pi}{2}, -\frac{\pi}{3}]$ and $[\frac{\pi}{3}, \frac{\pi}{2}]$. Given any $\e > 0$, there is an $E' \in (0,1)$ such that for all $E \in [0,E')$, we have $\frac{\pd \ell_E}{\pd a}(\xi) \geq 0$ on $[-\frac{\pi}{2}, -\frac{\pi}{3} - \e] \cup [\frac{\pi}{3} + \e, \frac{\pi}{2}]$. Since $m_E(-\frac{\pi}{3})|_{\gamma = 0}$ and $m_E(\frac{\pi}{3})|_{\gamma = \g|_{\text{min}}}$ are uniformly bounded above $0$ for all sufficiently small $E$, it suffices to show $\ell_E(\xi)|_{a = 0} > 0$ on $(-\frac{\pi}{2}, -\frac{\pi}{3}] \cup [\frac{\pi}{3}, \frac{\pi}{2})$. We have 
\begin{align*}
\ell_E(\xi)|_{a = 0} \,&=\, 2\cos\xi \cos\frac{\xi}{2} + 2\gamma \sin \xi \cos \xi - \frac{1}{2}\cos^2\xi
\\
&=:\, n(\xi). 
\end{align*}
On $(-\frac{\pi}{2}, -\frac{\pi}{3}]$, we have $n(\xi) \geq n(\xi)|_{\gamma = 0} > 0$. On $[\frac{\pi}{3}, \frac{\pi}{2}]$, we have $n(\xi) \geq n(\xi)|_{\gamma = -\frac{1}{2}} > 0$. This finishes the proof of the proposition for case (4). 

We now consider case (1): $\l < 0$ and $\kappa > 0$. We  use the barrier given by the line $\Omega = \frac{\xi}{2} - \frac{\pi}{2}$. For all sufficiently small $E$, we want to show that the slope field along this line is $> \frac{1}{2}$. Hence it suffices to show that $j_E(\xi) > 0$ for all $\xi$ where $j_E(\xi)$ is given by 
\begin{align*}
j_{E}(\xi) \,&=\, g_E\left(\xi, \frac{\xi}{2} - \frac{\pi}{2}\right) - \frac{1}{2}\cos^2\xi
\\
&=\, 2a\sin\xi\sin\frac{\xi}{2} -2\l\cos\xi\cos\frac{\xi}{2} + 2\gamma \sin \xi \cos \xi + (2\kappa - \frac{1}{2})\cos^2\xi - 2aE.
\end{align*}
 Since $\frac{\pd j_E}{\pd \l}(\xi) := -2\cos \xi \cos \frac{\xi}{2} \leq 0$ on $[-\frac{\pi}{2}, \frac{\pi}{2}]$, $j_E$ is greater than the maximum value $\lambda$ can take. Since $\kappa > 0$, we have $\lambda \leq -\frac{1}{2} - \kappa +a $. Therefore
\begin{align*}
j_{E}(\xi) \,&\geq\, 2a\sin\xi\sin \frac{\xi}{2} + (1 + 2\kappa -2a)\cos\xi\cos\frac{\xi}{2} + 2\gamma\sin\xi\cos\xi +(2\kappa-\frac{1}{2})\cos^2\xi -2aE
\\
&=:\, h_{E}(\xi). 
\end{align*}
Since $\frac{\pd h_{E}}{\pd \kappa}(\xi) = 2\cos \xi \cos\frac{\xi}{2} + 2\cos^2\xi \geq 0$ on $[-\frac{\pi}{2}, \frac{\pi}{2}]$, we have
\begin{align*}
j_E(\xi) \,&\geq\, 2a\sin\xi\sin\frac{\xi}{2} + 2(1-a)\cos\xi\cos\frac{\xi}{2} + 2\gamma \sin \xi \cos \xi + \frac{1}{2}\cos^2\xi- 2aE.
\end{align*}
The right hand side is greater than $\ell_E(\xi)$ from case (4). Therefore the proposition is proved for case (1) by following the sames steps as in case (4).

Lastly, consider case (3): $\l > 0$ and $\kappa > 0$. We will use the barrier given by the line $\Omega = -\frac{\xi}{2} - \frac{3\pi}{2}$. For all sufficiently small $E$, we want to show that the slope field along this line is $> -\frac{1}{2}$. Hence we want to show $j_E(\xi) > 0$ for all $\xi$ where $j_E(\xi)$ is given by 
\begin{align*}
j_E(\xi) \,&:=\, g_E\left(\xi, -\frac{\xi}{2} - \frac{3\pi}{2}\right) + \frac{1}{2}\cos^2\xi
\\
&=\, 2a\sin\xi \sin \frac{\xi}{2} + 2\l \cos \xi \cos \frac{\xi}{2} + 2\gamma\sin \xi \cos\xi + (2\kappa + \frac{1}{2})\cos^2\xi -2aE.
\end{align*}
Since $\frac{\pd j_E}{\pd \l}(\xi) = 2\cos \xi \cos \frac{\xi}{2} \geq 0$ on $[-\frac{\pi}{2}, \frac{\pi}{2}]$. Therefore $j_E$ is greater than the minimum value $\l$ can take. Since $\kappa > 0$, we have $\l \geq \frac{1}{2} + \kappa - a$. Therefore 
\begin{align*}
j_E(\xi) \,&\geq\, 2a\sin \xi \sin \frac{\xi}{2} + (1 + 2\kappa - 2a) \cos \xi \cos \frac{\xi}{2} + 2\gamma \sin \xi \cos \xi + (2\kappa + \frac{1}{2})\cos^2 \xi - 2aE.
\end{align*}
The right hand side is greater than $h_E(\xi)$ from case (1). Therefore the proposition is proved for case (3) by following the same steps as in case (1). \qed

\medskip
\medskip

Propositions \ref{Omega winding leq 0 prop} and \ref{Omega wind geq n + 1 prop} together with Proposition \ref{saddles conn prop} establishes the existence of saddles connectors.

\medskip
\medskip

\begin{prop}\label{Omega wind geq n + 1 prop}\: $\phantom{nix}$
\begin{itemize}
\item[$\bullet$] Let $\l < 0$. For any integer $N \geq 0$, there is an $E'' \in (0,1)$ such that $w_{E''} \geq N+1$. 
\item[$\bullet$] Let $\l > 0$. For any integer $N \geq 1$, there is an $E'' \in (0,1)$ such that $w_{E''} \geq N+1$. 
\end{itemize}
\end{prop}

\medskip
\medskip

We now set out to prove Proposition \ref{Omega wind geq n + 1 prop}. The proof will be an argument by contradiction which we briefly sketch: Suppose for all $E \in (0,1)$, the winding number of $\wt{\mc{W}}^-_E$ is between $0$ and $N$. Then $\wt{\mc{W}}^-_E$ lies within a compact region in the universal cover $\wt{\mc{C}}$ for all $E$. This will allow us to take a limit $\lim_{E \to 1}\wt{\mc{W}}^-_E(\tau)$ for each $\tau$. The resulting limit curve, call it $\wt{\mc{W}}^-_1(\tau)$, is shown to be an orbit for the dynamical system when $E = 1$. Since $\wt{\mc{W}}^-_E$ lies within a compact region for all $E$, it follows that $\wt{\mc{W}}^-_1$ lies within the same compact region. Consequently, the $\omega$-limit set of $\wt{\mc{W}}^-_1$ corresponds to the equilibrium point on the right side of the cylinder by the Poincar{\'e}-Bendixson Theorem. However, this contradicts the fact that $\wt{\mc{W}}^-_1$ has an empty $\omega$-limit set which is a consequence of the fact that the equilibrium point for $E = 1$ corresponds to a nilpotent singularity.

The next few propositions will make the above argument rigorous. Proposition \ref{in compact prop} shows that if the winding number of $\wt{\mc{W}}_E^-$ is less than or equal to some fixed integer $N$ for all sufficiently large $E$, then the unstable manifolds are contained within a compact region of the universal cover $\wt{\mc{C}}$. This implies that the unstable manifolds are bounded which will be used in Proposition \ref{limit orbit prop} where we prove that $\lim_{E \to 1}\wt{\mc{W}}_E^-(\tau)$ converges for each $\tau \in \R$.

\medskip
\medskip

\newpage

\begin{prop}\label{in compact prop} \: $\phantom{nix}$
\begin{itemize}
\item[$\bullet$] Let $\l < 0$. Suppose there is an integer $N \geq 0$ such that for all $E \in (0,1)$, the winding number of $\wt{\mc{W}}_E^-$ is less than or equal to $N$. Then the image of $\wt{\mc{W}}_E^-$ is contained in the compact set $[-\frac{\pi}{2}, \frac{\pi}{2}] \times [\frac{\pi}{2} - 2 \pi N, \frac{\pi}{2}]$ for all $E \in (0,1)$.
\item[$\bullet$] Let $\l > 0$. Suppose there is an integer $N \geq 1$ such that for all $E \in (0,1)$, the winding number of $\wt{\mc{W}}_E^-$ is less than or equal to $N$. Then the image of $\wt{\mc{W}}_E^-$ is contained in the compact set $[-\frac{\pi}{2}, \frac{\pi}{2}] \times [-\frac{\pi}{2} - 2 \pi N, -\frac{\pi}{2}]$ for all $E \in (0,1)$.
\end{itemize}
\end{prop}

\proof
Consider $\l < 0$. By Proposition \ref{horiz barr prop}, the horizontal line $\Omega = \frac{\pi}{2}$ is a barrier. Since $\dot{\Omega}$ is $2\pi$-periodic in $\Omega$, it follows that $\Omega = \frac{\pi}{2} - 2 \pi N$ is also a barrier. Since $\wt{\mc{W}}_E^-$ begins below $\Omega = \frac{\pi}{2}$, it remains below it. If $\wt{\mc{W}}_E^-$ crosses $\Omega = \frac{\pi}{2} - 2 \pi N$, then it will remain below it and consequently have a winding number greater than $N$ which contradicts the hypothesis. The $\l > 0$ case is analogous. 
\qed

\medskip
\medskip

\begin{prop}\label{limit orbit prop}
Assume the hypothesis of Proposition \emph{\ref{in compact prop}}. Then the limit $\displaystyle\lim_{E \to 1}\wt{\mc{W}}_E^-(\tau)$ exists for all $\tau \in \R$. Moreover, the set
\[
\left\{\lim_{E \to 1}\wt{\mc{W}}_E^-(\tau) \mid \tau \in \R\right\}
\]
is the image of an orbit of the dynamical system when $E = 1$. 
\end{prop}

\proof
Recall we are writing $\wt{\mc{W}}_E^-(\tau) = \big(\xi(\tau), \Omega_E(\tau)\big)$ where $\xi(\tau) = \tan^{-1}(\tau)$ so that $\xi(0) = 0$. Let $\phi_\tau$ denote the flow map of the dynamical system considering $E$ as a parameter. Hence we have several ways to write $\wt{\mc{W}}_E^-(\tau)$:
\[
\phi_\tau\big(\xi(0), \Omega_E(0), E \big) \,=\, \phi_t\big(0, \Omega_E(0), E\big) \,=\, \big(\xi(\tau), \Omega_E(\tau)\big) \,=\, \wt{\mc{W}}_E^-(\tau).
\]
By Proposition \ref{in compact prop}, for each $\tau$, we have $\Omega_E(\tau)$ is bounded in $E$. Moreover $\Omega_E(\tau)$ is monotone in $E$ by Lemma \ref{lem 7.3 in JMP paper}. Therefore $\lim_{E \to 1}\Omega_E(\tau)$ converges for each $\tau$. This proves the first part of the proposition. To prove the second part, define $\Omega^\tau_1 = \lim_{E \to 1}\Omega_E(\tau)$. Then it suffices to show 
\[
\lim_{E \to 1}\phi_\tau\big(0, \Omega_E(0), E\big) \,=\, \phi_\tau\big(0, \Omega_1^0, 1\big)
\]
for each $\tau$. To prove this, define 
\[
A = \left\{\tau \in \R \mid \lim_{E \to 1}\phi_t\big(0, \Omega_E(0), E\big) \,=\, \phi_\tau\big(0, \Omega_1^0, 1\big) \right\}. 
\]
It suffices to show $A = \R$. We will prove this by showing $A$ is nonempty, open, and closed. $A$ is nonempty since $0 \in A$. To prove $A$ is open, fix $s \in A$. Since $\Omega_E(s) \to \Omega_1^s$ as $E \to 1$, continuous dependence of solutions to ordinary differential equations on initial conditions and parameters (Theorem 2 in section 2.3 of \cite{Perko}) implies that there exists an $\e > 0$ such that for all $\tau \in (-\e, \e)$, we have
\[
\lim_{E \to 1}\phi_\tau\big(\xi(s), \Omega_E(s), E \big) \,=\, \phi_\tau\big(\xi(s), \Omega_1^s, 1\big). 
\]
Hence
\[
\lim_{E \to 1}\phi_{\tau+s}\big(0, \Omega_E(0), E\big) \,=\, \phi_\tau\big(\xi(s), \Omega_1^s, 1\big). 
\]
Let $\Omega^0_1(\tau)$ be given by $\big(\xi(\tau), \Omega_1^0(\tau)\big) = \phi_\tau\big(0, \Omega_1^0, 1\big)$. Since $s \in A$, we have $\Omega_E(s) \to \Omega_1^0(s)$ as $E \to 1$. Therefore $\Omega^0_1(s) = \Omega^s_1$. Thus
\[
\lim_{E \to 1}\phi_{\tau+s}\big(0, \Omega_E(0), E \big) \,=\, \phi_\tau\big(\xi(s), \Omega_1^s, 1\big) \,=\, \phi_\tau\big(\xi(s), \Omega^0_1(s), 1\big) \,=\, \phi_{\tau + s}\big(0, \Omega_1^0, 1\big). 
\]
Hence $A$ is open.  

Lastly, we show $A$ is closed. Fix $s \notin A$. Again, using continuous dependence of solutions on initial conditions and parameters, there exists an $\e > 0$ such that for all $\tau \in (-\e, \e)$, we have 
\[
\lim_{E \to 1}\phi_\tau\big(x(s), \Omega_E(s), E\big) \,=\, \phi_\tau\big(\xi(s), \Omega_1^s, 1\big). 
\]
Since $\Omega_1^s \neq \Omega^0_1(s)$ and $\phi_\tau$ is injective, we have 
\[
\phi_\tau\big(\xi(s), \Omega_1^s, 1\big) \,\neq\, \phi_\tau\big(\xi(s), \Omega^0_1(s), 1\big) \,=\, \phi_{\tau+s}\big(0, \Omega^0_1, 1\big),
\]
while
\[
\phi_\tau\big(\xi(s), \Omega_1^s, 1\big) \,=\, \lim_{E \to 1}\phi_\tau\big(\xi(s), \Omega_E(s), E\big) \,=\, \lim_{E \to 1}\phi_{\tau + s}\big(0, \Omega_{E}(0), E\big). 
\]
Thus 
\[
\lim_{E \to 1}\phi_{\tau+s}\big(0, \Omega_E(0), E\big) \neq \phi_{t + s}(0, \Omega_1^0, 1)
\]
for all $\tau \in (-\e, \e)$. Hence $A$ is closed. 
\qed

\medskip
\medskip

Next we prove Proposition \ref{Omega wind geq n + 1 prop}. The idea behind the proof is this: Proposition \ref{limit orbit prop} implies that an orbit for the dynamical system when $E = 1$ is contained in a compact set and hence its $\omega$-limit set is nonempty by the Poincar{\'e}-Bendexison theorem. However, we will show that any orbit for the dynamical system when $E = 1$ has an empty $\omega$-limit set. Here we sketch how this latter part works with specific parameter values $a = \frac{1}{4}$, $\l = -1$, $\g = -\frac{1}{4}$, and $\kappa = \frac{1}{2}$.  Shifting $\xi \mapsto \xi - \frac{\pi}{2}$ so that the equilibrium at $\xi = \frac{\pi}{2}$ occurs at the origin and keeping only second order terms, our dynamical system when $E = 1$ becomes 
\[
\left\{
 \begin{array}{ll}
      \dot{\xi} \,=\, \xi^2   \\
      \dot{\Omega} \,=\,  - \frac{1}{4}\Omega^2 + \frac{3}{4}\xi^2  + 2\xi \Omega + \frac{1}{2} \xi .
\end{array} 
\right.
\]
Compare this with the system
\[
\left\{
 \begin{array}{ll}
      \dot{\xi} \,=\, \xi^2   \\
      \dot{z} \,=\,    \frac{3}{4}\xi^2  + 2\xi z + \frac{1}{2} \xi .
\end{array} 
\right. 
\]

Any orbit for the $(\dot{\xi}, \dot{z})$-system is a barrier for the $(\dot{\xi}, \dot{\Omega})$-system, i.e. the slope field for the $(\dot{\xi}, \dot{\Omega})$-system is `more negative' than the slope of any orbit for the $(\dot{\xi}, \dot{z})$-system.  The $(\dot{\xi}, \dot{z})$-system has can be solved via an integrating factor. We have $dz/d\xi = \dot{z}/\dot{\xi} = \frac{3}{4} +  \frac{2z}{\xi} + \frac{1}{2\xi}$. The solution is $z(\xi) = c\xi^2 - \frac{3}{4}\xi - \frac{1}{4}$ where $c$ is an arbitrary constant. Therefore any orbit terminates at $(0,-\frac{1}{4})$. 

Let $\big(\xi(\tau), \Omega(\tau)\big)$ and $\big(\xi(\tau), z(\tau)\big)$ be orbits for the $(\dot{\xi}, \dot{\Omega})$-system and the $(\dot{\xi}, \dot{z})$-system, respectively,  with the same initial condition. The $\omega$-limit set of $\big(\xi(\tau), \Omega(\tau)\big)$ is either the equilibrium point $(0,0)$ or it is empty. Since $\big(\xi(\tau), z(\tau)\big)$ is a barrier for $\big(\xi(\tau), \Omega(\tau)\big)$ and $z(\tau) \to -\frac{1}{4}$ as $\tau \to \infty$, we have $\Omega(\tau) < -\frac{1}{8}$ for all sufficiently large $\tau$.  Hence the $\omega$-limit set of $\big(\xi(\tau), \Omega(\tau)\big)$ is empty.  

\medskip
\medskip

\noindent{\emph{Proof of Proposition \emph{\ref{Omega wind geq n + 1 prop}}}.}

\medskip

Assume the hypothesis of Proposition \ref{in compact prop}. By Proposition \ref{limit orbit prop}, there is an orbit $c(\tau) = \big(\xi(\tau), \Omega(\tau)\big)$ for the dynamical system when $E = 1$ such that $c(\tau) = \lim_{E \to 1}\wt{\mc{W}}_E^-(\tau)$. By Proposition \ref{monotone prop}, $\Omega(\tau)$ is monotone for all $\tau> \tau'$ for some $\tau'$. By Proposition \ref{in compact prop}, $\Omega(\tau)$ is bounded. Therefore $\lim_{\tau \to \infty}\Omega(\tau)$ converges. Since the boundary $\xi = \frac{\pi}{2}$ coincides with an orbit, $c(\tau)$ must converge to an equilibrium point. 

However this contradicts the fact that the $\omega$-limit set of any orbit is empty for $E = 1$. This follows since the equilibrium point for $E = 1$ corresponds to a nilpotent singularity since the Jacobian is a nilpotent matrix. This follows from applying Theorem 3.5 in \cite{QTPDS} with $x = \Omega$ and $y = \xi$.  The phase portrait about the equilibrium point is illustrated in \cite[Fig. 3.21(i)]{QTPDS} which shows that any orbit starting in the region $\xi < \frac{\pi}{2}$ does not limit to the equilibrium point. 

To show that Theorem 3.5 in \cite{QTPDS} holds, we shift our system $\xi \mapsto \xi - \frac{\pi}{2}$ so that the equilibrium point at $\xi = \frac{\pi}{2}$ occurs at the origin. Then our system expanded to 2nd order is
\[ 
\left\{
 \begin{array}{ll}
      \dot{\xi} \,=\, \xi^2 + \text{h.o.t.}  \\
      \dot{\Omega} \,=\,  -2\gamma \xi + (1-a)\xi^2 -a\Omega^2 -2\lambda\xi\Omega  +\text{h.o.t.}
\end{array} 
\right.
\]
To apply Theorem 3.5 in \cite{QTPDS}, we want the factor in front of $\xi$ to be equal to one in the expression for $\dot{\Omega}$, so we compress $\wt{\xi} = -2\gamma \xi$. Then $\dot{\wt{\xi}} = -2\gamma \dot{\xi} = -\frac{1}{2\gamma}\wt{\xi}^2$. Relabeling $\wt{\xi}$ by $\xi$, our system becomes
\[ 
\left\{
 \begin{array}{ll}
      \dot{\xi} \,=\, -\frac{1}{2\gamma}\xi^2 + \text{h.o.t.}  \\
      \dot{\Omega} \,=\,  \xi + \frac{1-a}{4\gamma^2}\xi^2 -a\Omega^2 + \frac{\lambda}{\gamma}\xi\Omega  +\text{h.o.t.}
\end{array} 
\right.
\]
Let $A(\xi, \Omega) = \frac{1-a}{4\gamma^2}\xi^2 -a\Omega^2 + \frac{\lambda}{\gamma}\xi\Omega$. Let $\xi = f(\Omega)$ be the solution to the equation $x + A(\xi, \Omega) = 0$. Then $f(\Omega) = a\Omega^2 + \text{h.o.t.}$. Define $F(\Omega) = B\big(f(\Omega)\big)$ where $B = \dot{\xi}$. Then $F(\Omega) = -\frac{1}{2\gamma}a^2\Omega^4 + \text{h.o.t.}$ Define 
\[G(\Omega) \,=\, \left(\frac{\pd A}{\pd \Omega} + \frac{\pd B}{\pd \xi}\right)\bigg|_{(f(\Omega),\Omega )} \,=\, -2a\Omega + \frac{\l}{\g}f(\Omega) - \frac{1}{\gamma}f(\Omega)+\text{h.o.t.} \,=\, -2a\Omega + \text{h.o.t.}
\]
Applying Theorem 3.5(4) in \cite{QTPDS} shows that $m = 4$ and $n = 1$; hence $(i2)$ is relevant. The phase portrait about the equilibrium point is illustrated in Figure 3.21(i) which shows that any orbit starting in the region $\xi < \frac{\pi}{2}$ does not limit to the equilibrium point. 
\qed

\medskip
\medskip

\begin{thm}\label{exist omega system saddle connectors thm}\: $\phantom{nix}$ 
\begin{itemize}
\item[$\bullet$] Let $\l < 0$. For each integer $N \geq 0$, there is a unique $E_N \in (0,1)$ such that $\wt{\mc{W}}^-_{E_N}$ is a saddles connector with winding number $w_{E_N} = N$. Moreover $E_0 \leq E_1 \leq E_2 \leq \dotsb$.
\item[$\bullet$] Let $\l > 0$. For each integer $N \geq 1$, there is a unique $E_N \in (0,1)$ such that $\wt{\mc{W}}^-_{E_N}$ is a saddles connector with winding number $w_{E_N} = N$. Moreover $E_1 \leq E_2 \leq E_3 \leq \dotsb$.
\end{itemize} 
\end{thm}

\proof
The existence of $E_N$ follows from Propositions \ref{Omega winding leq 0 prop}, \ref{Omega wind geq n + 1 prop}, and  \ref{saddles conn prop}. The uniqueness of $E_N$ follows from Proposition \ref{unique omega prop}. 

Now we prove the monotonicity condition. Consider $\l < 0$; the proof for $\l > 0$ is analogous. Seeking a contradiction, suppose $E_1 < E_0$.  Therefore Lemma \ref{lem 7.3 in JMP paper} implies that $\Omega_{E_1}(\tau) \geq \Omega_{E_0}(\tau)$ for all $\tau \in \R$. But this contradicts the fact that $\Omega_{E_1}(\tau) \to -2\pi - \cos^{-1}(E_1)$ and $\Omega_{E_0}(\tau) \to -\cos^{-1}(E_0)$ as $\tau \to \infty$. The same argument shows $E_N \leq E_{N+1}$ for each $N \geq 0$. 
\qed

\medskip
\medskip

\subsection{Iterating the $\Theta$ and $\Omega$ systems}

\medskip

In sections \ref{th sec} we showed how to obtain saddles connectors for the $\Theta$ system given an arbitrary energy $E$. In \ref{Om sec} we showed how to obtain saddles connectors for the $\Omega$ system given an arbitrary $\l$. If $\Psi$ is of the form (\ref{ontology}) constructed from solutions of (\ref{eq:Om}) - (\ref{eq:S}), then $\Psi$ is a bound state provided \emph{both} $E$ and $\l$ are values which jointly correspond to saddles connectors for the $\Theta$ system \emph{and} the $\Omega$ system. To prove that such joint values exist, we will iterate between the $\Theta$ and $\Omega$ systems and use the contraction mapping principle to establish the existence of $E$ and $\l$ (Theorem \ref{iteration thm}). Propositions \ref{lambda c1 prop} and \ref{E c1 prop} will be used to establish the contraction. 

\medskip
\medskip

\begin{prop}\label{lambda c1 prop} Fix an integer $N$. 
\begin{itemize}
\item[$\bullet$] Suppose $N \geq 0$. For each $E \in [0,1]$, let $\l_N(E)$ be given by Theorem \emph{\ref{exist theta system saddle connectors thm}}. Then $\l_N(E)$ is a $C^1$ function of $E$ and $|\l'_N(E)| < a$.
\item[$\bullet$] Suppose $N \leq -1$. For each $E \in [0,1]$, let $\l_N(E)$ be given by Theorem \emph{\ref{exist theta system saddle connectors thm for neg wind num}}. Then $\l_N(E)$ is a $C^1$ function of $E$ and $|\l'_N(E)| < a$.
\end{itemize}
\end{prop}

\proof  We first show that $\l_N(E)$ is a $C^1$ function of $E$. For each $E$ and $\l$, let $\wt{\mc{W}}_{\l,E}^\pm(\tau) = \big(\theta(\tau), \Theta^\pm_{\l,E}(\tau)\big)$ denote the orbits corresponding to the unstable and stable manifolds emanating from $S^-$ and $S^+ - 2\pi N$, respectively, with initial condition determined by $\theta(0) = \frac{\pi}{2}$ where $S^+ - 2\pi N := (\pi, - \pi - 2\pi N)$ if $\kappa > 0$ and $S^+ - 2\pi N := (\pi, -2\pi N)$ if $\kappa < 0$.

 By smooth dependence of solutions on parameters, the function 
\[
\Phi(\l,E) \,=\, \Theta^-_{\l,E}(0) - \Theta^+_{\l,E}(0)
\]
is smooth in $E$ and $\l$.  Since $\l_N(E)$ corresponds to a saddles connector with winding number $N$, we have $\Phi\big(\l_N(E),E\big) = 0$. By the implicit function theorem, if 
\[
\frac{\pd \Phi}{\pd \l}\big(\l_N(E),E\big) \,\neq\, 0,
\]
then it follows that $\l_N$ is a $C^1$ function of $E$. To show this, define $u^\pm_{\l,E}(\tau) \,=\, \frac{\pd}{\pd \l}\Theta^\pm_{\l,E}(\tau)$. Then $u^\pm_{\l,E}$ satisfies the following differential equation
\[
\frac{d}{d\tau}u^\pm_{\l,E}(\tau) \,=\, P^\pm_{\l,E}(\tau) u^\pm_{\l,E}(\tau) + Q(\tau)
\]
where
\begin{align*}
P^\pm_{\l,E} \,&=\, 2a\sin\theta \cos\theta \sin\Theta^\pm_{\l,E} + (2aE\sin^2\theta -2\kappa)\cos\Theta^\pm_{\l,E}
\\
Q \,&=\, 2\sin\theta.
\end{align*}
Setting $U^\pm_{\l,E}(\tau,\tau_0) = \exp(\int_{\tau_0}^\tau P^\pm_{\l,E}(s)ds )$, the solution to this differential equation is
\[
u^\pm_{\l,E}(\tau) \,=\, U^\pm_{\l,E}(\tau,\tau_0)u^\pm_{\l,E}(\tau_0) \,+\, \int_{\tau_0}^\tau U^\pm_{\l,E}(\tau,s)Q(s)ds. 
\]
Taking the limit $\tau_0 \to - \infty$, we find $P^-_{\l,E}(-\infty) = -1$ and so $U^-_{\l,E}(\tau,-\infty) = 0$. Similarly, taking the limit $\tau_0 \to \infty$, we find $P^+_{\l,E}(\infty) = 1$ and so $U^+_{\l,E}(\tau,\infty) = 0$. Therefore $U^\pm_{\l,E}(\tau,\tau_0) u^\pm_{\l,E}(\tau_0) \to 0$ as $\tau_0 \to \pm\infty$ (it's also the case that $u^\pm_{\l,E}(\pm \infty) = 0$, but this follows because the $\Theta$ system has special equilibrium points which don't depend on the parameter $\l$; this doesn't happen in general).  Therefore
\[
\frac{\pd \Phi}{\pd \l}(\l,E) \,=\, u^-_{\l,E}(0) - u^+_{\l,E}(0) \,=\, \int_{-\infty}^0U^-_{\l,E}(0,s)Q(s)ds + \int_0^\infty U^+_{\l,E}(0,s)Q(s)ds. 
\]
Since $\l_N(E)$ corresponds to a saddles connector with winding number $N$, we have $\Theta^-_{\l_N(E),E}(\tau) = \Theta^+_{\l_N(E),E}(\tau)$ for all $\tau \in \R$. Therefore $U^-_{\l_N(E),E} = U^+_{\l_N(E),E}$. Thus
\[
\frac{\pd \Phi}{\pd \l}\big(\l_N(E),E\big) \,=\, \int_{-\infty}^\infty U^-_{ \l_N(E),E}(0,s)Q(s)ds \,>\, 0.
\]

Now we prove the second part of the theorem: $|\l_N'(E)| < a$ for all $E$. Write $\Theta_{\l_N(E)} := \Theta^-_{\l_N(E),E} = \Theta^+_{\l_N(E),E}$. Define $v = \frac{\pd }{\pd E}\Theta_{\l_N(E)}$. Then $v$ satisfies the differential equation $\frac{d}{d\tau}v(\tau) = P(\tau)v(\tau) + R(\tau)$ where
\begin{align*}
P \,&=\, 2a \sin \theta \cos \theta \sin \Theta_{\l_N(E)} + (2aE \sin^2\theta -2\kappa) \cos\Theta_{\l_N(E)}
\\
R \,&=\, 2a\sin^2\theta\sin\Theta_{\l_N(E)} + 2 \l_N'(E)\sin \theta.
\end{align*}
Setting $U(\tau,\tau_0) = \exp(\int_{\tau_0}^\tau P(s)ds)$, the solution to this differential equation is 
\[
v(\tau) \,=\, U(\tau,\tau_0)v(\tau_0) + \int_{\tau_0}^\tau U(\tau,s)R(s)ds. 
\]
Using the same analysis as above, again we find $U(\tau,\tau_0) \to 0$ as $\tau_0 \to \pm \infty$. Therefore 
\[
v(0) \,=\, \int_{-\infty}^0U(0,s)R(s)ds \,=\, -\int_0^\infty U(0,s)R(s)ds. 
\]
Hence
\begin{align*}
0 \,&=\, \int_{-\infty}^\infty U(0,s)R(s)ds 
\\
&=\, 2a \int_{-\infty}^\infty U(0,s)\sin^2\theta(s)\sin\Theta_{\l_N(E)}(s)ds + 2\l_N'(E)\int_{-\infty}^\infty U(0,s)\sin \theta(s)ds. 
\end{align*}
Solving for $\l_N'(E)$, we find
\[
\l_N'(E) \,=\, a\left(\frac{\int_{-\infty}^\infty U(0,s)\sin^2\theta(s)\sin \Theta_{\l_N(E)}(s)ds }{\int_{-\infty}^\infty U(0,s)\sin\theta(s)ds }\right).
\]
Since the numerator in absolute value is strictly less than the denominator in absolute value, we have $|\l_N'(E)| < a$. 
\qed

\medskip
\medskip

\begin{prop}\label{E c1 prop}\:$\phantom{nix}$
\begin{itemize}
\item[$\bullet$] Let $\l < 0$. Let $E_N(\l)$ be given by Theorem  \emph{\ref{exist omega system saddle connectors thm}}. Then $E_N(\l)$ is a $C^1$ function of $\l$ and 
\[
|E_N'(\l)| \,<\, \frac{1}{a} \:\:\:\: \text{ for all } \:\:\:\: \l \,\leq\, \left\{
  \begin{array}{ll}
      -\frac{1}{2} - \kappa + a  & \text{ if } \kappa > 0   \\\\
      -\frac{1}{2} + \kappa -a & \text{ if } \kappa < 0    
\end{array} 
\right. \]

\item[$\bullet$] Let $\l > 0$. Let $E_N(\l)$ be given by Theorem  \emph{\ref{exist omega system saddle connectors thm}}. Then $E_N(\l)$ is a $C^1$ function of $\l$ and 
\[
|E_N'(\l)| \,<\, \frac{1}{a} \:\:\:\: \text{ for all } \:\:\:\: \l \,\geq\, \left\{
  \begin{array}{ll}
      \frac{1}{2} + \kappa - a  & \text{ if } \kappa > 0   \\\\
      \frac{1}{2} - \kappa -a & \text{ if } \kappa < 0    
\end{array} 
\right.\]
\end{itemize}
\end{prop}

\proof
We first show $E_N(\l)$ is a $C^1$ function of $\l$. For each $E$ and $\l$, let $\wt{\mc{W}}^\pm_{E,\l}(\tau) = \big(\xi(\tau), \Omega^\pm_{E,\l}(\tau)\big)$ denote the orbits corresponding to the unstable/stable manifolds emanating from $S^-_E$ and $(\frac{\pi}{2}, -\cos^{-1}(E) - 2\pi N)$ (this is just $S^+_E$ shifted down by $2\pi N$) with initial condition determined by $\xi(0) = 0$. 

By smooth dependence of solutions on parameters, the function 
\[
\Phi(E, \l) \,=\, \Omega^-_{E, \l}(0) - \Omega^+_{E, \l}(0)
\]
is smooth in $E$ and $\l$.  Since $E_N(\l)$ corresponds to a saddles connector with winding number $N$, we have $\Phi\big(E_N(\l), \l\big) = 0$. By the implicit function theorem, if 
\[
\frac{\pd \Phi}{\pd E}\big(E_N(\l), \l\big) \,\neq\, 0,
\]
then it follows that $E_N$ is a $C^1$ function of $\l$. To show this, define $u^\pm_{E,\l}(\tau) \,=\, \frac{\pd}{\pd E}\Omega^\pm_{E, \l}(\tau)$. Then $u^\pm_{E,\l}$ satisfies the following differential equation
\[
\frac{d}{d\tau}u^\pm_{E,\l}(\tau) \,=\, P^\pm_{E,\l}(\tau) u^\pm_{E,\l}(\tau) -2a
\]
where
\begin{align*}
P^\pm_{E,\l} \,&=\, -2a\sin\xi\sin \Omega^\pm_{E,\l} + 2\l\cos\xi \cos \Omega^\pm_{E,\l}.
\end{align*}
Setting $U^\pm_{E,\l}(\tau,\tau_0) = \exp(\int_{\tau_0}^\tau P^\pm_{E,\l}(s)ds )$, the solution to this differential equation is
\[
u^\pm_{E,\l}(\tau) \,=\, U^\pm_{E,\l}(\tau,\tau_0)u^\pm_{E,\l}(\tau_0) \,-\, 2a\int_{\tau_0}^\tau U^\pm_{E,\l}(\tau,s)ds. 
\]
Taking the limit $\tau_0 \to - \infty$, we find $P^-_{E,\l}(-\infty) = -2a\sqrt{1 - E^2}$ and so $U^-_{E,\l}(\tau,-\infty) = 0$. Similarly, taking the limit $\tau_0 \to \infty$, we find $P^+_{E,\l}(\infty) = 2a\sqrt{1-E^2}$ and so $U^+_{E,\l}(\tau,\infty) = 0$. Therefore $U^\pm_{E,\l}(\tau,\tau_0) u^\pm_{E,\l}(\tau_0) \to 0$ as $\tau_0 \to \pm\infty$ since $|u^\pm_{E,\l}(\pm \infty)| = |\frac{d}{dE} \cos^{-1}(E)| < \infty$.  Therefore
\[
\frac{\pd \Phi}{\pd E}(E, \l) \,=\, u^-_{E,\l}(0) - u^+_{E\l}(0) \,=\, -2a\int_{-\infty}^0U^-_{E,\l}(0,s)ds - 2a\int_0^\infty U^+_{E,\l}(0,s)ds. 
\]
Since $E_N(\l)$ corresponds to a saddles connector with winding number $N$, we have $\Omega^-_{E_N(\l),\l}(\tau) = \Omega^+_{E_N(\l), \l}(\tau)$ for all $\tau \in \R$. Therefore $U^-_{E_N(\l),\l} = U^+_{E_N(\l),\l} =: U_{E_N(\l)}$. Thus
\[
\frac{\pd \Phi}{\pd E}\big(E_N(\l), \l\big) \,=\, -2a\int_{-\infty}^\infty U_{E_N(\l)}(0,s)ds \,<\, 0.
\]

Now we prove the second part of the theorem: $|E_N'(\l)| < 1/a$ for all $\l$. Define $v^\pm_{E,\l} = \frac{\pd }{\pd \l}\Omega^\pm_{E_N(\l), \l}$. Then $v^\pm_{E,\l}$ satisfies the differential equation 
\[
\frac{d}{d\tau}v^\pm_{E,\l} \,=\, P^\pm_{E,\l}(\tau)v(\tau) + 2\cos \xi(\tau) \sin \Omega^\pm_{E,\l}(\tau).
\]
The solution to this differential equation is 
\[
v^\pm_{E,\l}(\tau) \,=\, U^\pm_{E,\l}(\tau,\tau_0)v^\pm_{E,\l}(\tau_0) + \int_{\tau_0}^\tau U^\pm_{E,\l}(\tau,s)\big(2\cos \xi(s) \sin \Omega^\pm_{E,\l}(s)\big)ds. 
\]
Using the same analysis as above, we have
\[
\frac{\pd \Phi}{\pd \l}\big(E_N(\l), \l\big) \,=\, \int_{-\infty}^\infty U_{E_N(\l)}(0,s)\big(2\cos\xi(s) \sin \Omega_{E_N(\l)}(s)\big) ds
\]
where $\Omega_{E_N(\l)} := \Omega^+_{E_N(\l), \l} = \Omega^-_{E_N(\l), \l}$.
Then $E_N'(\l)$ is given by implicit differentiation
\[
E_N'(\l) \,=\, -\frac{\pd \Phi/ \pd \lambda}{\pd \Phi/ \pd E}\big(E_N(\l), \l) \,=\, \frac{1}{a}\left(\frac{\int_{-\infty}^\infty U_{E_N(\l)}(0,s)\big(\cos\xi(s) \sin \Omega_{E_N(\l)}(s)\big) ds}{\int_{-\infty}^\infty U_{E_N(\l)}(0,s)ds}\right).
\]
Since the numerator in absolute value is strictly less than the denominator in absolute value, we have $|E_N'(\l)| < \frac{1}{a}$. \qed 

\medskip
\medskip

For the following theorem, we define the following closed intervals. Let $N_\Theta \geq 0$ be an integer for $I_1$ and $I_2$, and let $N_\Theta \leq -1$ be an integer for $I_3$ and $I_4$.
\medskip
\begin{align*}
I_1 \,&=\, \left[-(2N_\Theta+1)(\frac{1}{2} + \kappa) - 2a,\, -\frac{1}{2} - \kappa + a\right]
\\
I_2 \,&=\, \left[-(2N_\Theta+1)(\frac{1}{2} - \kappa) - 2a,\, -\frac{1}{2} + \kappa - a\right]
\\
I_3 \,&=\, \left[\,\frac{1}{2} + \kappa -a, \, -(2N_\Theta+1)(\frac{1}{2} + \kappa) + 2a\right]
\\
I_4 \,&=\, \left[\,\frac{1}{2} - \kappa-a,\, -(2N_\Theta+1)(\frac{1}{2} - \kappa) + 2a\right].
\end{align*}
\medskip

Intervals $I_1$ and $I_2$ correspond to $\l < 0$ with $\kappa > 0$ and $\kappa < 0$, respectively; their endpoints are determined by Theorem \ref{exist theta system saddle connectors thm}. Intervals $I_3$ and $I_4$ correspond to $\l > 0$ with $\kappa > 0$ and $\kappa < 0$, respectively; their endpoints are determined by Theorem \ref{exist theta system saddle connectors thm for neg wind num}.

\medskip
\medskip

\begin{thm}\label{iteration thm}\:$\phantom{nix}$
\begin{itemize}
\item[$\bullet$] Let $\l < 0$. For each pair of integers $N_\Omega \geq 0$ and $N_\Theta \geq 0$, there is a unique $E_* \in (0,1)$ and a unique $\l_* \in I_1$ if $\kappa > 0$ or unique $\l_* \in I_2$ if $\kappa < 0$ such that  $E_* = E_{N_\Omega}(\l_*)$ and $\l_* = \l_{N_\Theta}(E_*)$ where $E_{N_\Omega}$ and $\l_{N_\Theta}$ are the functions from Propositions \emph{\ref{E c1 prop}} and \emph{\ref{lambda c1 prop}}. There are no saddles connectors with $N_\Omega \leq -1$.

\item[$\bullet$] Let $\l > 0$. For each pair of integers $N_\Omega \geq 1$ and $N_\Theta \leq -1$, there is a unique $E_* \in (0,1)$ and a unique $\l_* \in I_3$ if $\kappa > 0$ or unique $\l_* \in I_4$ if $\kappa < 0$ such that $E_* = E_{N_\Omega}(\l_*)$ and $\l_* = \l_{N_\Theta}(E_*)$ where $E_{N_\Omega}$ and $\l_{N_\Theta}$ are the functions from Propositions \emph{\ref{E c1 prop}} and \emph{\ref{lambda c1 prop}}. There are no saddles connectors with $N_\Omega \leq 0$.
\end{itemize}
\end{thm}

\proof
We prove it for $\l < 0$; the proof for $\l > 0$ is analogous. That there are no saddles connectors with $N_\Omega \leq -1$ follows from Proposition \ref{horiz barr prop}. We set out now to establish existence for $N_\Omega \geq 0$. To simplify the notation, set $n = N_\Theta$ and $m = N_\Omega$. 

Define the function $\phi_{n,m} \colon [0,1] \to [0,1]$ by $\phi_{n,m}(E) = E_m\big(\l_n(E)\big)$. By Propositions \ref{lambda c1 prop} and \ref{E c1 prop}, we have $\phi_{n,m}$ is a $C^1$ function on $[0,1]$ and $|\phi_{n,m}'(E)| = |E_m'\big(\l_n(E)\big)\l_n'(E)| < \frac{1}{a}a = 1$. Therefore, since $[0,1]$ is compact, the mean value theorem implies that $\phi_{n,m}$ is a contraction. Thus there exists a unique $E_* \in [0,1]$ such that $E_* = E_m\big(\l_n(E_*)\big)$. Similarly, define $\psi_{n,m} \colon I \to I$ by $\psi_{n,m} = \l_n\big(E_m(\l)\big)$ where $I = I_1$ if $\kappa > 0$ and $I = I_2$ if $\kappa < 0$. The same argument shows that $\psi_{n,m}$ is a contraction, and thus there exists a unique $\l_* \in I$ such that $\l_* = \l_n\big(E_m(\l_*)\big)$. 

We have $E_m(\l_*) = E_m \big(\l_n\big(E_m(\l_*)\big)\big)$. By uniqueness it follows that $E_* = E_m(\l_*)$. Hence $E_* \in (0,1)$ by Theorem \ref{exist omega system saddle connectors thm}. 
\qed

\medskip
\medskip

Theorem \ref{iteration thm} completes the proof of Theorem \ref{main thm}.

\section{Connection with the hydrogenic spectra}

Based on the above results, for fixed $a$ and $\ga$, the discrete spectrum of our Dirac hamiltonian is indexed by {\em three} integers: $N_\Omega$, $N_\Theta$,  and $2\kappa$.  
By contrast, the energy spectrum of special relativistic hydrogen, i.e., the Dirac hamiltonian for a point-like electron in ordinary Minkowski space interacting with a point charge at the origin, is indexed by {\em two} integers only, namely the main (or Bohr's) quantum number, often denoted by $n$, and the spin-orbit quantum number,\footnote{called $\kappa_j$ in Thaller; it should not be confused with our $\kappa$, which is the eigenvalue of the $z$-component of angular momentum, for which Thaller uses the notation $m_j$.}
 i.e., the set of eigenvalues of the spin-orbit operator $K = \beta(2\bS\cdot\bL +1)$.

In this section, we classify the energy eigenvalues of our hamiltonian, relate the numbers $(N_{\Omega},N_{\Theta},\kappa)$ to the quantum numbers of the special relativistic hydrogen theory, and associate the corresponding energy eigenstates with the well-known {\em orbitals} of the hydrogen atom. In the next section we report on the results of several numerical experiments which provide evidence for our theoretical calculations and arguments.

In the limit $a \to 0$, the angular hamiltonian (\ref{eq:Tang}),  takes the simple form
\beq\label{eq:Tang0}
\fa_\ka \,:=\, \lim_{a\to 0} T_{ang} \,=\,  i\si_2 \pd_\theta + \frac{\ka}{\sin\theta} \si_1.
\eeq
From \cite{BSW}, if $\la$ is an eigenvalue of $T_{ang}$, then $k := \lim_{a \to 0} \la$ is an eigenvalue of $\fa_\ka$ since $\la$ is analytic in $a$. Note that $\fa_\ka$ is independent of $E$ unlike $T_{ang}$.

In the limit $a\to 0$, the formal limit of the radial hamiltonian (\ref{eq:Hrad}) coincides with the radial hamiltonian arising in the special relativistic Hydrogen problem (e.g. see \cite[eq. (7.105)]{ThallerBOOK}):
\beq\label{eq:Hrad0}
\fh_{k} \,:=\, \lim_{a\to 0} H_{rad} \,=\, \left(\begin{array}{cc} m  + \frac{\gamma}{r} & -\pd_r + \frac{k}{r} \\[20pt]
 \pd_r +\frac{k}{r} & -m + \frac{\gamma}{r}  \end{array}\right).
\eeq
Therefore $k$ can be identified with the \emph{spin-orbit coupling}.

The spectrum of $\fa_\ka$ is completely understood\cite[Appendix A]{BSW}.\footnote{Note that the sign of $\kappa$ in \cite[eq. (5)]{BSW} is opposite of ours appearing in eq. (\ref{eq:Tang0}). This ambiguity in the sign follows from choosing either $\pm (Et - \kappa \varphi)$ in eq. (\ref{chandra-ansatz}).} In particular, for all half-integers $\ka \in \mathbb{Z} + \half$, the operator $\fa_\ka$ is essentially self-adjoint and has a discrete spectrum indexed by a nonzero integer which we call $N$ (see \cite[eq. (7)]{BSW}):

\beq\label{kintermsofN}
k \,=\, -\sgn(N) \left( |N| + |\ka| - \half\right),
\eeq
as well as a complete set of eigenvectors $\vec{S}_{N,\ka}$ that are explicitly known and can be expressed in terms of Jacobi polynomials.
\begin{equation}
\vec{S}_{N,\ka}(\theta) \,:=\, \sin^{\ka+\half} \theta \left(\begin{array}{c} 
\qquad -\sqrt{\cot\frac{\theta}{2}} P_{|N|-1}^{\ka-\half,\ka+\half}(\cos\theta) \\
\sgn(N)\sqrt{\tan\frac{\theta}{2}} P_{|N|-1}^{\ka+\half,\ka-\half}(\cos\theta)
\end{array}\right).
\end{equation}
Note that the above holds for $\ka>0$.  To find the eigenvectors for $\ka<0$, recognize that if $\vec{S}$ is an eigenvector of $\fa_\ka$, then $i\si_2 \vec{S}$ is an eigenvector of $\fa_{-\ka}$. 

From the above and the definition of $\Theta$ given by eq. (\ref{eq:prufer}), it follows that the saddles connectors of the $\Theta$-system that correspond to the eigenvectors of the angular hamiltonian are given explicitly by the formula
\begin{equation}
\Theta_{N,\ka}(\theta) \,=\,  -2\,\sgn(N)  \tan^{-1}\left(\frac{P_{|N|-1}^{|\ka|+\half,|\ka|-\half}(\cos\theta)}{P_{|N|-1}^{|\ka|-\half,|\ka|+\half}(\cos\theta)}\tan \frac{\theta}{2}  \right) .
\end{equation}
The above formula holds for $\kappa > 0$. For $\kappa < 0$, the formula is the same except one switches the numerator and denominator. We choose our branch of $\tan^{-1}$ so that inital values of the saddles connectors agrees with the saddle equilibrium points within the fundamental domain chosen in the beginning of section \ref{th sec}. That is,

\begin{equation}
\Theta_{N,\ka}(0) \,=\, \left\{\begin{array}{cc} 0, & \ka>0\\ \pi, & \ka<0
\end{array}\right.
\end{equation}
Then the final values of those connectors are 
\begin{equation}\label{th end point mod 2pi}
\Theta_{N,\ka}(\pi) \,=\, \left\{\begin{array}{cc}\quad  -\sgn(N)\pi, & \ka>0 \\ \pi - \sgn(N) \pi, & \ka<0 \end{array}\right.\qquad [\mbox{mod} 2\pi]
\end{equation}
The branch of $\tan^{-1}$ needs to be chosen in such a way that $\theta_{N,k}$ is continuous on $[0, \pi]$. Since the Jacobi polynomials $P_{|N| - 1}^{\a,\b}(\cos \theta)$ have $|N|-1$ distinct roots within $(0,\pi)$, it follows that we need to add $-\sgn(N) 2\pi$ to eq. (\ref{th end point mod 2pi}) for each root to determine the actual endpoint of $\Theta_{N,\kappa}$. That is
\begin{equation}\label{Thetafinal}
\Theta_{N,\ka}(\pi) \,=\, -\sgn(N)(|N|-1)2\pi+\left\{\begin{array}{cc}\quad -\sgn(N)\pi, & \ka>0 \\ \pi - \sgn(N)\pi, & \ka<0 \end{array}\right.
\end{equation}
Thus we establish a correspondence between the integer $N$ and the winding number $N_\Theta$ of the $\Theta$-saddles connectors in the case $a=0$:
\beq\label{Thetaconn}
N_\Theta \,\sim\, 
\left\{\begin{array}{cc} N - 1, & N \geq 1 
\\ 
\quad N,\quad & N \leq -1, \end{array}\right.
\eeq
where `$\sim$' simply means a correspondence.

We now turn our attention to the radial hamiltonian $\fh_k =\lim_{a\to 0} H_{rad}$. This operator is closely related to the radial hamiltonian $h_k$ of the special-relativistic Hydrogen problem, as formulated by Dirac\cite{Dir28}:
\begin{equation}
h_k = -i\si_2 \pd_r + m \si_3 + \frac{k}{r} \si_1 + \frac{\ga}{r} I_{2\times 2}.
\end{equation}

The only difference between $h_k$ and $\fh_k$ is their domains:  Since $H_{rad}$ is defined on the double-sheeted Sommerfeld space, this is inherited by its $a\to 0$ limit $\fh_k$, which is still defined on two copies of Minkowski space glued together along a timelike line.  In particular the $r$ variable in $\fh_k$ has the range $(-\infty,\infty)$.  By contrast, the $r$ variable in $h_k$ goes from $0$ to $\infty$.
  The eigenvalue problem for $h_k$ was shown to be exactly solvable by Gordon \cite{Gor28}, who proved the discrete spectrum to coincide exactly with the Bohr-Sommerfeld spectrum, and found explicit formulas for the eigenfunctions in terms of generalized Laguerre polynomials.

 Consider the operator $h_k$. As described in detail in section 7.4 of Thaller's book\cite{ThallerBOOK}, for all $k \in \Z\setminus\{0\}$, $h_k$ is essentially self-adjoint on $C^\infty_c((0,\infty))$, has a discrete spectrum in $(0,1)$, and a complete set of eigenvectors,  provided $-\sqrt{3}/2 < \ga < 0$.  The discrete spectrum is indexed by two integers, $n\geq 1$ and $k = -n,\dots,-1,1,\dots,n-1$.  Let $M:= n - |k|$.  Then we have
\begin{equation}\label{eq:SommerfeldSpec}
E_{M,k} \,=\, \frac{m}{\sqrt{ 1+ \frac{\ga^2}{\left(M + \sqrt{ k^2 - \ga^2}\right)^2}}}
\end{equation}
Note that the case $k > 0$ and $M = 0$ is excluded. 

Gordon \cite{Gor28} computed the corresponding eigenfunctions in terms of generalized Laguerre polynomials:  Let $(\phi_1,\phi_2)$ be defined by 
\begin{equation}
u = \sqrt{1+E_{M,k}}(\phi_1+\phi_2),\qquad v = \sqrt{1-E_{M,k}}(\phi_1-\phi_2)
\end{equation}
where $(u,v)$ is an eigenfunction of $\fh_k$ with eigenvalue $E_{M,k}$ (and we have set $m=1$).  Then, using the abbreviations
\begin{equation}
\rho := \sqrt{k^2 - \ga^2},\qquad \eta := \sqrt{1-E_{M,k}^2},
\end{equation}
for all nonnegative integers $M$ and real constants $c_1,c_2$ such that
\begin{equation}
\mu := \frac{c_1}{c_2} = \frac{M}{k + \frac{\ga}{\eta}},
\end{equation}
we have
\begin{align}
\phi_1(r) \,&=\, c_1 e^{-\eta r} r^\rho F(-M+1, 2\rho+1, 2\eta r)\\
\phi_2(r) \,&=\, c_2 e^{-\eta r} r^\rho F(-M,2\rho+1,2\eta r),
\end{align}
where $F$ denotes Gauss\rq{}s {\em confluent hypergeometric function}
\begin{equation}
F(\al,\beta,x) \,=\, 1 + \frac{\al}{1!\beta}x+\frac{\al(\al+1)}{2!\beta(\beta+1)}x^2 + \dots
\end{equation}
Note that when $\al$ is a negative integer, the above series terminates, and $F$ is a polynomial of degree $-\al$, which (up to a numerical factor) is the generalized Laguerre polynomial $L_{-\al}^{(\beta-1)}(x)$.  

Accordingly, since $\Om = 2 \tan^{-1}\frac{v}{u}$, the corresponding solution to the $a \to 0$ limit of the $\Om$-equation (\ref{eq:Om}) will be
\beq\label{Omsol}
\Om(r) = 2 \tan^{-1}\left( \sqrt{\frac{1-E}{1+E}}\ \frac{\mu F(-M+1,2\rho+1,2\eta r) - F(-M,2\rho+1,2\eta r)}{\mu F(-M+1,2\rho+1,2\eta r) + F(-M,2\rho+1,2\eta r)}\right)
\eeq
The values of $\Om$ at $r=0$ and $r=\infty$ can thus be calculated (modulo $2\pi$):
\begin{align}
\Om(0) \,&=\, 2 \tan^{-1}\left( \sqrt{\frac{1-E}{1+E}}\ \frac{\mu-1}{\mu+1}\right) \,=  \sin^{-1}\left(\frac{-\ga}{k}\right)
\\
 \Om(\infty) \,&=\, -2\tan^{-1}\left( \sqrt{\frac{1-E}{1+E}}\right) \,=\, -\cos^{-1}E
\end{align}
 which is in agreement with the analysis of the equilibrium points of the corresponding dynamical system.  We choose our principal branch of $\sin^{-1}$ depending on the sign of $k$:
\begin{align}
\Om(0) \,&=\, \left\{\begin{array}{ll} 
\sin^{-1}(\frac{-\ga}{k}) & \mbox{ if }k<0
\\
-\pi - \sin^{-1}(\frac{-\ga}{k}) & \mbox{ if }k>0
\end{array}\right. .
\end{align}
On the other hand, from the properties of Laguerre polynomials, it follows that the denominator of the rational function in (\ref{Omsol}) has $M$ zeros in $(0, \infty)$ when $k < 0$ and $M-1$ zeros in $(0, \infty)$ when $k>0$. That is

\medskip

\begin{prop}\label{zeros of Om denom prop}
The polynomial 
$\mu F(-M + 1, 2\rho + r, 2\eta r) + F(-M, 2\rho + 1, 2\eta r)$ has
\begin{itemize}
\item[$\bullet$] $M \geq 0$ distinct zeros in $(0, \infty)$ when $k < 0$ and
\item[$\bullet$] $M - 1 \geq 0$ distinct zeros in $(0, \infty)$ when $k > 0$,  provided $|\g| < \half$.  
\end{itemize}
\end{prop}

\medskip

We prove Proposition \ref{zeros of Om denom prop} at the end of this section. Note that it explains why we must have $k \leq n - 1$: if $k = n$, then $M - 1 = -1$, but a polynomial can't have a negative number of zeros. The assumption $|\g| < \half$ is merely technical and used to simplify the proof of Proposition \ref{zeros of Om denom prop}. It's expected Proposition \ref{zeros of Om denom prop} should still hold even under the assumption $|\g| < \frac{\sqrt{3}}{2}$. However, note that $|\gamma| < \half$ is part of our assumptions in Theorem \ref{main thm}.

Thus in order for $\Om$ to be a continuous function of $r$, the branch of $\tan^{-1}$ needs to be chosen such that $-2\pi $ gets added to the value every time $r$ crosses one these zeros. Therefore
\begin{equation}
\Omega(\infty) \,=\, -2\pi M - \cos^{-1}E
\end{equation}
which holds for both $k < 0$ and $k > 0$ (this is why we chose $\Omega(0) = -\pi - \sin^{-1}(\frac{-\ga}{k})$ for $k > 0$). Thus we establish the following correspondence between the integer $M$ and the winding number $N_\Omega$ for the $\Omega$-saddles connectors in the case $a = 0$. 
\beq\label{Om Conn}
N_\Omega \,\sim\, M. 
\eeq

Now consider the case $a > 0$ (i.e. the Dirac hamiltonian on z$G$KN). The bound states are indexed by three integers $N_\Theta, N_\Omega$, and $2\kappa$ appearing in Theorem \ref{main thm}. Using (\ref{Om Conn}) and (\ref{Thetaconn}), we can define a correspondence between the integers $N_\Theta, N_\Omega, 2\kappa$ and the usual spectroscopic notation $n\ell_j$ of hydrogenic states. Given $N_\Theta$ and $N_\Omega$, define $N$ and $M$ via  
\beq
N \,:=\,
\left\{\begin{array}{cc} N_\Theta + 1, & N_\Theta \geq 0 
\\ 
\; N_\Theta,\; & N_\Theta \leq -1, \end{array}\right. \quad \text{ and } \quad M \,:=\, N_\Omega. 
\eeq
Then $n$, $\ell$, and $j$ appearing in $n\ell_j$ are given by 
\begin{align}
k \,&:=\, -N - \sgn(N)\left(|\kappa| - \half\right)
\\
n\,&:=\, M + |k|
\\
j\,&:=\, |k| - \half
\\
\ell \,&:=\, j + \sgn(k)\half
\\
m_j \,&:=\, \kappa. 
\end{align}

This establishes the desired correspondence between the winding numbers appearing in Theorem \ref{main thm} and the usual spectroscopic notation $n \ell_j$ of hydrogenic states. For example, suppose we ask: Which state in z$G$KN corresponds to $2p_{1/2}$ with $m_{1/2} = -\half$? Note $\ell = 1$ implies $1 = \half + \sgn(k)\half$. Therefore $k$ is positive and so $j = \half$ implies $k = 1$. Therefore $N = -1$ implies $N_\Theta = -1$. Lastly, $n = 2$ implies $M = 1$ and so $N_\Omega = 1$. Thus  the state $2p_{1/2}$ with $m_{1/2} = - \half$ corresponds to $N_\Theta = -1$, $N_\Omega = 1$, and $\kappa = -\half$. 


\begin{table}[h!]
  \begin{center}
    \label{tab:table1}
    \begin{tabular}{c|c|c|c} 
      Hydrogenic state & $N_\Theta$ & $N_\Omega$ & $\kappa$ \\
      [.1cm]\hline
      & & & \\ 
      $1s_{1/2}$,\: $m_{1/2} = \pm \half$ & $0$ & $0$ & $\pm \half$ \\& & & \\
      $2s_{1/2}$,\: $m_{1/2} = \pm \half$ & $0$ & $1$ & $\pm \half$\\& & & \\
      $2p_{1/2}$,\: $m_{1/2} = \pm \half$ & $-1$ & $1$ & $\pm \half$\\     & & & \\ $2s_{3/2}$,\: $m_{3/2} = \pm \half$ & $-2$ & $0$ & $\pm \half$\\& & & \\
      $2s_{3/2}$,\: $m_{3/2} = \pm \frac{3}{2}$ & $-1$ & $0$ & $\pm \frac{3}{2}$\\ [.1cm]
    \end{tabular}
  \end{center}
  \caption{\small{Correspondence between hydrogenic states and winding numbers.}}
\end{table}

\medskip

\noindent{\emph{Proof of Proposition \emph{\ref{zeros of Om denom prop}}}}.

\smallskip

Set $\b = 2\rho + 1$ and $x = 2 \eta r$. And set $P_M(x) = F(-M, \beta, x)$ and $Q_M = \mu P_{M-1} + P_M$. We first show that $Q_M$ has at least $M - 1$ real distinct roots in $(0, \infty)$. Since complex roots come in pairs, this will imply that the roots of $Q_M$ are all real. Seeking a contradiction, suppose $Q_M$ does not have $M - 1$ roots in $(0, \infty)$. Since $P_M$ are proportional to the generalized Laguerre polynomials, they form a system of orthogonal polynomials with respect to the weight function $w(x) = x^{\beta - 1}e^{-x}$ on $(0,\infty)$.  Since $\int_0^\infty Q_M(x)w(x)dx = 0$, it follows that $Q_M$ changes sign at least once in $(0, \infty)$; hence $Q_M$ has at least one zero of odd multiplicity in $(0, \infty)$. Let $x_1, \dotsc, x_k$ denote the distinct zeros of odd multiplicity located in $(0, \infty)$. By assumption $k < M -1$. Set $Z(x) =(x - x_1) \dotsb (x - x_k)$. Then $Z(x)Q_M(x)$ is a polynomial with no zeros of odd multiplicity in $(0, \infty)$ and therefore $\int_0^\infty Z(x) Q_M(x) w(x) dx \neq 0$. Since $Z(x)$ is a polynomial with degree $k$, we can write it as $Z(x) = \sum_{i = 0}^k a_i P_i(x)$. Therefore we get a contradiction: 
\[
0 \,\neq\, \int_0^\infty Z(x) Q_M(x) w(x) dx \,=\, \int_0^\infty \left(\sum_{i = 0}^k a_i P_i(x)\right)\big(\mu P_{M - 1}(x) + P_M(x)\big) w(x) dx \,=\, 0.   
\]
We get zero for the last equality since $k < M - 1$. Thus $Q_M$ has at least $M - 1$ simple roots in $(0, \infty)$. 

Since $Q_M$ has at least $M - 1$ distinct real roots in $(0, \infty)$, by Descartes' rule of signs, it suffices to show that the signs of the coefficient for $Q_M$ (ordered by ascending variable exponent) alternate $M$ times when $k < 0$ and alternate $M - 1$ times when $k > 0$. An induction argument shows that the coefficients of $P_M$ form Pascal's triangle with alternating signs. For example $P_3(x) = 1 - \frac{3}{\beta} x + \frac{3}{\beta(\beta + 1)}x^2 - \frac{1}{\beta(\beta + 1)(\beta + 2)}x^3$. So for $M = 4$, we have
\begin{align*}
Q_4(x) \,&=\, \mu P_3(x) + P_4(x)
\\
&=\,  (\mu + 1) -(3\mu + 4) \frac{x}{\beta} + (3\mu + 6)\frac{x^2}{\beta(\beta + 1)} - (\mu + 4)\frac{x^3}{\beta(\beta + 1)(\beta + 2)} + \frac{x^4}{\beta \dotsb (\beta + 3)}
\end{align*} 
In general, the zeroth order term for $Q_M$ will be $(\mu + 1)$, the first order term will be proportional to $(\mu + \frac{M}{M-1})$, and each term after that will be proportional to something greater than $(\mu + \frac{M}{M-1})$ in magnitude. Moreover each term alternates in sign. Thus
\begin{itemize}
\item[(1)]for $k < 0$, it suffices to show that $\mu + 1 > 0$ since this implies that the signs for $Q_M(x)$ go like $+-+-+ \dotsb$, and
\item[(2)] for $k > 0$, it suffices to show that $\mu + 1 < 0$ and $\mu + \frac{M}{M-1} > 0$ since this implies that the signs for $Q_M(x)$ go like $--+-+\dotsb$. Note this only applies for $M \geq 2$; the case $M = 1$ can be verified independently.
\end{itemize}
Using $k = \sgn(k) \sqrt{\rho^2 + \g^2}$ and $\sqrt{1 - E^2} = -\g/\sqrt{(M + \rho)^2 + \g^2}$, we have
\[
\mu \,=\, \frac{M}{k + \g/\sqrt{1 - E^2}} \,=\, \frac{M}{\sgn(k) \sqrt{\rho^2 + \g^2} - \sqrt{(M + \rho)^2 + \g^2}}.
\]
From this expression, it follows that $-\mu < 1$ if $k < 0$;  this proves (1). Now we prove (2). Since $k > 0$, rationalizing the denominator gives
\[
-\mu \,=\, \frac{\sqrt{\rho^2 + \g^2} + \sqrt{(M + \rho)^2 + \g^2}}{M + 2\rho} \,<\, \frac{\rho + |\g| + M + \rho + |\g|}{M + 2\rho} \,=\, 1 + \frac{2|\gamma|}{M + 2\rho}. 
\]
Therefore $\mu + 1 < 0$. Lastly, since $\frac{M}{M-1} = 1 + \frac{1}{M-1}$, it suffices to show $\frac{2|\gamma|}{M + 2\rho} < \frac{1}{M -1}$. Indeed
\[
\frac{2|\gamma|}{M + 2\rho} \,=\, \frac{2|\gamma|}{M + 2\sqrt{k^2 - \g^2}} \,<\, \frac{2|\gamma|}{M + 2\sqrt{1-\gamma^2}} \,<\, \frac{1}{M-1}  
\]
where the last inequality holds whenever $|\g| < \half$. \qed


\section{Numerical results}\label{num}
The following numerical simulations are performed {using MATLAB \cite{matlab}.}  The algorithm for finding the energies of saddles connectors is as follows: The two ODEs \eqref{eq:Theta} and \eqref{eq:Om} are numerically integrated (using Runge-Kutta 4-5) over a suitably large interval symmetric about zero, with initial data corresponding to the saddle nodes on the left boundary of the corresponding cylinders. The two desired winding numbers $N_\Theta$ and $N_\Omega$ are fixed, as well as the other parameters $\kappa$, $a$ and $\gamma = -Z\al_S$.  A shooting method is used iteratively on the parameters $\lambda$ for the $\Theta$ equation and $E$ for the $\Omega$ equation, to fine tune the values for these parameters via a binary search, until solutions with the desired winding numbers are found for both equations using the same (up to a specified tolerance) pair of $(\lambda, E)$ values.  MATLAB's graphic routines are then used to generate the plots.
\subsection{Comparison of computed eigenvalues for the angular hamiltonian with previously known results}
The angular hamiltonian \eqref{eq:Tang} is identical to the one obtained for Dirac's equation on the Kerr-Newman spacetime.  It has been extensively studied, see \cite{BSW} and references therein.\footnote{We note that a different sign convention for the Chandrasekhar Ansatz \eqref{chandra-ansatz} was used in \cite{BSW}.  As a result, their expression for the angular hamiltonian looks different from ours.  It is easy to see that in order to get the same expression, one needs to change the signs of the parameters $a$, $\om$ (which corresponds to our $E$), and $\kappa$ in eq. (5) of \cite{BSW}.\label{fn:bsw}} It is known to be essentially self-adjoint, its spectrum is discrete, and consists of simple eigenvalues $\{\lambda_N\}_{N\in \mathbb{Z}^*}$ that, {as can be seen in \eqref{eq:Tang} and \eqref{opdefsB},} depend on the parameters $\mu := am$ and $\nu := aE$, as well as on the half-integer $\kappa$.  
It was shown in \cite{BSW} that this dependence is analytic, and that $\la_N(\kappa; \mu,\nu)$ has a double series expansion in $\mu$ and $\nu$, with coefficients that can be computed recursively.\footnote{The index $j$ used in \cite{BSW} to enumerate the angular eigenvalues has the opposite sign of our $N$, i.e. their $\la_j$ is our $\la_{-N}$.}  In particular the authors of that paper
 showed that, setting $\alpha := \nu-\mu$ and $\beta := \nu+\mu$, one has
\beq\label{expansion}
\lambda_N(\kappa; \alpha, \beta) = \sum_{m,n=0}^\infty c_{m,n}\alpha^m\beta^n
\eeq
Moreover, they showed that\footnote{Correcting for the sign of $\kappa$, see footnote \ref{fn:bsw}.}
\beq\begin{array}{lll}
c_{0,0} = -\sgn(N)(|\kappa|-\half+|N|) =: k, &
c_{1,0} = \frac{-\kappa}{2k+1}, &  c_{0,1} = \frac{-\kappa}{2k-1}, \\[5pt]
c_{2,0} = \frac{(2k+1)^2 - 4\kappa^2}{4(2k+1)^3},&  c_{1,1} = 0, &
c_{0,2} = \frac{(2k-1)^2 - 4\kappa^2}{4(2k-1)^3}
\end{array}
\eeq
and that all other coefficients can be computed using these and the recursion relations they had provided.  

Here we have demonstrated a different way to compute the eigenvalues $\lambda$ of the angular hamiltonian \eqref{eq:Tang}, namely, we compute the two eigenvalues $(\lambda,E)$ of the angular and radial hamiltonians in an iterative procedure, by finding saddles connectors for the corresponding flows on finite cylinders.  It is interesting to compare these two methods for computing $\lambda_N$.
\begin{figure}[h!]
\centering
\includegraphics[scale=0.4]{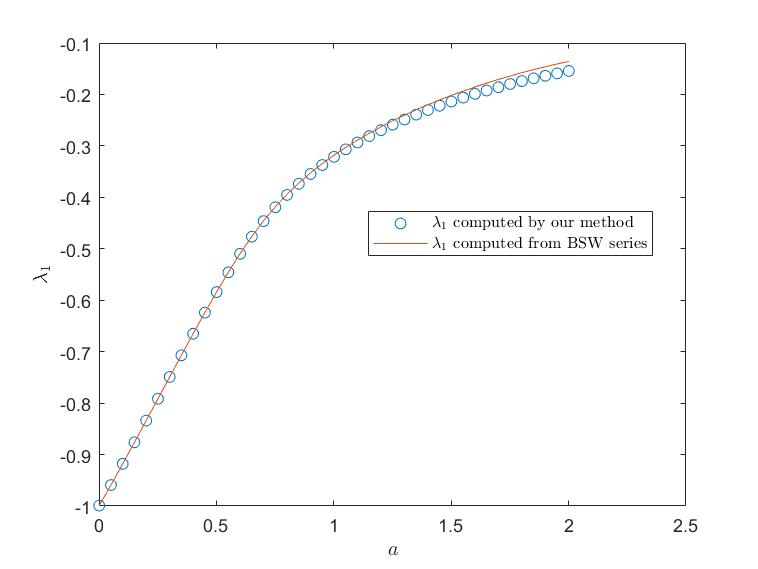}
\caption{Comparison of angular eigenvalues computed in two different ways. 
}
\label{fig:comparison}
\end{figure}
In Figure~\ref{fig:comparison} the blue circles are values of $\la_1$ computed using our method, with the one given by the first few terms in the expansion \eqref{expansion} (corresponding to $m+n\leq 2$), in which we have substituted for $E$ its approximate value computed using our method.  We see that there is pretty good agreement between the two all the way to $a\approx 1$. The minor discrepancies that show for $a>1$ we expect to 
go away when more terms of the analytic expansion are taken into account.

\subsection{Breaking of degeneracies in energy eigenvalues}
\begin{figure}[h!]
\centering
\includegraphics[scale=0.4]{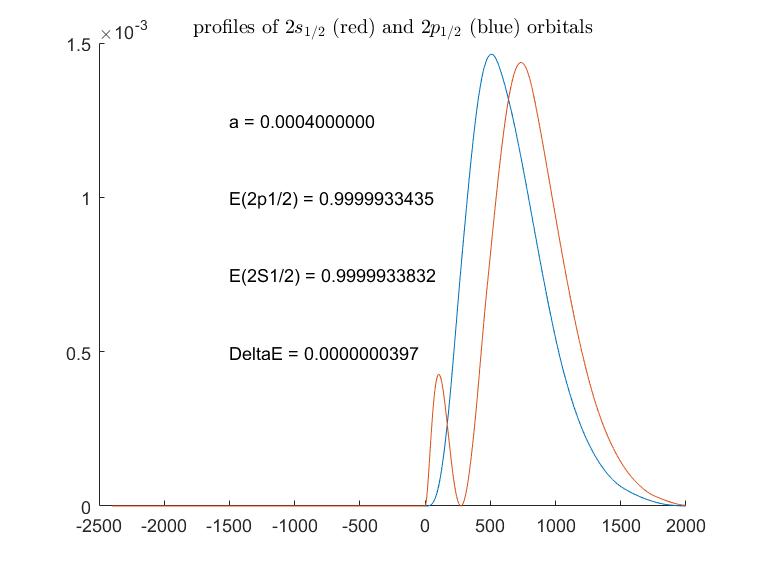}
\caption{Breaking of accidental degeneracy for $a = 0.0004$. }
\label{fig:lambshift}
\end{figure}

The discrete spectrum of the Dirac equation on Minkowski space dressed with a static point charge (the coulombic Dirac hamiltonian)
is well-known to have accidental degeneracies, since energy eigenvalues do not depend on the sign of the spin-orbit coupling (see \eqref{eq:SommerfeldSpec}) while the eigenfunctions of course do.  This degeneracy is broken when $a\ne 0$, leading to effects that mimic the Lamb-shift (of course they do so only qualitatively) as depicted in Figure~\ref{fig:lambshift}.  In this figure the horizontal axis is the variable $r$, in units of $\hbar/m_ec$, and the vertical axis is $|\Psi|^2$.

\begin{figure}[h!]
\centering
\includegraphics[scale=0.4]{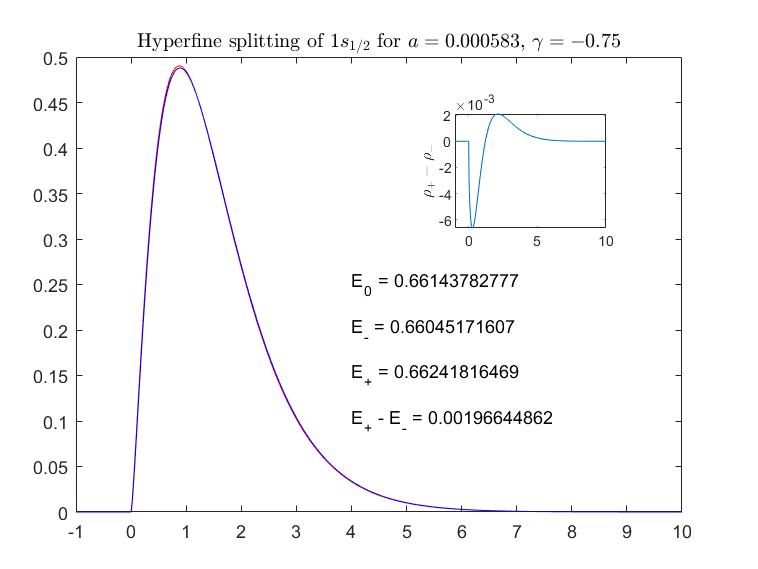}
\caption{Hyperfine-like structure.}
\label{fig:hyperfine}
\end{figure}

Another degeneracy present in the spectrum of the coulombic Dirac operator on Minkowski space, is the lack of dependence of the energy levels on the eigenvalue of the $z$-component of the angular momentum, usually denoted by $m_j$ (see the previous section for a dictionary), while the corresponding eigenfunctions do depend on $m_j$ through their dependence on spherical harmonics.  This degeneracy is also broken in our model, since the $\Omega$ equation manifestly depends on $\kappa$, which is the corresponding quantity to $m_j$ in this context.  The breaking of this degeneracy results in effects that are akin to {\em hyperfine splitting}, as can be seen in Figure~\ref{fig:hyperfine}.
This figure shows the splitting of the two identical $a=0$ ground state profiles into two distinct
profiles with the values for $a>0$ and $\gamma$, as indicated, but with $\kappa = \half$ (corresponding to parallel spins for the electron and the nucleus) and with $\kappa = -\half$ (the anti-parallel case).  The two curves are almost indistinguishable at the scale of this figure (the inset shows the difference between the two,) but their energies are clearly different, as indicated.

\subsection{Dependence of the spectrum on the atomic number}

The dependence of the spectrum of the coulombic Dirac hamiltonian on $\gamma = -Z\al_S$, the strength of the Coulomb potential, is well-documented, e.g. \cite{ThallerBOOK}, as well as the interesting behavior of the energies of the excited states when a term corresponding to the anomalous magnetic moment of the electron is added to the hamiltonian ({\em ibid.}, p. 219).  As discussed earlier, anomalous magnetic moment effects are already included in our Dirac on z$G$KN hamiltonian, so one expects a similar pattern of behavior to emerge in our case as well.  As we shall see, there are some similarities, but there are also some big surprises.

\begin{figure}[h!]
\centering
\includegraphics[scale=0.4]{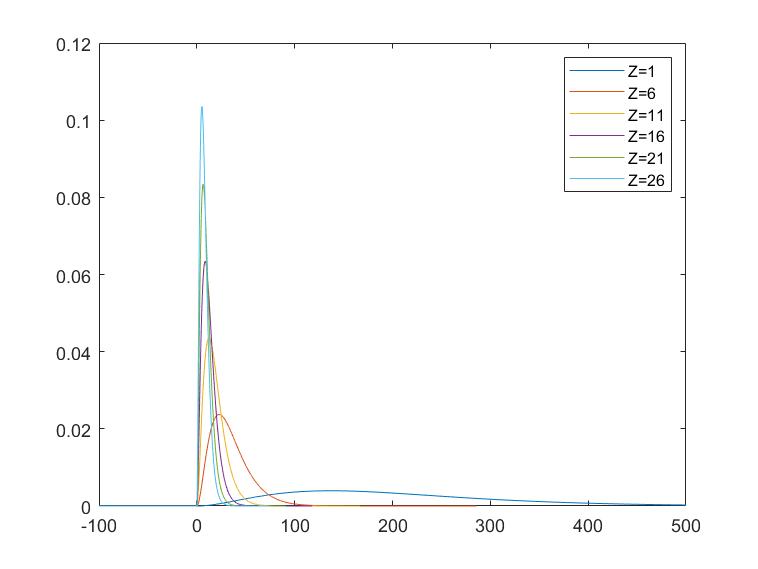}
\caption{Ground state profile for different $Z$.}
\label{fig:Zprofile}
\end{figure}

But first, it stands to reason that the profile of the ground state should show that as $Z$ increases, the electron would tend to hover closer to the ring singularity.  This is easily confirmed by the numerics, as depicted in Figure~\ref{fig:Zprofile}.

\begin{figure}[h!]
\centering
\includegraphics[scale=0.4]{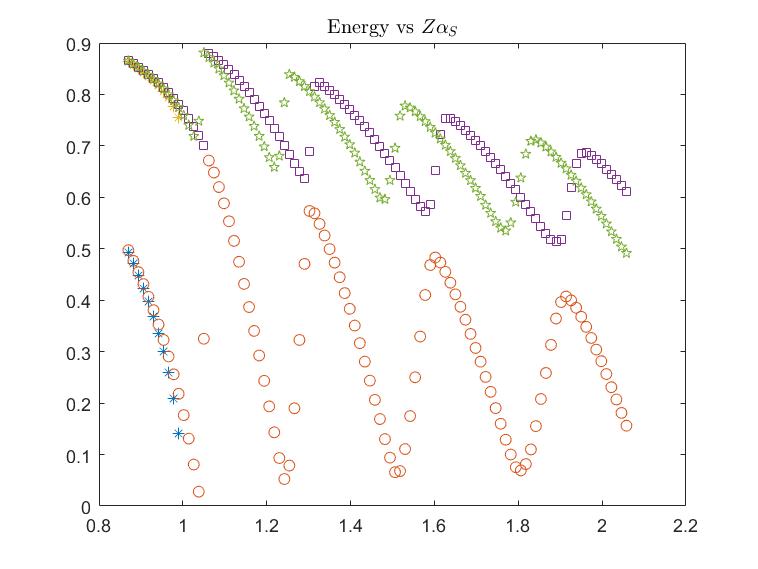}
\caption{Onset of oscillatory behavior of energy eigenvalues for $Z\alpha_S>1$.}
\label{fig:Zdependence}
\end{figure}

Second, in the case of the coulombic Dirac hamiltonian, the energy of the ground state has the following behavior: it decreases to 0 as $-\gamma = Z\al_S \to 1$, and at $-\gamma = 1$ the ground state energy curve (represented by blue stars in Fig.~\ref{fig:Zdependence}) appears to have a vertical tangent, with no further values found for the ground state energy when $-\gamma$ is pushed beyond 1. 

The ground state energies of our Dirac on z$G$KN hamiltonian, represented by the circles in Fig.~\ref{fig:Zdependence}, also follow this pattern, but then begin to deviate from it  in two ways: they continue to exist past $-\ga = 1$, and the energy value starts to oscillate 
as $-\ga$ is increased further and further, apparently with a damped amplitude, thus perhaps converging to a limiting positive value as 
$|\gamma|\to\infty$ (assuming a discrete spectrum exists for all $|\gamma|>1$).

The same pattern of behavior is observed for the excited state energies as $-\gamma$ is pushed beyond 1.  In Fig.~\ref{fig:Zdependence} the energies of the first excited states of the coulombic Dirac hamiltonian, i.e. $2s_{1/2}$ (which because of the accidental degeneracy is the same as the energy of $2p_{1/2}$) are depicted by the orange stars, which cease to exist past the critical value of $\gamma$.  By contrast, in our z$G$KN case these two excited states continue to exist in the supercritical range, and exhibit a similar oscillatory behavior as the ground state did.  The energies of $2s_{1/2}$ orbitals are represented by purple squares, and that of $2p_{1/2}$ by green 5-point stars.  Due to this oscillation, these two energy levels criss-cross one another, resulting in a Lamb-shift-like value that is sometimes positive and sometimes negative.

\subsection{The $a$-dependence of the spectrum}
At a formal level, the limit $a\to 0$ of the radial hamiltonian \eqref{eq:Hrad} coincides with the coulombic Dirac hamiltonian on Minkowski space.  This is of course purely formal, since the domain of these two hamiltonians is vastly different.  Nevertheless, one may expect that for small values of the parameter $a$, the spectrum of Dirac hamiltonian on z$G$KN is not that different from that of the standard coulombic Dirac problem, except that it must be doubled up since we know that the former is symmetric with respect to zero.  Because the parameter $a$ shows up in many different places in \eqref{eq:Hrad}, it is possible that changing this parameter causes many different effects to happen at the same time, so that our initial intuition
about $ea$ representing the anomalous magnetic moment is conceivably to be revised.  
 In particular, we recall that $a$ is hidden in the definition of $\varpi$, and we also know that $\lambda$ must depend on $a$.

\begin{figure}[h!]
\centering
\includegraphics[scale=0.4]{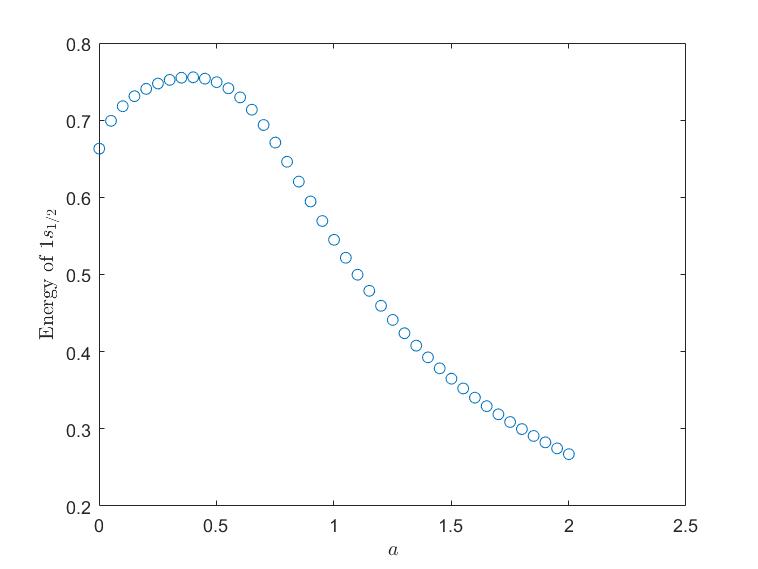}
\caption{Ground state energy plotted against $a$.}
\label{fig:Evsa}
\end{figure}

\begin{figure}[h!]
\centering
\includegraphics[scale=0.4]{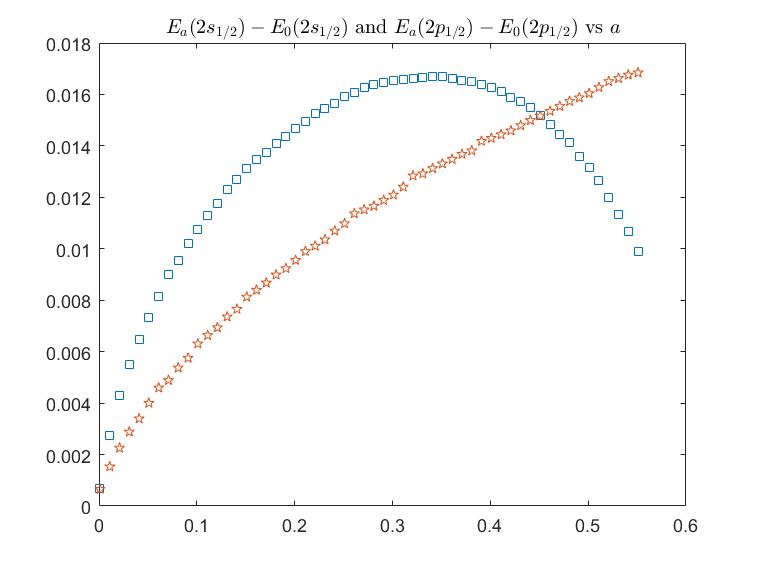}
\caption{Difference between energy of the first two excited states for $a>0$ and for $a=0$, plotted against $a$.}
\label{fig:exciteda}
\end{figure}

We first study the dependence of the ground state energy on $a$.  In Fig.~\ref{fig:Evsa} we have plotted the ground state energy for $0<a<2$. The limiting value at $a=0$ appears to coincide with the ground state energy of the coulombic Dirac problem.  The dependence on $a$ seems to follow a power law with exponent $< 1$ as $a \searrow 0$. In the other direction, as $a$ increases, a maximum appears, after which the energy decreases.   
We can also perform a similar study for the differences of the energies of the excited states $2s_{1/2}$ and $2p_{1/2}$ computed for $a>0$, from their coulombic Dirac values (i.e. for $a=0$).  The result is shown in Fig.~\ref{fig:exciteda}. Note that the corresponding energy levels for the coulombic Dirac problem are the same, due to the aforementioned  accidental degeneracy, so the same number is subtracted from both of these.   The fact that both curves appear to go through the origin supports our speculations about the $a\to 0$ limiting.  Moreover, we observe that the two curves cross and therefore the corresponding Lamb-shift-like difference between these two energy levels changes sign as $a$ is increased.

\section{Summary and Outlook}

In this paper we have supplied a complete characterization of the discrete spectrum of a Dirac electron in the zero-$G$ Kerr--Newman
spacetime, though only under certain smallness conditions on the electric coupling constant and on the radius of the ring singularity.
 This improves on our previous works \cite{KTZzGKNDa}, \cite{KTZzGKNDb}, \cite{KTZzGKNDc} where it was shown that the discrete
spectrum is non-empty for more restrictive conditions on the parameters of the problem. 
 The discrete spectrum is associated with a triplet of integers.
 One set of integers is the familiar enumeration of the azimuthal modes, associate with the axisymmetry. 
 The other two integer families are associated with topological winding numbers of a coupled system of dynamical equations
on finite cylinders.
 We have provided a dictionary for how our characterization maps into the familiar hydrogenic Dirac spectra in the small radius
regime of the ring singularity. 
 We have also provided some numerical illustrations of the low-lying eigenvalues as functions of the coupling constant and of the
radius of the ring singularity, obtained by implementing the algorithm of our constructive proof. 
 Empirically, the algorithm converges outside of the estimated convergence regime, demonstrating that our estimates are far from sharp.
 Interestingly, the numerically obtained results show some never-before-seen oscillatory behavior of the Dirac eigenvalues for large
coupling constants beyond the currently assessible regime of experimental data. 
 
Future work should improve on our assumptions on the parameters of the problem and obtain sharp results.
 
 Although our constructive proof relies on the iteration of two dynamical systems that determine the pairs $(\lambda, E)$ jointly, 
and which is instrumental in identifying the doubly integer labels for $(\lambda,E)$, from the computational point of view it is of
interest to also try to compute the energy eigenvalues $E$ directly from the dynamical $\Omega$ system, with the angular 
eigenvalues $\lambda$ expressed as analytical functions of $E$ in the manner determined in \cite{BSW}. 
 Whether this has computational advantages or not is an interesting question.
 
 In the future we plan to rigorously investigate the $a \to 0$ limit and hopefully show that the $a \to 0$ limit of the z$G$KN spectrum converges to the Bohr-Sommerfeld spectrum as suggested in Figures \ref{fig:Evsa} and \ref{fig:exciteda}. In fact, due to the symmetry of the spectrum, we expect to get two copies of the Bohr-Sommerfeld spectrum. One copy corresponds to a negative point charge $-e$ in the $r > 0$ sheet, while the other copy corresponds to a positive point charge $+e$ in the $r < 0$ sheet. An interpretation of this negative spectrum is given in \cite{KTZzGKNDc}. 

The KN electromagnetic field can also be generalized to exhibit what elsewhere we have called a \emph{KN-anomalous magnetic moment},
\beq\label{def:AKNanomalous}
\bAKNanom = - \frac{r}{r^2+a^2\cos^2\theta} \left(\textsc{q}dt - \frac{\textsc{i}\pi a^2}{c^2} \sin^2\theta\, d\varphi\right),
\eeq
with which one can decorate the zero-$G$ Kerr spacetime of the same ring radius $a$; here,
$\textsc{i}$ is the electrical current which produces a magnetic dipole moment $\textsc{i}\pi a^2/c$ when viewed from
spacelike infinity in the $r>0$ sheet.
 Although no spacetime with such a generalized KN field has yet been found for when $G>0$, it is an interesting open question
whether one exists. 
 In any event, in the zero-$G$ limit this family of spacetimes does exist, and it is not yet known whether the Dirac operator is
essentially self-adjoint or not on its minimal domain. 
 Clearly there are plenty of intriguing questions that await a rigorous mathematical treatment.
 
\medskip

\section*{Acknowledgments} 
Eric Ling gratefully acknowledges being supported by the Harold H. Martin Postdoctoral Fellowship at Rutgers University. Shadi Tahvildar-Zadeh thanks Prof. Siavash Shahshahani for pointing out Ref. \cite{QTPDS}.

 \newpage
 \appendix
 
\section{Errata for reference \cite{KTZzGKNDa}}
While working on this follow-up paper to \cite{KTZzGKNDa} we found several sign slip-ups and some other blunders
that we correct in this list of errata:
\begin{itemize}
\item[-] P.16, Eq.(95): $e$ should be replaced with $-e$.
\item[-] P.16, Eq.(101): $m$ should be replaced with $-m$.
\item[-] P.16, Eq.(102): change the signs of all four matrix entries.
\item[-] P.16, Eq.(103): $i$ should be replaced with $-i$. 
\item[-] P.16, Eq.(104): change $\mp$ into $\pm$.
\item[-] P.17, first two unnumbered equations: change $\la$ to $-\la$.
\item[-] P.17, Eq.(108): exchange $u$ and $v$, and as a result:
\item[-] P.17, Eq.(109): the r.h.s. should be replaced with $\frac{1}{\sqrt{2}}\begin{pmatrix} 1 & -i\\ 1 & i \end{pmatrix}$.
\item[-] P.18, Eq.(114): change $R = \half\sqrt{u^2+v^2}$ to $R = \frac{1}{\sqrt{2}}\sqrt{u^2+v^2}$.
\item[-] P.18, text after Eq.(114): change to $R_1 = Re^{-i\Om/2}$ and $R_2 = Re^{i\Om/2}$.
\item[-] P.18, Eq.(115): the r.h.s. should be replaced with eq. (\ref{ontology}) in this paper.
\item[-] P.19, Proposition 7.2.: change $|\kappa|\geq \half$ to $\kappa\geq \half$. (The case of $\kappa<0$ requires a different boundary condition than (123)).
\item[-] P.20, in the THEOREM: replace $(-\infty,b]\cup[a,\infty)$ by $(-\infty,b)\cup (a,\infty)$.
\item[-] P.21, above Eq.(131): the fundamental domain is fixed to be $[x_-, x_+] \times [-\pi, \pi)$. In this paper we let the fundamental domain be arbitrary $[x_-, x_+] \times [y_0, y_0 + 2\pi)$. 
\item[-] P.21, below Eq.(133): replace $s_- < n_-$ with $s_- > n_-$ so that it agrees with assumption (c) in section \ref{assumptions on flow section}. 
\item[-] P.21, Eq.(137): purge ``$(\text{mod}\: 2\pi)$''.
\item[-] P.24, below Eq.(142): it should read $-\frac{\pd g/\pd \mu}{\pd g/ \pd y} \leq 0$. 
\item[-] P.27, the expression for $q(T)$: the first $-$ should be a $+$ and the last $+$ should be a $-$. The conclusion ``$\lambda < -a"$ on page 28 is no longer true given this correction.
\item[-] P.32, section 10: $n_-$ should be replaced with $-\pi - \cos^{-1}(E)$. 
\item[-] P.36, the second $\leq$ after $h(\xi)$: this is not valid since $\l$ is negative. Proposition \ref{horiz barr prop} in this paper replaces Proposition 7.10 in \cite{KTZzGKNDa}.

\item[-] P.36, when examining $j'(\xi_0)$: $\frac{\pi}{2}$ needs to be added to $\xi_0$. Proposition \ref{Omega winding leq 0 prop} in this paper replaces Proposition 7.11 in \cite{KTZzGKNDa}.

\end{itemize}

\newpage
\providecommand{\bysame}{\leavevmode\hbox to3em{\hrulefill}\thinspace}
\providecommand{\MR}{\relax\ifhmode\unskip\space\fi MR }
\providecommand{\MRhref}[2]{%
  \href{http://www.ams.org/mathscinet-getitem?mr=#1}{#2}
}
\providecommand{\href}[2]{#2}


\begin{thebibliography}{10}


\bibitem{Appell}
  P. Appell, 
  \textit{Quelques remarques sur la th\'eorie des potentiels multiforms.}
         {Math. Ann.} \textbf{30} ({1887}), {155--156}.
\bibitem{BSa}
   D. Batic and H. Schmid, 
   \textit{The {D}irac propagator in the {K}err--{N}ewman Metric.}
   {Prog. Theor. Phys.} \textbf{116} (2006), 517--544.
\bibitem{BSb}
   D. Batic and H. Schmid, 
   \textit{Chandrasekhar Ansatz and the generalized total angular momentum operator for the {D}irac equation in the {K}err--{N}ewman metric.}
   {Revista Colomb. Mat.} \textbf{42} (2008), 183--207. 
\bibitem{BSW}
 D.~Batic, H.~Schmid, and M.~Winklmeier, 
   \textit{On the eigenvalues of the {C}handra\-sekhar--{P}age angular equation.}
   {J. Math. Phys.} \textbf{46} (2005),  no.~1,  012504 (35).
\bibitem{Bel98}
  F. Belgiorno, 
  \textit{Massive Dirac fields in naked and in black hole Reissner--Nordstr\"om manifolds.}
  Phys. Rev. D \textbf{58} (1998), 084017 (8).
\bibitem{BelMar99}
  F. Belgiorno and M. Martellini, 
  \textit{Quantum properties of the electron field in Kerr--Newman black hole manifolds.} 
  Phys. Lett. B \textbf{453} (1999), 17--22.
\bibitem{BMB00}
  F. Belgiorno, M. Martellini, and M. Baldicchi,
  \textit{Naked Reissner-Nordstr\"om singularities and the anomalous magnetic moment of the electron field.}
   Phys. Rev. D \textbf{62} (2000), 084014. 
\bibitem{BeSaBOOK}
  {H. A. Bethe and E. E. Salpeter},
  \textit{Quantum {M}echanics of {O}ne- and {T}wo-{E}lectron {A}toms.}         {Plenum Press}, {1977}.
\bibitem{BrillCohen66}
D.~R. Brill and J.~M. Cohen, \emph{Cartan frames and the general relativistic
  {D}irac equation}, J. Math. Phys. \textbf{7} (1966), no.~2, 238--243.
\bibitem{Car68}
  B. Carter, 
  \textit{Global structure of the {K}err family of gravitational fields},
  {Phys. Rev.}, \textbf{174} ({1968}), {1559--1571}.
\bibitem{CarMcL82}
  B. Carter and R. G. McLenaghan, 
  \textit{Generalized master equations for wave equation separation in a {K}err or {K}err--{N}ewman black hole background},
  pp. {575--585} in ``{Proceedings of the 2nd {M}arcel {G}rossmann meeting on general relativity},''
  R. Ruffini (ed.), {North-Holland Publishing Company}, {1982}.
\bibitem{CacDor}
 \! A.\! C\'aceres\! and\! C.\! Doran,\!
  \textit{Minimax\! determination\! of\! the\! energy\! spectrum\! of\! the\! Dirac\! equation\! in\! a\! Schwarzschild\! background.}\!
  Phys.\! Rev.\! A\textbf{72}\! (2005),\! 022103\! (7).
\bibitem{Chandra76a}
  S. Chandrasekhar,
  \textit{The solution of {D}irac's equation in {K}err geometry.}
         {Proc. Roy. Soc. London Ser. A} \textbf{349} (1976), {571--575}.
\bibitem{Chandra76b}
  S. Chandrasekhar,
  \textit{Errata: The solution of {D}irac's equation in {K}err geometry.}
         {Proc. Roy. Soc. London Ser. A} \textbf{350} (1976), 565.
\bibitem{ChandraBHbook}
  S. Chandrasekhar,
  \textit{The mathematical theory of black holes},
         Oxford Univ. Press, New York (1983).
\bibitem{CohPow}
  J. M. Cohen and R. T. Powers,
  \textit{The general relativistic Hydrogen atom},
         {Comm. Math. Phys.} \textbf{86} (1982), {69--86}.

\bibitem{Dir28}
  Dirac, P.~A.~M.,
  \textit{The Quantum Theory of the Electron},
  Proc. Roy. Soc. Lond., A \textbf{117}, 610--624 (1928); 
  \textit{The Quantum Theory of the Electron. Part II},
  Proc. Roy. Soc. Lond., A \textbf{118}, 351--361 (1928).  


\bibitem{QTPDS}
F. 
Dumortier, 
J. 
Llibre, 
and J. 
~C. Art\'{e}s, \emph{Qualitative theory of planar differential systems}, Universitext, Springer-Verlag, Berlin, 2006.

\bibitem{Evans}
  G. C. Evans, 
  \textit{Lectures on Multiple-Valued Harmonic Functions in Space},
         {Univ. of California Press}, {Berkeley and Los Angeles}, (1951).

\bibitem{FKSYinCPAM2000a}
  F. Finster, N. Kamran, J. Smoller, and S.-T. Yau,
  \textit{Nonexistence of time-periodic solutions of the {D}irac equation in an axisymmetric black hole geometry,}
         {Comm. Pure Appl. Math.} \textbf{53} (2000), {902--929}.
\bibitem{FKSYinCPAM2000b}
\bysame,
	  \textit{Erratum: Nonexistence of time-periodic solutions of the {D}irac equation in an axisymmetric black hole geometry.}
	  {Comm. Pure Appl. Math.} \textbf{53} (2000), {1201}.
\bibitem{FKSYinCMP2002}
\bysame,	  \textit{Decay rates and probability estimates for massive {D}irac
              particles in the {K}err--{N}ewman black hole geometry.}
	  {Comm. Math. Phys.} \textbf{230} (2002), {201--244}.
\bibitem{FKSYinATMP2003}
\bysame,
	  \textit{The long-time dynamics of {D}irac particles in the {K}err--{N}ewman black hole geometry.}
	 {Adv. Theor. Math. Phys.} \textbf{7} ({2003}), {25--52}.
\bibitem{Fock29}
 V. Fock, \textit{Geometrisierung der Diracschen Theorie des Elektrons},
  {Z. Phys.} \textbf{57} ({1929}), {261--277}.

\bibitem{GairPRIZE}
	J. Gair, \textit{An investigation of bound states in the {K}err--{N}ewman potential},
	Prize Essay, Cambridge Univ. (2001).
\bibitem{Geroch}
	 R. Geroch,
	\textit{Limits of spacetimes}
	Commun. Math. Phys. \textbf{13} (1969), 180--193.

\bibitem{Gor28}
W. Gordon, 
\textit{Die {E}nergieniveaus des {W}asserstoffatoms nach der {D}iracschen {Q}uantentheorie des {E}lektrons},
Z. Phys. {\bf 48} (1928), 11--14.

\bibitem{GreinerETalBOOK}
	W. Greiner, B. M\"uller, and J. Rafelski, 
	  \textit{Quantum electrodynamics of strong fields},
	{Springer}, 1985.
\bibitem{HeineckeHehl}
  C. Heinecke and F. W. Hehl,
  \textit{Schwarzschild and {K}err solutions to {E}instein's field equations: an introduction},
  Int. J. Mod. Phys. D \textbf{24} (2015) 1530006 (78 pages).
\bibitem{Israel}
	 W. Israel, 
	\textit{Source of the {K}err metric},
	{Phys. Rev. D} \textbf{2} (1970), {641--646}.
\bibitem{Kaiser}
  G. Kaiser, 
  \textit{Distributional sources for {N}ewman's holomorphic {C}oulomb field},
         {J. Phys. A} \textbf{37} (2004), {8735--8745}.
\bibitem{KalMil91}
E. G. Kalnins
and
W. Miller Jr.,
 \textit{Series solutions for the Dirac equation in Kerr--Newman space‐time},
         {J. Math. Phys.} \textbf{33} (1992), {286--296}.
\bibitem{KeppelerBOOK}
  S. Keppeler, 
  \textit{Spinning Particles - Semiclassics and Spectral Statistics},
         {Springer}, {2003}.
\bibitem{KTZzGKNDa}
  M. K.-H. Kiessling and A. S. Tahvildar-Zadeh, 
  \textit{The Dirac point electron in zero-gravity Kerr--Newman spacetime.}
         {J. Math. Phys.} \textbf{56} (2015), 042303 (43pp).
\bibitem{KTZzGKNDb}
\bysame,
 \emph{Dirac's point electron in the zero-gravity {K}err--{N}ewman world}, 
  Quantum {M}athematical {P}hysics (Regensburg, Sept. 29--Oct. 2, 2014) (F.~Finster, J.~Tolksdorf, and
  E.~Zeidler, eds.), Birkh\"auser, Basel (2016).
\bibitem{KTZzGKNDc}
\bysame,
  \textit{A novel quantum-mechanical interpretation of the Dirac equation.}
              {J. Phys. A} \textbf{49} (2016), 135301 (57pp).
\bibitem{Lam55}
W. E. Lamb Jr \& R. C. Retherford,
\textit{ Fine structure of the hydrogen atom by a microwave method}.
 Physical Review {\bf 72}:3 (1947), 241 .
\bibitem{LeeSmoothMan}
J. 
~M. Lee, \emph{Introduction to smooth manifolds}, second ed., Graduate Texts in Mathematics, vol. 218, Springer, New York, 2013.

\bibitem{LedZofBic98}
T. Ledvinka, M. \v{Z}ofka, and J. Bi\v{c}\'ak, 
\textit{Relativistic Disks as Sources of {K}err--{N}ewman fields}, in:
{Proceedings of the 8th {M}arcel {G}rossmann meeting on general relativity, Jerusalem (1997)},
R. Ruffini (ed.), {World Scientific, Singapore}, 1998.
\bibitem{LiebLossBOOK}
  E. H. Lieb and M. Loss,
  \textit{Analysis}, 2nd ed.,  Amer. Math. Soc. (2010).
\bibitem{LB}
  D. Lynden-Bell, 
  \textit{Electromagnetic magic: The relativistically rotating disk},
 {Phys. Rev. D} \textbf{70} (2004),  105017 (10).

\bibitem{matlab}
{The Mathworks, Inc.},
{\em MATLAB version 9.10.0.1710957 (R2021a) Update 4}, 
Natick, Massachusetts,
(2021)

\bibitem{McVittie}
  G. C. McVittie, 
  \textit{Dirac's equation in general relativity.}
         {Mon. Not. Roy. Astron. Soc.} \textbf{92} (1932), 868--877.
\bibitem{Mel00}
  F. Melnyk, 
  \textit{Wave operators for the massive charged linear Dirac field on the Reissner--Nordstr\"om metric.}
         {Class. Quantum Grav.} \textbf{17} (2000), 2281-2296.
\bibitem{NewmanETal}
{E. T. Newman, E. Couch, K. Chinnapared, A. Exton, A. Prakash, and R. Tor\-rence},
 \textit{Metric of a Rotating, Charged Mass.}
        {J. Math. Phys.} {\textbf{6}} ({1965}), {918--919}.
\bibitem{Page76}
 D. Page, 
 \textit{Dirac equation around a charged, rotating black hole},
        {Phys. Rev. D} \textbf{14} (1976), {1509--1510}.
\bibitem{Pap56}
 A. Papapetrou, 
 \textit{Rotverschiebung und Bewegungsgleichungen.}
        {Annalen Phys.} \textbf{452} (1956), 214--224.
\bibitem{Pek87}
  C. L. Pekeris, 
  \textit{The nucleus as a source in Kerr--Newman geometry},
         {Phys. Rev. A} \textbf{35} (1987), {14--17}.
 
\bibitem{Perko}
L. 
Perko, \emph{Differential equations and dynamical systems}, 3rd ed.,
  Texts in Applied Mathematics, vol.~7, Springer-Verlag, New York, 2001.

\bibitem{Pru26}
H.~Pr{\"u}fer, \emph{Neue {H}erleitung der {S}turm-{L}iouvilleschen {R}eihenentwicklung stetiger {F}unktionen}, Math. Ann. \textbf{95} (1926),
  no.~1, 499--518.
\bibitem{Schmid}
   H. Schmid,  \textit{Bound state solutions of the Dirac equation in the extreme Kerr geometry},
   Math. Nachr. \textbf{274--275} (2004), 117--129. 
\bibitem{Erwin32}
 E. Schr\"odinger, 
 \textit{Diracsches {E}lektron im {S}chwerefeld. I.}
        {Sitzber. Preuss. Akad. Wiss.} \textbf{11-12} (1932), 105--128.
\bibitem{Som97}
  A. Sommerfeld, 
  \textit{{\"U}ber verzweigte {P}otentiale im {R}aum},
  {Proc. London Math. Soc.} \textbf{s1-28} ({1896}), {395--429}.
\bibitem{Stueckelberg42}
  {E. C. G. St\"uckelberg},
   \textit{La m\'ecanique du point mat\'eriel en th\'eorie de relativit\'e et en th\'eorie des quanta.}
  {Helv. Phys. Acta} \textbf{15} ({1942}), {23--37}.
\bibitem{SufFacCos83}
K.~G. Suffern, E.~D. Fackerell, and C.~M. Cosgrove, \emph{Eigenvalues of the {C}handrasekhar–-{P}age angular functions}, J. Math. Phys. \textbf{24}
  (1983), no.~5, 1350--1358.
\bibitem{TZeEMBI}
  A. S. Tahvildar-Zadeh, 
  \textit{On the static spacetime of a single point charge.}
         {Rev. Math. Phys.} \textbf{23} (2011), {309--346}.
\bibitem{TZzGKN}
  A. S. Tahvildar-Zadeh, 
  \textit{On a zero-gravity limit of the Kerr--Newman spacetime and its electromagnetic fields.}
         {J. Math. Phys.} \textbf{56} (2015), 042501 (19 pages).

\bibitem{ThallerPAP}
  B. Thaller, 
  \textit{Potential scattering of {D}irac particles.}
         {J. Phys. A: Math. Gen.} \textbf{14} ({1981}), {3067--3083}.
\bibitem{ThallerBOOK}
  B. Thaller, 
  \textit{The {D}irac equation}.
  {Springer-Verlag}, {Berlin}, ({1992}).
  \bibitem{Toop76}
  N. Toop, \textit{The Thermal Radiation of Electrons from a Charged Spinning Black Hole in a Cosmological background} (DAMTP,
Cambridge, 1976).
\bibitem{Wei82}
 J. Weidmann,
  \textit{Absolut stetiges {S}pektrum bei {S}turm-{L}iouville-{O}peratoren und {D}irac-{S}ystemen},
         {Math. Z.} \textbf{180} ({1982}), {423--427}.
\bibitem{Weyl29}
 H. Weyl, \textit{Elektron und Gravitation. I.},
  {Z. Phys.} \textbf{56} ({1929}), {330--352}. 
\bibitem{WinYamA}
  M. Winklmeier and O. Yamada,
  \textit{A spectral approach to the {D}irac equation in the non-extreme {K}err--{N}ewman metric.}
  {J. Phys. A} \textbf{42} (2009), 295204 (15).
\bibitem{WinYamB}
  M. Winklmeier and O. Yamada,
  \textit{Spectral analysis of radial {D}irac operators in the {K}err--{N}ewman metric 
    and its applications to time-periodic solutions.}
   {J. Math. Phys.} \textbf{47} (2006), 102503 (17).
\bibitem{Zecca}
  A. Zecca, 
  \textit{Dirac equation in Schwarzschild geometry.}
         {Nuovo Cim. B} \textbf{113} (1998), 1309--1315.
\bibitem{Zipoy}
  D. M. Zipoy,
  \textit{Topology of Some Spheroidal Metrics.}
         {J. Math. Phys.} \textbf{7} (1966), {1137--1143}.
\end{thebibliography}
\end{document}